%% file: paper.tex
\documentclass[]{fairmeta}
\title{SAM Audio: Segment Anything in Audio}

\author[*]{Bowen Shi}
\author[*]{Andros Tjandra}
\author[*]{John Hoffman}
\author[*]{Helin Wang}
\author[*]{Yi-Chiao Wu}
\author[*]{Luya Gao}
\author[\dagger]{Julius Richter}
\author[\dagger]{Matt Le}
\author[\dagger]{Apoorv Vyas}
\author[\dagger]{Sanyuan Chen}
\author[\ddagger]{Christoph Feichtenhofer}
\author[\ddagger]{Piotr Dollár}
\author[\ddagger]{Wei-Ning Hsu}
\author[\ddagger]{Ann Lee}

\affiliation{Meta Superintelligence Labs}

\contribution[*]{Core contributors (random order from second author onward)}
\contribution[\dagger]{Contributors (random order)}
\contribution[\ddagger]{Project leads (random order)}

\input{macro}

\abstract{
General audio source separation is a key capability for multimodal AI systems that can perceive and reason about sound.  Despite substantial progress in recent years, existing separation models are either domain-specific, designed for fixed categories such as speech or music, or limited in controllability, supporting only a single prompting modality such as text. In this work, we present \samaudio{}, a foundation model for general audio separation that unifies text, visual, and temporal span prompting within a single framework. Built on a diffusion transformer architecture, \samaudio{} is trained with flow matching on large-scale audio data spanning speech, music, and general sounds, and can flexibly separate target sources described by language, visual masks, or temporal spans. The model achieves state-of-the-art performance across a diverse suite of benchmarks, including general sound, speech, music, and musical instrument separation in both in-the-wild and professionally produced audios, substantially outperforming prior general-purpose and specialized systems. Furthermore, we introduce a new real-world separation benchmark with human-labeled multimodal prompts and a reference-free evaluation model that correlates strongly with human judgment. 
}
\date{\today}
\correspondence{Bowen Shi \email{bshi@meta.com}, Andros Tjandra \email{androstj@meta.com}}

\metadata[Demo]{\url{https://aidemos.meta.com/segment-anything/editor/segment-audio}}
\metadata[Code]{\url{https://github.com/facebookresearch/sam-audio}}
\metadata[Website]{\url{https://ai.meta.com/samaudio/}}

\begin{document}

\maketitle

\input{intro}

\input{method}

\input{data}

\input{evaluation}

\input{results}

\input{conclusion}

\input{acknowledgement}

\clearpage
\newpage
\bibliographystyle{assets/plainnat}
\bibliography{paper}

\clearpage
\newpage
\beginappendix

\input{appendix}

\end{document}

%% file: macro.tex
\usepackage{etoolbox}
\usepackage{xspace}
\usepackage{adjustbox}
\usepackage[table,dvipsnames]{xcolor}  % in preamble
\usepackage{amssymb}
\usepackage{tabularx}
\usepackage{longtable}
\usepackage{booktabs}
\usepackage{threeparttable}
\usepackage{array}
\usepackage{pifont}
\usepackage{multirow}
\usepackage{wrapfig}
\usepackage{mdframed}

\newcommand{\cmark}{\ding{51}} % ✓
\newcommand{\xmark}{\ding{55}} % ✗

\newcommand{\samaudio}{\textsc{SAM Audio}\xspace} % name for full system
\newcommand{\samaudiobench}{\textsc{SAM Audio-Bench}\xspace}
\newcommand{\samaudiojudge}{\textsc{SAM Audio-Judge}\xspace}
\newcommand{\R}{\mathbb{R}}
\newcommand{\xtgt}{x_\text{tgt}}
\newcommand{\xres}{x_\text{res}}
\newcommand{\xmix}{x_\text{mix}}
\newcommand{\samsmall}{\textsc{SAM Audio-Small}\xspace}
\newcommand{\sammid}{\textsc{SAM Audio-Base}\xspace}
\newcommand{\samlarge}{\textsc{SAM Audio-Large}\xspace}

\newif\ifdraft  % set \draftfalse for submission
\drafttrue
\ifdraft

    \newcommand{\bsr}[1]{\textcolor{blue}{\sout{#1}}}
    \newcommand{\bsc}[1]{\textcolor{blue}{[BS: #1]}}
    \newcommand{\atj}[1]{\textcolor{purple}{(ATJ: #1)}}

\else

    \newcommand{\bsr}[1]{}
    \newcommand{\bsc}[1]{}
    \newcommand{\atj}[1]{}
\fi

\newcommand{\tablestyle}[2]{%
    \fontfamily{ptm}\selectfont%
    \let\itold\it%
    \def\it{\itold \fontfamily{ptm}\selectfont}%
    \setlength{\tabcolsep}{#1}\renewcommand{\arraystretch}{#2}\centering\kindatiny%
    \let\citeold\cite%
    \renewcommand{\cite}[1]{\normalfont\fontfamily{ptm}\selectfont\tiny\citeold{##1}}%
}
\newcommand{\bigtablestyle}[2]{%
    \fontfamily{ptm}\selectfont%
    \let\itold\it%
    \def\it{\itold \fontfamily{ptm}\selectfont}%
    \setlength{\tabcolsep}{#1}\renewcommand{\arraystretch}{#2}\centering\footnotesize%
    \let\citeold\cite%
    \renewcommand{\cite}[1]{\normalfont\fontfamily{ptm}\selectfont\footnotesize\citeold{##1}}%
}

%% file: intro.tex
\section{Introduction}
\label{section:intro}

Audio source separation aims to decompose a complex sound mixture into individual source tracks corresponding to distinct sound events.
% 
%. 
Separation systems play an essential role in a broad range of real-world applications. In the production and creative domains, they enable sound engineers to isolate and remix individual stems, restore archival recordings, or remove unwanted background noise. In education and accessibility, separation can highlight key sounds or voices for learners or assistive listening devices. From a research perspective, it also serves as a valuable testbed for studying audio understanding in multimodal AI systems, as effective separation requires identifying the individual sources, an essential element of AI models understanding audio.

Audio separation has been studied extensively, and existing approaches can be broadly grouped into \textit{promptless} and \textit{prompted} methods.  
Promptless systems aim to decompose an audio mixture into a fixed set of predefined sources and have shown strong performance in specialized tasks such as speech enhancement, speaker separation, and music demixing~\citep{mdx2021,mdx2023,zhao2024mossformer2,kong2023universal}.  
However, these methods assume a fixed output configuration and rely on predefined taxonomies of sound categories. As a result, they struggle to adapt to open-domain mixtures or user-defined sound types, where boundaries between sound classes are ambiguous and highly context dependent.
Recent progress has shifted the field toward \textit{prompted separation}, in which the target source is specified through an external signal.  
Text prompts~\citep{liu2023separate,flowsep,CLAPSep,wang2025soloaudio,jiarui2024dpmtse} allow users to describe arbitrary sound events (e.g., \textit{``dog barking''}, \textit{``female speech''}) and thereby remove the need for fixed taxonomies.  
Visual prompts~\citep{zhao2018sound,Huang2024high,IIANet,ephrat2018looking,dong2023clipsep} complement text by providing instance-level grounding, enabling disambiguation when multiple similar sources appear in the same scene.  

The prompted separation paradigm greatly expands the applicability of separation, but important challenges remain.
First, current text-prompted systems are largely benchmarked on general sound effects and often struggle in specialized domains such as music or speech. For example, text-based instrument separation significantly lags behind domain-specific methods like Demucs~\citep{rouard2022hybrid,defossez2021hybrid}. While these specialized models perform well within their own domain, they cannot generalize beyond fixed source types.  
Second, visual-prompted systems are comparatively underexplored and typically tested on small or synthetic datasets. Real-world videos contain a mix of on-screen and off-screen sources and ambiguous sound–vision correspondences, making it unclear whether simply selecting a visual region is sufficient for reliable separation.  
Finally, existing models struggle to distinguish subtle but perceptually important differences between similar sounds such as variants of movie sound effects, which are hard to express precisely with text alone.
These challenges point to the need for a more flexible and universally high-performing prompting framework that integrates cues of various modalities to guide separation.

% .  
% .

% % 
% 

% . 
% , 
% % . 
% .
% %  

%.}
Meanwhile, progress in audio separation research has also been slowed by the \textit{lack of} unified \textit{benchmarks} and reliable evaluation \textit{metrics}. Existing text-prompted models are typically evaluated on disparate test sets, often focusing narrowly on environmental sound separation, with limited coverage of speech and music. In contrast, traditional benchmarks such as MUSDB~\citep{musdb18} for instrument separation are restricted to a few predefined stems and are incompatible with prompting. Furthermore, widely used objective metrics such as Signal-to-Distortion Ratio (SDR) rely on synthetic mixtures with clean reference stems, which are scarce or nonexistent in real recordings. Reference-free metrics like CLAP similarity~\citep{wu2023clap} offer some insight into audio–text alignment but correlate poorly with human judgment of separation quality (see Section~\ref{sec:exp_sam_audio_judge}). These limitations make it difficult to compare models fairly and to assess true performance in open-domain scenarios.

In this work, we introduce Segment Anything Model Audio (\samaudio{}), a foundation model for general audio separation that unifies text, visual, and temporal prompting within a single framework. \samaudio{} allows users to specify \textit{what} to separate using a \textit{text} description, where to separate using a \textit{visual} mask via positive/negative clicks, and \textit{when} to separate using a \textit{temporal span }prompt. These modalities can be used independently or jointly, enabling flexible interaction — for example, a user can describe “piano playing” in a music video and highlight the corresponding region to isolate it precisely. At the core of \samaudio{} is a diffusion transformer trained with flow matching on large-scale audio mixtures spanning speech, music, and general sound events. 
During inference, the model simultaneously produces the target stem along with a residual stem capturing all remaining audio content.

% .

\textbf{Contributions.}
Our main contributions are threefold.
\begin{itemize}
    \item We propose \samaudio{}, the first foundation model that supports \textit{multimodal prompting} (text, visual, span) -- used either individually or in combination, for open-domain audio separation, achieving state-of-the-art results for both in-the-wild and professional audios of general and specialized domains (e.g., speech, music).
    \item We introduce \textit{span prompting}, a novel form of temporal conditioning that enables precise frame-level control for audio separation.
    \item We develop a unified and comprehensive separation benchmark, \samaudiobench, that covers major audio domains (speech, music, and sound effects), includes human-labeled multimodal prompts, and provides a reference-free automatic evaluation model with strong correlation to human perception.
\end{itemize}

\section{Related Work}
\label{sec:related_work}

\subsection{Audio Separation Models}
\textbf{Speaker separation.}
Conventional source separation aims to decompose an audio mixture into individual \textit{sources}. 
In speech, this typically involves isolating the utterances of multiple speakers from an audio mixture. 
A dominant line of work formulates this as a discriminative regression problem, where the model predicts a latent-domain mask for each source, and the clean signals are reconstructed by applying these masks. 
% .
Early deep learning models relied on spectrogram masking~\citep{takahashi2018mmdenselstm, choi2021tasnetstft}, while time-domain approaches such as Conv-TasNet~\citep{luo2019conv} demonstrated that directly operating on waveforms yields superior perceptual quality.
% . 
Subsequent dual-path architectures such as DPRNN~\citep{luo2020dualpath}
% , 
and TF-Locoformer~\citep{guso2022tflocoformer} improved efficiency and long-range context modeling by combining intra- and inter-chunk processing.
Transformer-based models further extended these ideas: SepFormer~\citep{subakan2021sepformer} used dual-path self-attention to capture long-term dependencies, while MossFormer2~\citep{zhao2024mossformer2} integrated multi-scale attention and mask refinement to improve robustness and latency in practical deployments.

Another line of research frames separation as a \emph{generative} modeling problem.
Instead of predicting deterministic masks, generative models learn the underlying distribution of source signals conditioned on the mixture.
Early work includes GAN approaches~\citep{subakan2018generative,pascual2017segan}, as well as normalizing-flow models~\citep{zhu2022music}.
Recent methods~\citep{mariani2024multisource,scheibler2024flowmatchingsep,dong2025edsep} leverage diffusion and flow matching~\citep{flow-matching} to handle complex speech overlaps.
SEP-Diff~\citep{chen2023sepdiff} employs a denoising diffusion process in the mel-spectrogram domain for speech separation, while DiffSep~\citep{scheibler2023diffsep} formulates a mixture-conditioned score model in waveform space.
Post-processing models such as Diffiner~\citep{sawata2023diffiner} and Fast-GeCo~\citep{wang2024noise} refine regression-based separators via diffusion.

Beyond blind separation, \emph{target} or \emph{promptable} speaker extraction leverages additional information to identify the desired speaker.
For instance, SpeakerBeam~\citep{delcroix2020speakerbeam}, VoiceFilter~\citep{wang2019voicefilter}, SpEx+~\citep{xu2020spexplus}, and SoloSpeech~\citep{wang2025solospeechenhancingintelligibilityquality} condition on an enrollment utterance to extract speech from a specific speaker, achieving instance-level disambiguation but requiring clean reference samples.
There has been little work on \textit{text-prompted} speaker separation, where the target speaker is specified by semantic attributes (e.g., “female speaker” or “child voice”). 
Such an approach is practically valuable, as it enables flexible and interpretable control without requiring reference audio.

% }.
% }.

\textbf{Audio enhancement.}
Audio enhancement is another domain where separation methods have been widely applied, aiming to remove additive noise or interference while preserving the underlying signal. 
Speech enhancement, in particular, has long served as a front-end for automatic speech recognition and hearing-aid systems~\citep{loizou2013speech,wang2018supervised}. 
Similar to speaker separation, generative modeling has been extensively explored for speech enhancement, improving perceptual quality across diverse acoustic conditions—from early GAN-based approaches~\citep{pascual2017segan,fu2019metricgan} to more recent diffusion-based models~\citep{lu2022conditional,richter2023speech}. 
Music enhancement~\citep{scaffer2022music,Kandpal2022MusicEV} has also been studied to restore noisy or degraded recordings, or as a post-processing step for instrument separation systems, with techniques often adapted from speech enhancement and source separation. 
While these enhancement models typically focus on \emph{restoration}, our objective differs. 
In this work, we focus on \emph{faithful separation}, where the separated track is a content-preserving extraction of the target source without altering intrinsic recording attributes such as echo.

\textbf{Musical instrument separation.}  
Separating a full-mix music audio into individual instrument tracks has been another widely studied problem in the separation literature, drawing many techniques from general source separation.  
For example, Demucs~\citep{defossez2019music} introduced a time-domain U-Net architecture with strided convolutions and learned synthesis, and was later extended with hybrid spectro-temporal paths and transformer modules to better capture long-range harmonic structure and transients~\citep{defossez2021hybrid,rouard2022hybrid}.  \textsc{KUIELab-MDX-Net}~\citep{kim2021kuielab}, features a two-stream architecture in the time-frequency domain blends their outputs to provide a strong accuracy-vs-computational‐cost trade-off.  
More recently, \cite{Lu2023MusicSS} proposes splitting the input complex spectrogram into sub-bands, then applying hierarchical Transformer layers to model both intra-band and inter-band dependencies.  

Nevertheless, most music instrument separation methods remain restricted to a fixed ontology. A commonly used benchmark, \textsc{MUSDB18}~\citep{musdb18}, defines output stems as vocals, drums, bass, and other. Accordingly high-performing models~\citep{rouard2022hybrid,Lu2023MusicSS} typically support only up to 5 stems. This limitation constrains open-domain usability and highlights the need for more flexible, prompt-able separation frameworks.

\textbf{Target sound separation.}  
Most prior work on source separation focuses on specialized domains such as speech or music, while relatively fewer efforts aim to generalize separation to open-domain sound events. Universal Sound Separation~\citep{kong2023universal} explored ontology-driven separation at scale, using weak labels and hierarchical conditioning to disentangle predefined sound classes derived from AudioSet~\citep{gemmeke2017audioset}. However, its reliance on fixed class ontologies limits flexibility and control.  
To overcome this limitation, recent approaches incorporate natural language prompting, enabling open-vocabulary separation. AudioSep~\citep{AudioSep} scales sound separation by leveraging CLAP-aligned audio–text embeddings and large-scale in-the-wild mixtures. CLIPSep~\citep{dong2023clipsep} trains text-prompted separation models using unlabeled videos, conditioning on CLIP~\citep{Radford2021Learning} embeddings to align audio with textual descriptions.  
Parallel to these discriminative approaches, generative formulations have also gained traction. FlowSep~\citep{flowsep} employs rectified flow matching within a latent space conditioned on FLAN-T5 text embeddings~\citep{Raffel2020t5}, while SoloAudio~\citep{wang2025soloaudio} adopts a diffusion-transformer architecture and augments training with synthetic audios.  
Although prompted separation greatly improves flexibility, it still struggles in specialized domains such as professional music demixing or multi-speaker scenes, where domain-specific systems~\citep{rouard2022hybrid} remain superior. Moreover, in many practical cases, text prompts alone are insufficient to disambiguate target sounds—for example, differentiating between subtle sound effects or co-occurring events. While prior work~\citep{wen2025promptsep,xu2020spexplus} explored audio as an alternative conditioning modality, we instead introduce \emph{span prompting}, a temporal conditioning mechanism that enables models to separate sound events without requiring external reference audio.

\textbf{Visual-prompted separation.}  
Beyond text prompting, several works have explored visual inputs as separation cues, allowing models to identify and extract sounds corresponding to specific visual objects or regions. Early work by \citet{zhao2018sound} showed that spatial location can guide separation by associating visual regions with corresponding sound components, effectively separating instruments in music videos. In the speech domain, \citet{ephrat2018looking} demonstrated separating target speaker audio by conditioning on the speaker’s facial regions.  
Recent work such as IIANet~\citep{IIANet} integrates multi-level audio–visual cues through intra- and inter-modality attention mechanisms, improving cross-modal fusion for speaker separation.  
In the general sound and music domains, DAVIS~\citep{huang2025davis} formulates visual-prompted separation as a conditional generative task, employing a diffusion-based framework to synthesize target sounds conditioned jointly on mixture audio and video inputs.  
Despite these advances, visual-prompted separation remains underexplored in real-world, open-domain scenarios. Most prior systems are domain-specific—focusing narrowly on speech or instrument separation—and are typically evaluated on synthetic or small-scale benchmarks.

\subsection{Audio Separation Benchmark and Evaluation}

\textbf{Separation Benchmarks}
A large portion of audio source separation research relies on \emph{synthetic mixtures} with available stems to enable reference-based metrics (SDR/SI-SDR). While these datasets have catalyzed progress (e.g., WSJ0-2mix and WHAM!/WHAMR! for speech)~\cite{hershey2016deep,wham,whamr}, they are limited in realism and domain diversity. Other suites broaden conditions (LibriMix)~\cite{cosentino2020librimix} or emphasize \emph{reference-free} conversational evaluation (AMI, CHiME-5/6, DIHARD, LibriCSS) via DER/WER~\cite{carletta2005ami,barker2018chime5,watanabe2020chime6,ryant2019dihard,chen2020librics}, but lack prompt modality coverage.
For music, MUSDB18/HQ and SiSEC/MDX focus on stem separation (vocals, drums, bass, other)~\cite{musdb18,mdx2021,mdx2023}, with strong systems like HT-Demucs~\cite{rouard2022hybrid}, while Slakh2100/MedleyDB/URMP offer instrument diversity under synthetic or smaller-scale multitracks~\cite{slakh2100,bittner2014medleydb,urmp2018}. General sound benchmarks (FUSS, AudioSep) aim at universal separation but are primarily synthetic~\citep{wisdom2021fuss,AudioSep}. 

Multimodal AV approaches leverage video (AVSpeech, LRS2/LRS3, VoxCeleb; MUSIC/URMP)~\cite{ephrat2018looking,afouras2018lrs3,voxceleb2,zhao2018sound,urmp2018}, and AVS-style benchmarks (AVSBench, LU-AVS) add temporal sound on/off and spatial masks~\cite{avsbench2022,liu2024benchmarking}. Generative systems (diffusion/flow) further advance AV separation~\cite{huang2024davis,flowsep}. However, most benchmarks (i) focus on narrow taxonomies, (ii) rely on synthetic/studio stems, and (iii) evaluate a \emph{single} prompt modality.

\textbf{Separation Evaluation} Traditional evaluation of audio separation systems relies on distortion-based metrics such as Signal-to-Distortion Ratio (SDR), Scale-Invariant SDR (SI-SDR), Signal-to-Interference Ratio (SIR), and Signal-to-Artifacts Ratio (SAR)~\citep{vincent2006performance,leroux2019sdr}. These measures quantify energy differences between separated outputs and reference signals, and have been widely adopted in benchmark datasets such as WSJ0-2mix~\citep{hershey2016deep} and MUSDB18~\citep{stoter2019open}. However, they often fail to reflect perceptual quality: two outputs with similar SDR values can sound drastically different~\citep{leroux2019sdr}, and their correlation with human Mean Opinion Scores (MOS) is known to be weak~\citep{cartwright2018crowdsourced,cano2016evaluation}.

Subjective listening tests remain the gold standard for evaluation~\citep{ITU_P800}, but they are expensive, time-consuming, and difficult to scale. This creates a persistent gap between easily computed distortion errors and perceptually meaningful assessments. To narrow this gap, perceptual metrics originally designed for speech coding and transmission, such as POLQA~\citep{beerends2013perceptual}, attempt to model auditory mechanisms. While effective in their intended domains, they generalize poorly to the diverse artifacts produced by modern separation systems~\citep{delgado2024towards}.

More recently, data-driven quality predictors have gained traction, which have shown stronger alignment with human judgments~\citep{huang2022voicemos,chinen2020visqol,mittag2021nisqa,manocha2021cdpam,reddy2021dnsmos}. Nonetheless, most of these efforts focus on speech synthesis, enhancement, or audio generation. In source separation, evaluation still largely relies on SDR-like metrics, with listening studies~\citep{jaffe2025musical} confirming their poor correspondence to perceptual judgments. This gap underscores the need for evaluation frameworks that move beyond distortion errors and more faithfully reflect human listening experience.

% .}  
% .  
% .  
% .

% 
% .  
% % . 
% .  
% . 
% . 

% .  

% }
% .
% }.
% [
% ]

% 
% .

% }.
% .

% .} 
% . 
% .
% .
% .
% .
% .
% 

% %
% }
% .
% . 
% .
% ,
% .

%%%%
% }.

% .
% }.
% }.
%
% .

% }
% }

% } 
% }
%  .
%  

%
% .

% 
% .

% 

%% file: method.tex
\section{Approach}
\label{sec:method}

\begin{figure}[h!]
  \centering
 \includegraphics[width=\linewidth]
  {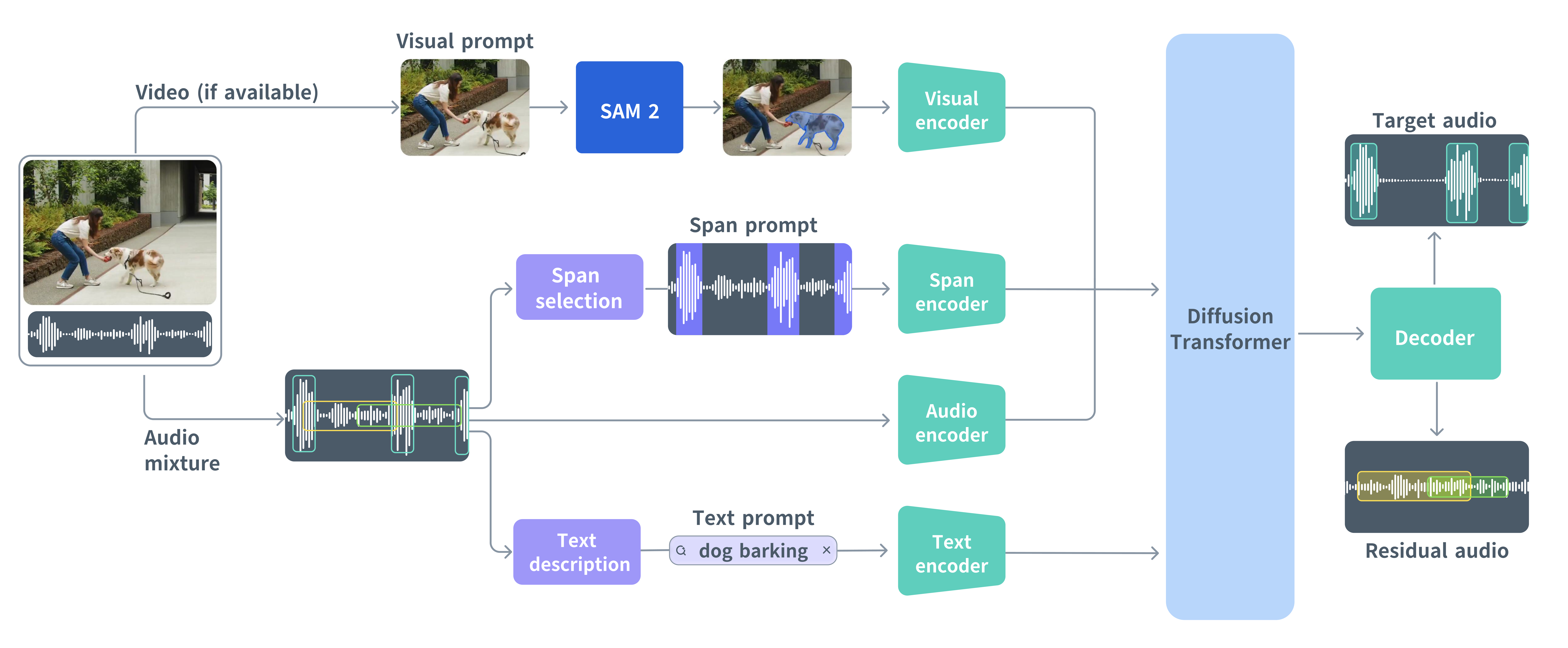}
  % ).}
  \caption{Overview of \samaudio{}. Given an audio mixture, \samaudio{} separates it into target and residual stems, conditioned on any combination of text descriptions (text prompts), visual masks (visual prompts), and temporal intervals (span prompts).}
  \label{fig:sam_audio_model}
\end{figure}

\samaudio{} is a generative separation model that extracts both target and residual stems from an audio mixture conditioned on text, visual and temporal span as prompts (see Figure~\ref{fig:sam_audio_model}). 
% .  
At its core, \samaudio{} employs a flow-matching model built on a Diffusion Transformer~\citep{dit} and operates in a DAC-VAE~\citep{Polyak2024MovieGA} latent space to generate target and residual audio jointly.  
% .  
% . 
% .

% :
%  
%  

\subsection{Model Architecture}
\subsubsection{Flow-Matching with Diffusion Transformer}
Traditionally, audio separation has been formulated as a discriminative learning problem, where a model predicts a mask over the audio spectrogram to extract the target signal from a mixture~\citep{liu2023separate,dong2023clipsep}. However, the task of isolating a target sound is \textit{one-to-many} by nature--given an audio mixture, there often exist multiple plausible sounds, particularly in noisy or overlapping acoustic scenes. 

% .
In \samaudio{}, we adopt a generative modeling approach. Instead of masked prediction, the model learns the target sound distribution conditioned on multimodal prompts, which better captures the underlying variability of sound sources. Specifically,
\samaudio{} follows the flow matching paradigm~\citep{flow-matching}, similar to recent works in text-to-audio generation~\citep{Le2023VoiceboxTM, vyas2023audiobox} and video-to-audio generation~\citep{Polyak2024MovieGA}. 

Our model learns a continuous vector field that transports a Gaussian prior sample $x_0 \sim \mathcal{N}(0, I)$ to the data distribution in latent space over a time variable $t \in [0, 1]$. At each step $t$, the model predicts the instantaneous velocity field $u(x_t, c, t; \theta)$, which is integrated to obtain the sample $x_1$ at $t=1$, where $c$ and $\theta$ are conditions and model parameters respectively. 
% } 
As is shown in recent works for audio generation, flow matching offers better efficiency and performance than diffusion models~\citep{Le2023VoiceboxTM,mehta2024matcha,vyas2023audiobox,prajwalmusicflow,Polyak2024MovieGA}. 
% 
% .

%  

% 

We use a Diffusion Transformer (DiT) based architecture~\citep{dit}, where each transformer block is modulated by the flow time embedding via scale-and-shift operations applied to normalization layers and residual connections. A shared multi-layer perceptron (MLP) maps the scalar $t$ to six modulation parameters (four scales, two biases) used across all transformer blocks, with layer-specific biases added to capture depth-dependent effects. This parameter sharing reduces model size without compromising performance.

We represent audio as a compact sequence of latent features using a separately trained DAC-VAE~\citep{Polyak2024MovieGA}, which adopts a DAC-like~\citep{kumar2024high} autoencoder architecture but replace the residual vector quantizater (RVQ) with a variational autoencoder (VAE)~\citep{kingma2013auto} bottleneck layer.
% }
Each audio clip is encoded into a $T \times C$ sequence at 25 Hz with $C=128$, which offers higher reconstruction quality and lower frame rate compared to commonly used Encodec features~\citep{Defossez2022HighFN}. The residual vector quantizer is removed and the model is trained as a VAE~\citep{kingma2013auto}, yielding smooth Gaussian latents that are well-suited for continuous flow-matching in latent space.

For \samaudio{}, we jointly model the target and residual sounds by concatenating their DAC-VAE features $x_{\text{tgt}}$ and $x_{\text{res}}$ along the channel dimension, forming $x = [x_{\text{tgt}}, x_{\text{res}}] \in \mathbb{R}^{T \times 2C}$. The model then gradually denoises this joint representation, allowing simultaneous prediction of both components in a single pass.

\subsubsection{Prompt Types}

\samaudio{} supports three types of prompts: \textit{text}, \textit{span}, and \textit{visual}, which can be used either individually or jointly to specify the target sound to extract.

\textbf{Text prompt.}  
The text prompt is a free-form natural language description of the target sound. For audio separation, we find that concise noun–verb phrase (NP/VP) descriptions (e.g., \textit{“woman speaking”}) are more effective and natural for users than full sentences (e.g., \textit{“a woman is delivering a speech”}). This format aligns better with common usage in audio editing.

\textbf{Visual prompt.}  
Video is a common source of multimodal audio data, and specifying a visual region offers an intuitive way to isolate a sound source. For example, to extract the sound of a barking dog in a video, the user can simply indicate the region of the dog in the visual frames (see Figure~\ref{fig:sam_audio_model}).  
We adopt SAM 2~\citep{ravi2024sam2} to obtain the target visual mask. Given a video $V \in \mathbb{R}^{T \times H \times W \times 3}$, the user provides a binary mask $M \in \{0,1\}^{T \times H \times W\times 3}$ by interacting with SAM2 through clicks or bounding boxes. \samaudio{} takes $(V, M)$ as input and, for simplicity, only processes the masked visual frames $V \odot M$.

\textbf{Span prompt.}  
Describing arbitrary audio events purely with text can be challenging, particularly for complex soundscapes such as movie soundtrack. To address this, we introduce \textit{span prompting}, a form of temporal conditioning that specifies the time intervals during which the target sound occurs.  
The key intuition is that timestamps can disambiguate overlapping sound events. For example, in an audio mixture containing female speech from 0–6 s and dog barking from 1–2 s, the time interval alone suffices to isolate the barking sound. Formally, a span prompt is defined as a set of time intervals $S \in \mathbb{R}^{k \times 2}$ marking the start and end times of the target event.  
Note span prompting can be inherently ambiguous when multiple sound events occur simultaneously. In practice, it performs particularly well for foreground sounds or when combined with text or visual prompts (see Section~\ref{exp:text_span}).

% .

% .

% .

% .

\subsubsection{Audio Mixture and Prompt Encoding}

To generate the target audio, \samaudio{} conditions on both the input mixture and three prompts. The visual and span prompts are first encoded into frame-aligned feature sequences, which are concatenated with the noisy latents before being passed into the DiT backbone. In contrast, the text prompt is encoded into a global textual embedding that is integrated via cross-attention layers within the DiT backbone. The combination of frame-level and global semantic information help capture heterogeneous cues for separation.

\textbf{Audio encoder.}
The input mixture is encoded with a separately trained DAC-VAE~\citep{Polyak2024MovieGA}, yielding a latent feature sequence $x_{\text{mix}} \in \mathbb{R}^{T \times C}$ at 25~Hz. We directly operate in the DAC-VAE space to minimize information loss in audios, which ensures that the separated output remains faithful to the original audio.

\textbf{Text encoder.}
% . 
We encode the text using a pre-trained T5-base encoder~\citep{Raffel2020t5}, producing a token sequence of features $c_{\text{text}} \in \mathbb{R}^{N_{\text{txt}} \times d_{\text{txt}}}$, with $d_{\text{txt}}=768$. These features are injected into the DiT backbone through cross-attention layers, enabling the model to focus on content that matches the description. As the prompts are short and unambiguous, we do not need to fine-tune the text encoder. %, .

\textbf{Video encoder.}
To provide visual grounding, we condition on frame-level video features extracted by the Perception Encoder (PE)~\citep{bolya2025PerceptionEncoder}, a state-of-the-art vision encoder trained with large-scale contrastive vision-language learning that achieves best-in-class zero-shot accuracy on image and video benchmarks. Compared to CLIP~\citep{Radford2021Learning} or MetaCLIP~\citep{xu2023demystifying}, which are widely used in prior works for video-conditioned audio generation~\citep{Polyak2024MovieGA} or separation~\citep{dong2023clipsep}, PE produces more robust and semantically rich representations, particularly for actions and scene context~\citep{bolya2025PerceptionEncoder}. We extract frame-level PE features and resample them to match the audio frame rate. A gated linear projection layer then maps these features to the DiT dimension, and the resulting sequence is concatenated with the audio representation on a frame-by-frame basis. 
% .

\textbf{Span encoder.}
% . 
% . 
% 
Analogous to the phoneme sequence used in text-to-speech (TTS), we convert the span prompt, originally expressed as a set of time intervals $S \in \mathbb{R}^{k \times 2}$, into a frame-synchronous token sequence $S'_{1:T}$. Each token $s'_t \in \{\texttt{<sil>}, \texttt{+}\}$ denotes whether the target event is silent or active at frame $t$.
The token sequence is embedded using a learnable embedding table and concatenated channel-wise with the audio features. This provides an explicit temporal prior that guides the model to focus on the relevant regions for extraction.

Not all training examples have all three prompt modalities. In such cases, we provide dummy conditions: an empty string for text, an all-zero vector sequence 
% $ 
for video, and a sequence of special null tokens (\texttt{<null>}) for span. To improve robustness, we also \textit{randomly drop} each available prompt during training with probabilities $p_{\text{video}}, p_{\text{text}}, p_{\text{span}}$ and replace them with the corresponding dummy conditions. This encourages the model to robustly handle missing prompts at inference time.
% }

\subsection{Training Objective}

\textbf{Flow Matching Objective.}
At each flow step $t$, the model receives the noisy latent $x_t$ and conditioning set
\[
c = \{x_{\text{mix}}, c_{\text{text}}, c_{\text{vid}}, c_{\text{span}}\},
\]
where $c_\text{text}$ representing text features, $c_\text{vid}$ representing masked video features, and $c_\text{span}$ is the target audio span embedding. The model predicts the velocity field $u(x_t, c, t; \theta)$ used to update $x_t$ toward the clean target-residual representation
\begin{align}
\mathcal{L}_{FM} = \|u(x_t,t,c;\theta)-(x_1-(1-\sigma_{min})x_0)\|,
\end{align}
where $x_1$ and $x_0\sim\mathcal{N}(0, 1)$ are respectively target audio and random noise.
% .

% }

% . 
% 

% . 

\textbf{Audio Representation Alignment.}
%  
% . 
In order to separate correct sound events based on the user prompt, the model  must first infer \emph{what} to extract and \emph{when} it occurs in the mixture. 
To encourage this behavior, we introduce auxiliary MLP layers following \citet{yu2025repa} that project intermediate joint representations, formed by conditioning the mixture audio with the input prompt, into the embedding space of an external audio event detection (AED) model.  
We then apply an auxiliary alignment loss that forces these projected representations to match the AED embedding of the ground-truth target audio.  
By explicitly guiding the hidden states to align with event-level semantics and temporal structure, this objective complements the flow-matching loss and encourages the model to internalize ``what and where to extract'' during training.

For the alignment objective, we first extract the target audio embedding $a_{tgt} = \text{AED}(x_{tgt}) \in \R^{T \times F}$, where $F$ is the feature dimension of the AED module.
Then, with $h_t \in \R^{T \times D} $ being the hidden DiT representation, where $T$ is the input audio length and $D$ is the transformer's channel dimensionality,  we project it to $\hat{a}_t = \phi(h_t) \in \R^{T \times F}$ where $\phi(\cdot)$ is a small MLP. Then, we maximize the cosine similarity between $a_{tgt}$ and $\hat{a}_t$ by minimizing
\begin{align}
    \mathcal{L}_{aux} = \mathbb{E}_{t}\left[ 1 - \text{sim}(\hat{a}_t, a_{tgt}) \right],
\end{align} where $\text{sim}(A,B) = \frac{A \cdot B}{\|A\| \|B\|}$. 

We combine both FM loss and aux loss into: 
\[
\mathcal{L} = \mathcal{L}_{FM} + \lambda * \mathcal{L}_{aux}
\]
where $\lambda$ is a hyperparameter to control the influence of the alignment loss.
We use the AED model in~\citep{kong2020panns} to extract the target audio representation $a_{tgt}$. For auxiliary MLP $aux_{\phi}$, we deployed a 3-layer MLP with 2048 hidden dimensions, GeLU activation function, and layer normalization.

\subsection{Boosting text prompting with span prediction}

Text prompts provide the most accessible interface for specifying the target source, as users can simply describe the desired sound. 
% .
Despite its simplicity, span annotations provide frame-level control when the target sound is active, resulting in more accurate separations.
Empirically, joint text and span conditioning outperforms text-only conditioning for general sound event separation (see Section~\ref{exp:text_span}). 
However, obtaining ground-truth spans requires manual boundary labeling, which can be time-consuming.
To bridge this gap, we propose a simple method for approximating span annotations at inference time. To this end, we use a helper model called $\text{PE}_\text{A-Frame}$ to predict the spans.

Specifically, $\text{PE}_\text{A-Frame}$~\citep{vyas2025peav} is a language-queried sound event detection model that learns to detect the precise time frames in an audio signal where a sound described by free-form text is active.
Given an audio clip and a textual description (e.g.,  ``a dog barking''), the model predicts frame-level probability scores indicating when that event occurs.
It extends CLAP-style audio–text embeddings by introducing a frame-level loss function, enabling fine-grained temporal grounding of language in continuous audio.

Given an audio mixture $x_{\text{mix}}$ and a text prompt $c_\text{text}$, we employ $\text{PE}_\text{A-Frame}$~\citep{vyas2025peav} to estimate the frame-wise activity of the sound event described by the text description. 
By thresholding the frame-wise activity, we obtain an approximate span sequence $\hat{c}_{\text{span}}$.
Similar to~\citet{wang2022improving}, we then condition the separation model on both the predicted span $\hat{c}_\text{span}$ and the original text prompt $c_\text{text}$. 
Formally, given a SAM audio model $\mathcal{M}$ and text conditioning $c_{\text{text}}$, we perform text prompting through $\mathcal{M}(x_{\text{mix}}, c_{\text{text}}, \text{PE}_{\text{A-Frame}}(x_{\text{mix}}, c_{\text{text}}))$
instead of using it $\mathcal{M}(x_{\text{mix}}, c_{\text{text}})$.

\subsection{Longform audio separation with Multi-diffusion}
Processing arbitrarily long audio in a single forward pass is infeasible due to GPU memory limits. A naive chunk-wise approach: splitting the mixture into disjoint segments and applying \samaudio{} independently, introduces boundary artifacts and discontinuities, especially for sustained or slowly varying sounds. To ensure temporal coherence, we adopt the \emph{multi-diffusion} approach~\citep{bar2023multidiffusion,Polyak2024MovieGA} and adapt it for audio separation.

Specifically, we divide the mixture into overlapping windows.  
For window $j$, we construct window-specific conditioning $c^{(j)}=\bigl\{x_{\mathrm{mix}}^{(j)},\, c_{\mathrm{vid}}^{(j)},\, c_{\mathrm{span}}^{(j)},\, c_{\mathrm{text}}\bigr\}$.
At each flow-matching step $t_i$, we solve the ODE in parallel for all windows using the shared time schedule $\{t_i\}$: 
$\tilde{x}_{t_{i+1}}^{(j)}
= x_{t_i}^{(j)} + \Delta t_i\, u\!\left(x_{t_i}^{(j)},\, t_i,\, c^{(j)}\right)$.

We then merge the local predictions into the global latent using normalized soft masks:
\[
x_{t_{i+1}}
= \sum_{j}
\mathrm{pad}\!\left(m^{(j)} \odot \tilde{x}_{t_{i+1}}^{(j)},\, j\right),
\qquad
\sum_j \mathrm{pad}\!\left(m^{(j)},\, j\right) = \mathbf{1},
\]
where $m^{(j)}$ is a triangular window applied to segment $j$, and \texttt{pad} zero-pads it back to global length.

This iterative procedure allows information to propagate across overlapping regions at every diffusion step, producing smooth, globally coherent long-form separations without boundary artifacts.

%% file: data.tex
\section{Data}
\label{section:data}

The training data for \samaudio{} consists of pairs of the form $(x_{\text{mix}}, x_{\text{tgt}}, x_{\text{res}}, c)$, where $x_{\text{mix}}$ is an audio mixture, $x_{\text{tgt}}$ is the target stem, $x_{\text{res}}$ is the residual stem, and $c \in \{\text{text}, \text{video}, \text{span}\}$ is one or more conditioning prompts. This section describes our strategy for constructing such training tuples. We first introduce the audio data used to form $(x_{\text{mix}}, x_{\text{tgt}}, x_{\text{res}})$, followed by how we generate prompts for $c$. As \samaudio{} covers all audio sub-modalities including speech, music and general sound effects, we devise data strategy for each audio sub-modality.
% . }

\subsection{Audios}

\begin{table}[h]
    \centering
    \adjustbox{max width=0.8\textwidth}{
    \begin{tabular}{lccc}
        \toprule
        \textbf{Category} 
         & \textbf{Input Audio Mixture} $\boldsymbol{x_{\text{mix}}}$
         & \textbf{Target Audio} $\boldsymbol{x_{\text{tgt}}}$
         & \textbf{Residual Audio} $\boldsymbol{x_{\text{res}}}$ \\
         \midrule
         Fully-real triplets      & \cmark & \cmark & \cmark \\
         Synthetic mixtures    & \xmark & \cmark & \cmark \\
         Pseudo-labeled stems  & \cmark & \xmark & \xmark \\
         \bottomrule
    \end{tabular}
    }
    \caption{Overview of training data types used in \samaudio{}. 
    Each row indicates whether the mixture, target, and residual signals originate from real audio 
    (\cmark) or are synthetic/pseudo-labeled (\xmark).}
    \label{tab:sam_audio_data_strategy}
\end{table}

Broadly, our training data come from (1) a large-scale, medium-quality audio–video corpus (about $\sim$1M hours) and  
(2) a collection of small- to medium-scale high-quality audio datasets (see Table~\ref{tab:audio_data_source}).  
Only a small subset includes ground-truth stems, which can be directly used for model training. The rest are used to construct synthetic mixtures or pseudo-labeled data depending on whether we synthesize the mixture ($x_{\text{mix}}$) or synthesize the target/residual stems ($x_{\text{tgt}}, x_{\text{res}}$).

Table~\ref{tab:sam_audio_data_strategy} summarizes the three data construction regimes used in \samaudio{}.  
Each regime differs in whether the mixture, target, and residual signals originate from real recordings or are obtained via synthesis or pseudo-labeling.

% .

\begin{table}[h]
    \centering
    \adjustbox{max width=\textwidth}{
    \begin{tabular}{l r c c c c c r r}
        \toprule
         \textbf{Data Source} 
         & \textbf{Modality} 
         & \textbf{w/ stem} 
         & \textbf{Quality} 
         & \textbf{Sound} 
         & \textbf{Music} 
         & \textbf{Speech} 
         & \textbf{\#Samples (M)} 
         & \textbf{\#Hours (K)} \\
         \midrule
         General Video         & audio, video & \xmark & Medium & \cmark & \cmark & \cmark & $\mathcal{O}(100)$ & $\mathcal{O}(1000)$ \\
         General Audio         & audio        & \xmark & Medium & \cmark & \cmark & \cmark & $\mathcal{O}(1)$   & $\mathcal{O}(1)$ \\
         Speech Conversation   & audio        & \cmark & High   & \xmark & \xmark & \cmark & $\mathcal{O}(10)$  & $\mathcal{O}(10)$ \\
         HQ Music              & audio        & \xmark & High   & \xmark & \cmark & \xmark & $\mathcal{O}(10)$  & $\mathcal{O}(10)$ \\
         Multi-track Music     & audio        & \cmark & High   & \xmark & \cmark & \xmark & $\mathcal{O}(0.1)$ & $\mathcal{O}(1)$ \\
         HQ SFX                & audio        & \xmark & High   & \cmark & \cmark & \cmark & $\mathcal{O}(10)$  & $\mathcal{O}(10)$ \\
         HQ Video              & audio, video & \xmark & High   & \cmark & \cmark & \cmark & $\mathcal{O}(0.1)$ & $\mathcal{O}(0.1)$ \\
         \bottomrule
    \end{tabular}
    }
    \caption{Sources of training data for \samaudio{}.  
    ``w/ stem'' indicates whether ground-truth target/residual stems are provided.}
    \label{tab:audio_data_source}

\end{table}

\subsubsection{Fully-real triplets}
\label{sec:fully_real_triplets}
The ideal form of training data $(x_{\text{mix}}, x_{\text{tgt}}, x_{\text{res}})$ consists of real, perfectly isolated stems that sum to a natural mixture:
$x_{\text{mix}} = x_{\text{tgt}} + x_{\text{res}}$.
Such data provides the cleanest possible supervision, as the model can directly learn to map from a mixture to its constituent sources without synthetic artifacts.

Fully-real triplets are available in music and speech domains.

\textbf{Fully-real music.}
We utilized internal high-quality music data with clear instrument stems (i.e., \textit{Multi-track Music} in Table~\ref{tab:audio_data_source}) to introduce instrument extraction capabilities. This datasets contains 10,610 unique music compositions over a total of 536 hours.
% .}
Each music composition consists of multiple instruments (e.g. drums, bass, guitars) and vocal tracks. We create the triplet $x_{mix}, x_{tgt}, x_{res}$ by mixing those instruments within same composition. Assume $\xmix=\sum_{i=1}^N{x_{i}}$ where $x_{i}$ denotes the audio of each instrument stem, then we create $N$ triplets of different targets, where for each instrument audio $x_i$, we define $\xtgt = x_{i}$ and residual $\xres=\sum_{j=1, j \neq i}^{N} x_j$. To increase the robustness, instead of simply adding each stem, we rescale each instrument stem by SNR $\pm 5$ respect to the target audio.
% 2) . 

\textbf{Fully-real speech.}
We use a conversational speech corpus (i.e., \textit{Speech Conversation} in Table~\ref{tab:audio_data_source}) containing 21{,}910 hours of audio.  
Each conversation provides two clean speaker stems, denoted $x_{\text{spk1}}$ and $x_{\text{spk2}}$.  
We form the mixture as $x_{\text{mix}} = x_{\text{spk1}} + x_{\text{spk2}}$ and construct two training triplets:  
$(x_{\text{mix}}, x_{\text{spk1}}, x_{\text{spk2}})$ and $(x_{\text{mix}}, x_{\text{spk2}}, x_{\text{spk1}})$.  
To improve robustness, we rescale the residual speaker with a randomly sampled SNR offset of $\pm 15$\,dB relative to the target speaker.

% .
% .}

\subsubsection{Synthetic audio mixtures}
\label{subsec:synthetic_audio_mixture}
While real stem data is highly valuable, it is relatively scarce and often domain-specific (e.g., multi-speaker speech data or instrument dataset). Synthetic mixtures consists of mixing two audios randomly, which have been a common training data strategy used in prior works~\citep{AudioSep,flowsep,CLAPSep} for general sound effects. As \samaudio{} covers both special and audio domains, we design noise mixing strategy tailored for each domain.

\textbf{Synthetic noisy music.}
We utilized internal data that has $\sim 20K$ hours of clean music (i.e., \textit{HQ Music} in Table~\ref{tab:audio_data_source}).
%  
% % . 
To create the synthetic mixture, we mix this dataset with non-music audio from the general sound data with SNR $\pm 15$. To improve vocal and background music separation, we categorized our music datasets into vocal and non-vocal datasets with AED~\citep{koutini22passt} and set text prompts accordingly (see Section~\ref{sec:text_prompt_generation}).

\textbf{Synthetic noisy speech.}
For speech extraction, we use the same conversational speech corpus described in Section~\ref{sec:fully_real_triplets}.  
To construct synthetic mixtures, we directly add the two speaker stems from each conversation into a single-channel mixture, producing $x_{\text{speech}} = x_{\text{spk1}} + x_{\text{spk2}}$, which naturally contains both single-speaker and multi-speaker segments.  
For the noise component $x_{\text{noise}}$, we randomly sample a non-speech audio clip from the general sound pool.  
The final mixture is then formed as $x_{\text{mix}} = x_{\text{speech}} + x_{\text{noise}}$.
% }
% }

\textbf{Synthetic general sound.}
We categorize our sound effects data into two primary types: \emph{in-the-wild} sound (i.e., \textit{General Video} and \textit{General Audio} in Table~\ref{tab:audio_data_source})and high-quality sound (i.e., \textit{HQ-SFX} in Table~\ref{tab:audio_data_source}).  
In-the-wild audio consists of recordings captured in uncontrolled environments and typically contains multiple overlapping sound events as well as ambient sound.  
In contrast, pro sound recordings are produced in controlled settings as source material for professional audio creation and they usually contain a single sound event with no background noise.  
For audio mixing, we generate training examples by mixing in-the-wild clips with other in-the-wild clips, and pro sound clips with either other pro sound clips or in-the-wild clips.

% .

% :

% 
    
% : 
% }

% .

% 

\subsubsection{Pseudo-labeling Data Engine}
\label{sec:pl_pipeline}
% \}

\begin{figure}[h!]
  \centering
  \label{fig:pl_pipeline}
    \includegraphics[width=0.93\linewidth]{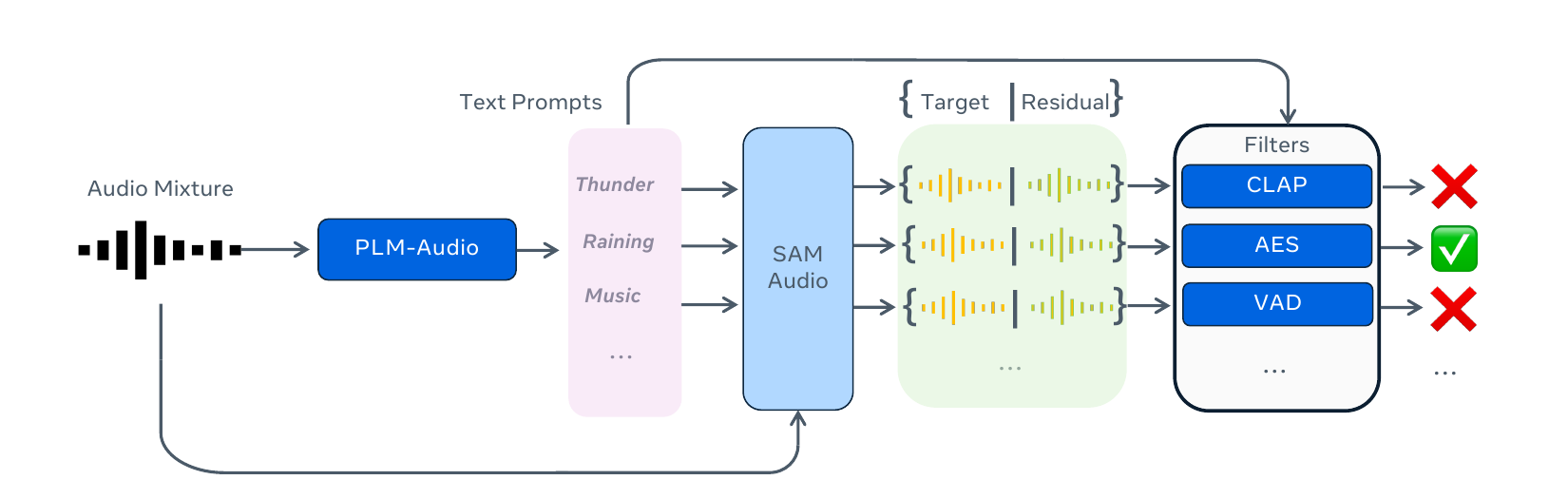}
  \caption{Illustration of our pseudo-labeling data synthesis pipeline. PLM-Audio generates text prompts from mixtures, which guide SAM Audio to produce target/residual stems.  A filter stage retains only high-quality pseudo-labeled stems.}
 
\end{figure}

A limitation of the random mixing strategy described above is that it often produces mixtures that are unrealistic and poorly reflect real-world audio mixture. For example, mixing crowd cheering in a stadium with bird chirps recorded in a forest yields an unnatural combination that rarely occurs in the wild. Training on such mixtures can circumvent the model learning meaningful separation cues.

To address this issue, we synthesize more realistic training tuples by using an intermediate checkpoint of \samaudio{}, trained with the identical recipe but without pseudo-labeled data. We generate target and residual stems from natural audio recordings using this early verison of \samaudio{}, essentially bootstrapping new training data from unlabeled mixtures. 

The data synthesis pipeline consists of two stages: \emph{Separation} with  \samaudio{}  and \emph{filtering} various characteristics of the separated audio using specialist models.  

\textbf{Separation.}
In the separation stage, we run the  intermediate \samaudio{} checkpoint as a data engine to generate target and residual stems from large-scale unlabeled audio recordings, forming a pool of pseudo-labeled training candidates.

We focus primarily on text-based synthesis, given the scalability of text-prompted separation.  
We first apply the audio captioning model PLM-Audio~\cite{vyas2025peav} (see Section~\ref{sec:text_prompt_generation}) to extract textual descriptions for the 1M-hour \emph{in-the-wild} general sound data.  
The output of PLM-Audio is a list of audio event descriptions and we use each one as the text prompt for \samaudio{} pseudo-labeling.
% .

\textbf{Filtering.}
In practice, this preliminary checkpoint exhibits significant variability in separation quality.  
To avoid degrading the final model, we apply strong filtering to remove low-quality candidates using a set of criteria covering various quality aspects.  
We measure text–audio alignment using CLAP~\citep{wu2023clap}, assess audio cleanliness via the production-complexity (PC) axis of the Audiobox-aesthetic model~\citep{tjandra2025meta}, and detect overly silent outputs using a voice-activity detector~\citep{pydub}.
To enhance visual-prompt training, we further curate a subset with strong audio–visual correspondence.  
Given the same text prompt used for separation, we apply an in-house text-prompted video segmentation model to obtain visual masks, and calculates the audio-visual alignment score and mask coverage.
For the former metric, we leverage ImageBind~\citep{girdhar2023imagebind}, a large-scale audio–video–text contrastive model, to compute the cosine similarity between the embeddings of the audio track and masked visual frames.
A pseudo-labeled sample is retained only if all of the following conditions in Table~\ref{tab:pseudo_label_filtering} hold.

\begin{table}[h]
\centering
\small
\begin{tabular}{l c}
\toprule
\textbf{Criterion} & \textbf{Threshold} \\
\midrule
\multicolumn{2}{c}{\textit{Text--Audio Filtering (all must pass)}} \\
\midrule
CLAP(text, target audio) & $> 0.35$ \\
CLAP(text, residual audio) & $< 0.0$ \\
Aesthetic PC score (target audio) & $< 2.5$ \\
Silence ratio (target audio) & $< 95\%$ \\
\midrule
\multicolumn{2}{c}{\textit{Additional Visual-Audio Filtering}} \\
\midrule
Mask coverage ratio (masked region) & $> 0.02$ \\
ImageBind(target audio, masked region) & $> 0.2$ \\
\bottomrule
\end{tabular}
\caption{Filtering criteria for synthesized training data. Samples are kept only if all text–audio criteria are satisfied; visual-prompt samples require additional visual-related constraints.}
\label{tab:pseudo_label_filtering}
\end{table}

Pseudo-labeling is applied to the general video portion of our training data 
(Table~\ref{tab:audio_data_source}). 
We only run pseudo-labeling on mixtures that contain multiple sound events. 
After multi-stage filtering, the resulting pseudo-labeled set is 
substantially smaller than the high-quality audio used for synthetic mixtures.
Table~\ref{tab:audio_pl_data_source} summarizes the final pseudo-labeled datasets 
used for \samaudio{} training.

\begin{table}[h]
    \centering
    \adjustbox{max width=.8\textwidth}{
    \begin{tabular}{l c c c c c r r}
        \toprule
         \textbf{Data Source} 
         & \textbf{Modality} 
         & \textbf{Synthetic Stems} 
         & \textbf{Sound} 
         & \textbf{Music} 
         & \textbf{Speech} 
         & \textbf{\#Samples (M)} 
         & \textbf{\#Hours (K)} \\
         \midrule
         PL-Audio  & audio         & \cmark & \cmark & \cmark & \cmark 
                   & $\mathcal{O}(1)$ & $\mathcal{O}(1)$ \\
         PL-Video  & audio, video  & \cmark & \cmark & \cmark & \cmark 
                   & $\mathcal{O}(0.1)$ & $\mathcal{O}(0.1)$ \\
         \bottomrule
    \end{tabular}
    }
    \caption{Pseudo-labeled training data used in \samaudio{}. 
    Both audio-only and audio--video inputs are processed by an intermediate \samaudio{} 
    checkpoint to produce synthetic stems.}
    \label{tab:audio_pl_data_source}
\end{table}

% 

% .

\subsection{Prompt Generation}

Given an audio tuple $(x_{\text{mix}}, x_{\text{tgt}}, x_{\text{res}})$, we generate a corresponding prompt $c$ that specifies the target audio $x_{\text{tgt}}$. 
This section describes how we generate textual, visual, and span-based prompts.

\subsubsection{Text}
\label{sec:text_prompt_generation}

Generating a text prompt for the target audio is equivalent to producing a concise and semantically faithful description of its content. In the context of separation, we find that prompts expressed in NP/VP form are more intuitive and effective than full natural-language sentences. For example, \textit{``dog barking''} is preferred over \textit{``a dog is barking''} because it more directly specifies the sound event to extract. 

To obtain high-quality prompts at scale, we train an in-house audio–language model, \textit{PLM-Audio}, a visually aware audio-LLM fine-tuned for audio captioning. Whenever raw metadata is available, we merge it with the PLM output, and we discard captions with low audio–text correspondence.

% .

% .}

\textbf{PLM-Audio.}
PLM-Audio is an audiovisual version of the Perception Language Model (PLM)~\citep{cho2025PerceptionLM} which instead of the visual Perception Encoder (PE)~\cite{bolya2025PerceptionEncoder}, uses an audiovisual encoder, PE-AV~\citep{vyas2025peav}. PE-AV extracts frame-level audio-visual representations which are decoded with an 8B-parameter LLaMA decoder~\citep{llama3} for language output. 

We train PLM-Audio using a three-stage process, closely mirroring the PLM~\citep{cho2025PerceptionLM}  training recipe:

\begin{enumerate}
    \item \emph{Warm-up.} We freeze the backbone encoder and fine-tune only a lightweight projection MLP that maps PE-Audio embeddings into the LLM embedding space. This aligns the feature spaces while preserving the pre-trained representations.
    \item \emph{Mid-training.} We unfreeze the full model and fine-tune on a large corpus of synthetic captions generated by PE-Audio, including both audio-only and audio-visual samples. We prioritize utterances with high audio-video alignment scores (e.g., high ImageBind similarity), ensuring that training focuses on well-grounded examples.
    \item \emph{Post-training.} We fine-tune on a curated mixture of tasks that encourage prompt generation in NP/VP format and improve coverage of downstream separation-relevant attributes. 
    % }.
\end{enumerate}

% }

\textbf{Incorporating metadata when available.}
A subset of our training data consists of professionally recorded ambience and sound-effects datasets. These datasets often include metadata such as titles, descriptions, and keyword tags. However, such metadata can be noisy, containing irrelevant, or overly broad information. Meanwhile, PLM-Audio typically produces relevant audio captions but may occasionally emit false-positive sound events. To obtain reliable NP/VP-style prompts, we apply a three-step rewriting pipeline:
\begin{enumerate}
    \item Run PLM-Audio on each audio clip to obtain an \textit{initial caption}.
    \item Use an LLM~\citep{llama3} to merge all available metadata (title, keywords, description) with the PLM-Audio caption and generate a cleaned,\textit{ detailed description} while discarding irrelevant content.
    \item Instruct the same LLM to extract only the \textit{NP/VP-style sound-event phrases }from the detailed description.
\end{enumerate}

\textbf{CLAP filtering.}
Because most of our captions are automatically generated rather than manually annotated, we perform a filtering step to remove low-quality captions. To this end, we compute text--audio similarity using CLAP~\citep{wu2023clap} and discard samples with similarity below $0.28$, which corresponds to approximately the 25th percentile in preliminary experiments. We evaluated several percentile thresholds ($p=\{0.05, 0.10, 0.25, 0.50\}$) and found that filtering at $p=0.25$ achieves the best balance between data scale and caption quality.

% }

% .

Note a small subset of our training data contains domains where quasi–ground-truth text descriptions can be obtained without the above pipeline. For the \textit{fully-real music} subset (Section~\ref{sec:fully_real_triplets}), we derive text prompts using simple templates applied to instrument labels (e.g., \textit{piano} \textrightarrow{} \textit{piano playing}). For the \textit{fully-real speech} subset (Section~\ref{sec:fully_real_triplets}), we apply a pretrained gender classifier~\citep{huh2024voicegender} to identify the speaker's gender and then map it to a templated description (e.g., \textit{female} \textrightarrow{} \textit{woman speaking}).

\begin{figure}[htp]
  \centering
  % }
    \includegraphics[width=0.9\linewidth]{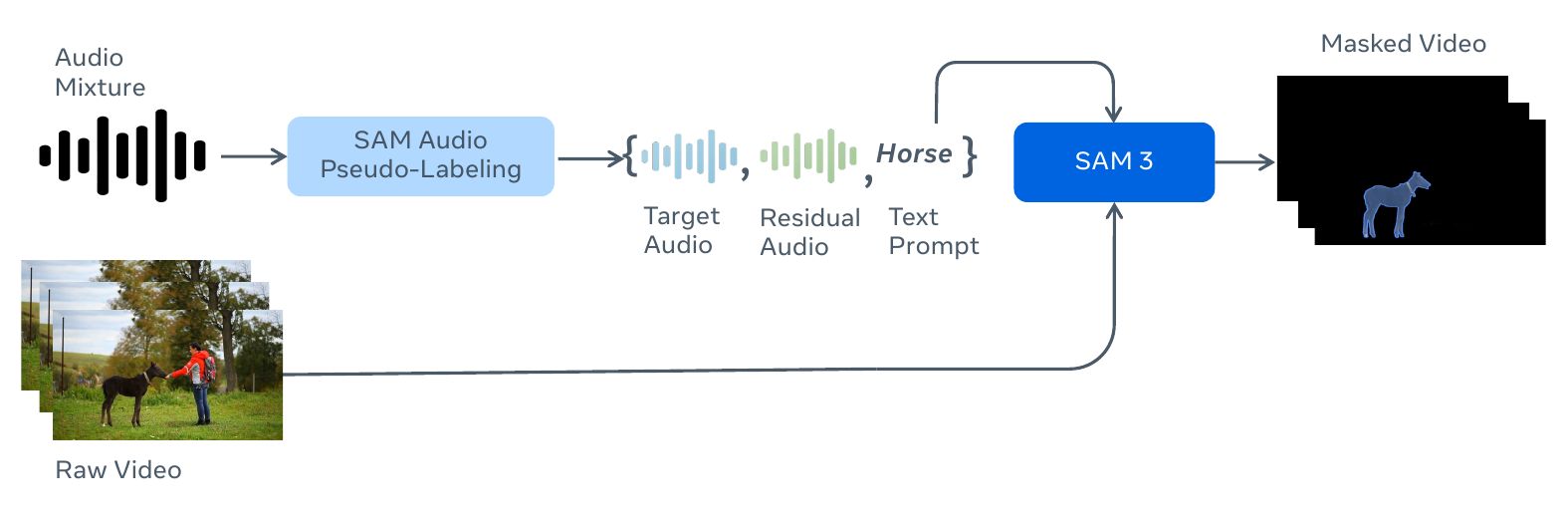}
  \caption{Illustration of pseudo-labeled visual data. The pseduo-labeling pipeline produces a text caption of the target audio, which is used to prompt SAM3 to obtain the visual mask.}
  \label{fig:visual_mask_gen_in_pl}

 \end{figure}
 
\subsubsection{Video}

During inference, visual prompting allows users to specify a target source by providing a visual mask over the video, and the model is required to separate the corresponding sound. The visual mask is generated using SAM 2~\citep{ravi2024sam2}, and covers either a single object or a region of interest across all frames. The primary goal of model training is to encourage the model to associate visually grounded sources to the audio so that, at test time, users can perform instance-level separation by simply selecting an object in the video.

Synthesizing videos for arbitrary audio is challenging, as it requires generating both spatio-temporally coherent frames. We therefore rely exclusively on natural videos with paired audio tracks. Our visual training data construction recipe consists of two complementary components: (a) whole video, (b) visual masks.

\textbf{Whole-video.}
For most of our training data, we directly use full video clips without relying on explicit segmentation masks, allowing the model to learn weakly supervised associations between visual content and audio events. However, many real-world videos contain sound that is not visually grounded—for example, background music added during post-production or off-screen narrations. Training on such samples can be detrimental, as it encourages spurious associations between visual frames and unrelated audio.  
To mitigate this, we compute the ImageBind score~\citep{girdhar2023imagebind} for every video and retain only those exceeding a threshold, biasing the dataset toward diegetic, visually grounded sound sources. This filtering is applied to our $\sim$1M-hour general sound corpus.

\textbf{Visual mask.}
For pseudo-labeled audio–visual data, we explicitly construct a visual mask corresponding to the target sound. Unlike whole-video conditioning, pseudo-labeled clips often contain multiple sounding objects, only a subset of which matches the target audio. We therefore generate masks by prompting SAM3~\citep{carion2025sam3segmentconcepts} with the target sound’s text caption derived from PLM-Audio (see Figure~\ref{fig:visual_mask_gen_in_pl}).  
In practice, we observe substantial variance in mask quality due to factors such as audio–visual mismatch (e.g., the caption describes a sound that is not visually present) and SAM3 prediction errors. Accordingly, the ImageBind-based filtering pipeline described in Section~\ref{sec:pl_pipeline} is critical to ensure that only samples with reliable audio–visual correspondence remain in the final training set.

% % }

\subsubsection{Span}

\begin{figure}[h!]
  \centering
  \includegraphics[width=\linewidth]{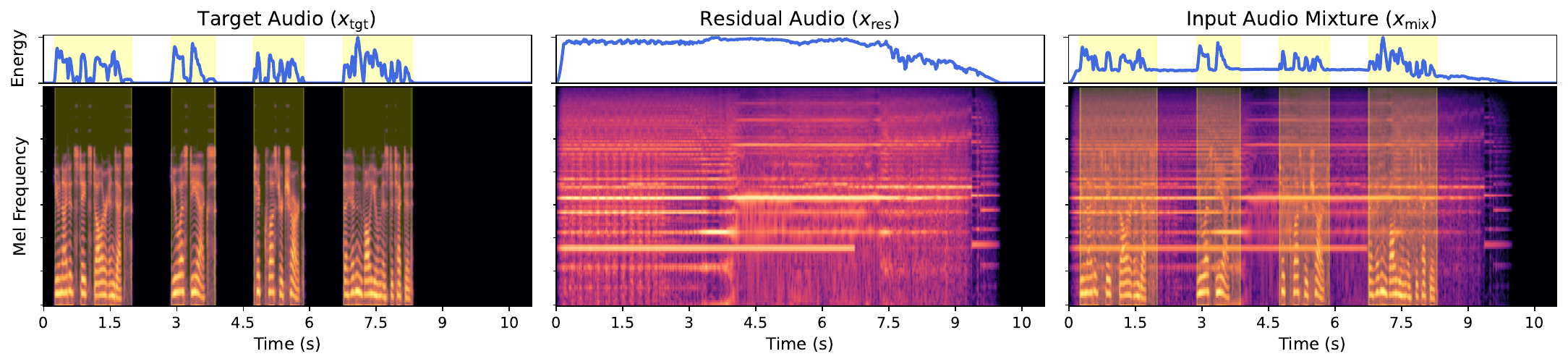}
  \caption{Illustration of span generation. RMS energy (top) and Mel-spectrograms (bottom) for the target $x_{\mathrm{tgt}}$, residual $x_{\mathrm{res}}$, 
and mixture $x_{\mathrm{mix}}$. 
Yellow intervals denote detected spans corresponding to active sound events.}
  \label{fig:span_generation}
 
\end{figure}

Not all audio mixtures are suitable for span prompting. In particular, mixtures dominated by long, continuous ambience (e.g., rain, traffic) offer little temporal structure, making span information uninformative. Our key observation is that span prompting is most effective for \emph{spiky}, discrete sound events—such as \textit{door slam} or \textit{dog bark}—whose temporal locations provide strong cues for separation.

To this end, we construct audio mixtures only from audios that satisfy this property. For general sound, the target audio is sampled from \textit{HQ SFX} sound data (Section~\ref{subsec:synthetic_audio_mixture}), which predominantly contains clean, isolated sound events, while noise audio is sampled from long-duration clips ($>10$~s) that typically contain ambient or continuous backgrounds. For music and speech, we additionally incorporate the fully-real music and fully-real speech datasets to ensure broad domain coverage. 

Given a target audio, we apply VAD~\citep{pydub} with a silence threshold of \(-40\,\mathrm{dBFS}\) and a minimum sounding duration of \(250\,\mathrm{ms}\) to obtain a binary frame-level mask indicating sounding versus silent regions.
Consecutive sounding segments are then converted into time intervals, which we treat as the span prompts supplied to \samaudio{}. An overview of generating span-prompted audio data is shown in Figure~\ref{fig:span_generation}.

% .

% .

% }

%% file: evaluation.tex
\section{Evaluation}
\label{section:evaluation}
\subsection{\samaudiobench}

\subsubsection{\samaudiobench: Unifying Modalities, Domains, and Realism}

We observe three persistent gaps in existing separation benchmarks: (i) realism vs.\ references (in-the-wild acoustics vs.\ availability of stems), (ii) limited \emph{prompt modality} coverage across text/visual/time, and (iii) \emph{cross-domain} breadth under a unified protocol (speech, music, instruments, general sounds).

We introduce a \textbf{real, in-the-wild, multi-modal separation benchmark} addressing these gaps:
\begin{enumerate}
  \item \textbf{Ecological validity}: All items are sourced from in-the-wild audio/video or production-quality video: \textbf{AudioSet}~\citep{gemmeke2017audioset}, \textbf{VGGSound}~\citep{chen2020vggsound}, \textbf{MUSIC}~\citep{zhao2018sound}, \textbf{MUSIC-AVQA}~\citep{musicavqa2022}, \textbf{AVSpeech}~\citep{ephrat2018looking} and \textbf{CondensedMovies}~\citep{rong2020condensedmovies}.
  \item \textbf{Multi-modal prompting}: Each 10\,s test instance includes human annotated \emph{visual} \textbf{SAM masklets}~\citep{kirillov2023sam} (when sounding object is on-screen), \emph{temporal} \textbf{positive/negative spans}, and \emph{language} \textbf{text descriptions}.
  \item \textbf{Taxonomic coverage}: Stem taxonomy is seeded from \textbf{AudioSet} ontologies~\citep{gemmeke2017audioset}, with annotator-extendable classes. Dedicated tasks: \textbf{speech cleaning}, \textbf{speaker separation}, \textbf{music cleaning/removal}, \textbf{instrument stems} (37 classes), and \textbf{general sounds}.

\end{enumerate}
\paragraph{Annotation protocol.} Annotators enumerate sound events with text descriptions (that we subsequently normalize to concise noun/verb phrases), draw visual masklets for visible sounding sources -- these masklets are available for every frame in the video, and every video is standardized to 24fps -- and mark \emph{temporal} presence/absence spans. Each (target sound, video) pair yields interchangeable text/vision/time prompts, enabling controlled ablations across modality combinations.

\subsubsection{Summary Comparison to Existing Evaluation Sets}
Table~\ref{tab:benchmark-comparison} summarizes modality coverage, realism, domain scope, and reference availability across representative benchmarks versus our \samaudiobench.
\begin{table}[htp]
  \centering
  \setlength{\tabcolsep}{3pt}
  \renewcommand{\arraystretch}{1.1}
  \footnotesize
  \begin{threeparttable}
  \caption{Comparison of representative audio separation benchmarks vs.\ \samaudiobench. 
Real: Y/N/M (mixed real+synthetic). Prompts: T=Text, V=Visual, S=Temporal spans.}
  \label{tab:benchmark-comparison}
  \begin{tabularx}{\linewidth}{
    >{\raggedright\arraybackslash}X
    c
    *{5}{c}
    c
    >{\raggedright\arraybackslash}X
  }
    \toprule
    & & \multicolumn{5}{c}{\textbf{Task coverage}} & & \\
    \cmidrule(lr){3-7}
    \textbf{Benchmark} 
      & \textbf{Real} 
      & \textbf{Sp.\ clean} 
      & \textbf{Spkr sep.} 
      & \textbf{General} 
      & \textbf{Mus.\ clean} 
      & \textbf{Instr.\ stems} 
      & \textbf{Prompt(s)} 
      & \textbf{Source / mixtures} \\
    \midrule

    WSJ0-2mix / WHAM! / WHAMR!\tnote{a}
      & N
      & --
      & \checkmark
      & --
      & --
      & --
      & --
      & Synthetic 2-spk mixtures from WSJ0; WHAM/WHAMR add real noise \\
    \midrule

    LibriMix\tnote{b}
      & N
      & --
      & \checkmark
      & --
      & --
      & --
      & --
      & Synthetic speech mixtures from LibriSpeech + noise \\
    \midrule

    DNS / VoiceBank+DEMAND\tnote{c}
      & M
      & \checkmark
      & --
      & --
      & --
      & --
      & --
      & Noisy speech from VoiceBank, DNS challenge (synthetic + some real) \\
    \midrule

    FUSS\tnote{d}
      & N
      & --
      & --
      & \checkmark
      & --
      & --
      & --
      & Synthetic mixtures from diverse sound event stems \\
    \midrule

    MUSDB18 / MDX\tnote{e}
      & M
      & --
      & --
      & --
      & --           % 
      & \checkmark
      & --
      & Studio multitrack music; vocals/drums/bass/other stems \\
    \midrule

    Slakh2100\tnote{f}
      & N
      & --
      & --
      & --
      & --
      & \checkmark
      & --
      & Fully synthetic MIDI-rendered multitrack music \\
    \midrule

    MedleyDB / URMP\tnote{g}
      & M
      & --
      & --
      & --
      & --
      & \checkmark
      & --
      & Real and studio multitrack ensembles, instrument stems \\
    \midrule

    AudioSep (AudioSet / AudioCaps / Clotho mixtures)\tnote{h}
      & N          % 
      & \checkmark
      & --
      & \checkmark
      & \checkmark
      & \checkmark
      & T
      & Synthetic mixtures from AudioSet, AudioCaps, Clotho, etc. \\
    \midrule

    DCASE lang-queried source sep.\tnote{i}
      & M
      & --
      & --
      & \checkmark
      & --
      & --
      & T
      & Synthetic mixtures from captioned / labeled sound events (e.g., FSD50K/AudioSet-style) \\
    \midrule

    CLAP-based text-queried sep.\tnote{j}
      & N
      & --
      & --
      & \checkmark
      & --
      & --
      & T
      & Synthetic mixtures from AudioSet / FSD50K-style datasets \\
    \midrule

    AVSpeech-based AV speech sep.\tnote{k}
      & M
      & \checkmark
      & \checkmark
      & --
      & --
      & --
      & V
      & Synthetic 2-spk mixtures, sources from AVSpeech, LRS2/3, VoxCeleb \\
    \midrule

    MUSIC\tnote{l}
      & M
      & --
      & --
      & --
      & --
      & \checkmark
      & V
      & Real online performance videos; synthetic or curated multi-instrument mixtures \\
    \midrule

    AVSBench / LU-AVS\tnote{m}
      & M
      & --
      & --
      & \checkmark
      & --
      & \checkmark
      & V,S
      & Real YouTube-like videos with object/instrument sounds, AV masks/spans \\
    \midrule

    \textbf{\samaudiobench (ours)}
      & \textbf{Y}
      & \checkmark
      & \checkmark
      & \checkmark
      & \checkmark
      & \checkmark
      & \textbf{T,V,S}
      & \textbf{Real in-the-wild A/V from AudioSet, VGGSound, MUSIC, AVSpeech, CondensedMovies} \\
    \bottomrule
  \end{tabularx}

  \begin{tablenotes}[flushleft]
    \scriptsize
    \item[a] WSJ0-2mix; WHAM!/WHAMR!\ \citep{hershey2016deep,wham,whamr}.
    \item[b] LibriMix \citep{cosentino2020librimix}.
    \item[c] VoiceBank+DEMAND; DNS challenge datasets \citep{valentini2016voicebank,dns2020}.
    \item[d] FUSS \citep{wisdom2021fuss}.
    \item[e] MUSDB18/HQ; SiSEC/MDX \citep{musdb18,mdx2021,mdx2023}.
    \item[f] Slakh2100 \citep{slakh2100}.
    \item[g] MedleyDB; URMP \citep{bittner2014medleydb,urmp2018}.
    \item[h] AudioSep benchmark from mixtures of AudioSet, AudioCaps, Clotho, etc.\ \citep{AudioSep,audiocaps2019,drossos2020clotho,gemmeke2017audioset}.
    \item[i] DCASE language-queried source separation benchmark (e.g., DCASE 2024 Task 8).
    \item[j] CLAP-based text-queried separation benchmarks \citep{CLAPSep}.%,wu2023clap}.
    \item[k] AV speech separation on AVSpeech, LRS2/LRS3, VoxCeleb \citep{ephrat2018looking,afouras2018lrs3,voxceleb2}.
    \item[l] MUSIC is a canonical AV instrument separation dataset, used by \citet{zhao2018sound}.
    \item[m] AVSBench; LU-AVS \citep{avsbench2022,liu2024benchmarking}.
    \item[] Abbrev.: Real Y/N/M; Sp.\ clean = speech cleaning; Spkr sep.\ = speaker separation; 
    General = open-domain / general sound separation; Mus.\ clean = music cleaning; 
    Instr.\ stems = instrument stem separation; Prompts: T=text, V=visual, S=temporal spans.
  \end{tablenotes}
  \end{threeparttable}
\end{table}

As Table~\ref{tab:benchmark-comparison} shows, \samaudiobench uniquely combines (i) \emph{real} in-the-wild audio/video, (ii) \emph{multi-modal prompts} (text, visual masklets, temporal spans) on the \emph{same} items, (iii) \emph{cross-domain} coverage (speech, music, instruments, general sounds), and (iv) \emph{reference-free} \emph{human} evaluation.

% }
% 

\begin{figure}
\centering
\includegraphics[width=0.9\textwidth]{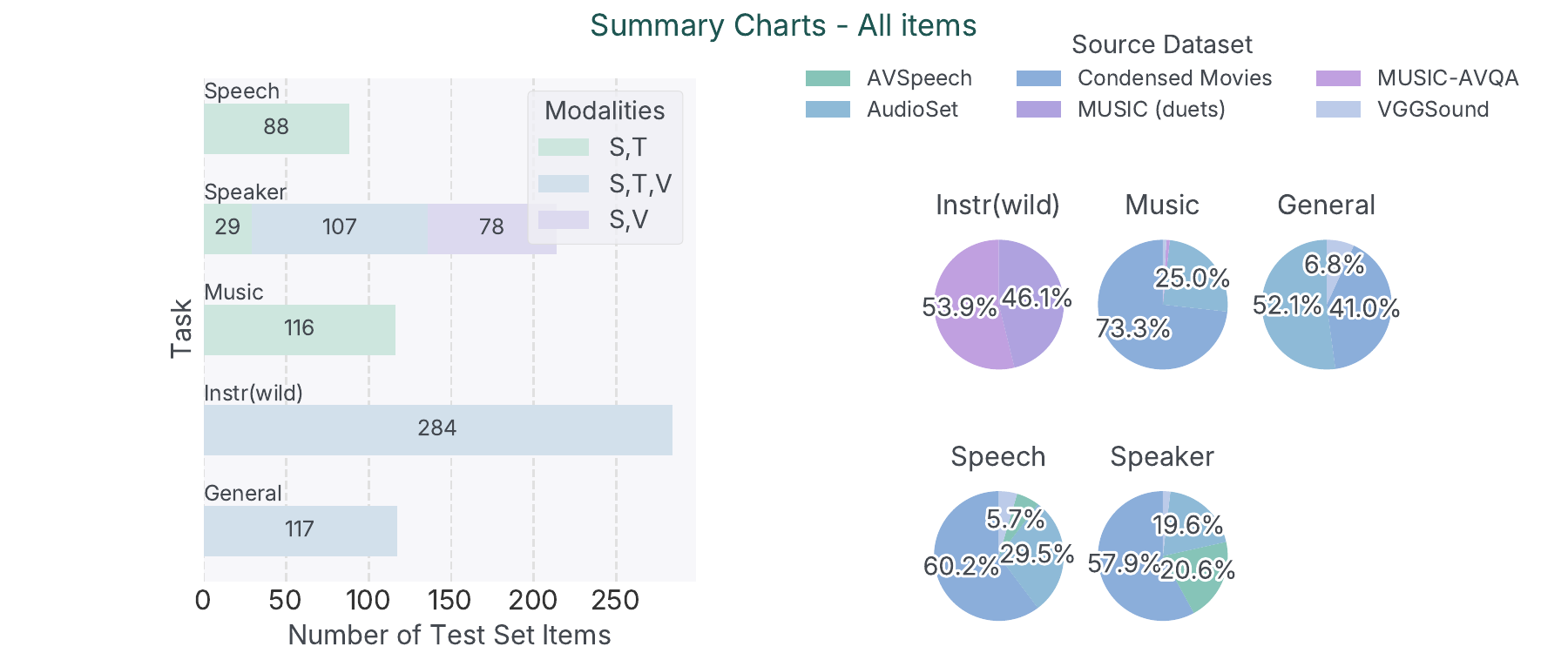}
\caption{\label{fig:samaudiobench_summary} Summary of task, modality, and dataset coverage in \samaudiobench. The modality abbreviations are as follows: ``T'' indicates the item can be used with a text-only prompt (e.g. for speaker separation this implies that the text description can be unambiguously associated with a single speaker), ``V'' indicates that the target sound is on-screen and that we have a SAM masklet provided and ``S'' denotes that there are event boundaries for the target sound.}
\end{figure}

Figure \ref{fig:samaudiobench_summary} shows summary statistics of the entire released \samaudiobench dataset, including how many test set items are available for each task, and which prompt modalities are supported for each; we also show a breakdown of which datasets \samaudiobench videos originate from.

Additional figures in Appendix \ref{appx:samaudiobenchcharacterization} show additional breakdowns for each task, for both the entire set and also for a more balanced subset used for human evaluations.

\subsection{SAM Audio Judge Model}

To develop SAM Audio judge, we conduct a systematic investigation into perceptually aligned evaluation for audio separation. Specifically, we aim to design an evaluation framework that (i) operates without requiring reference signals, making it suitable for real-world applications, (ii) enables fine-grained perceptual assessment of separated audio, and (iii) exhibits stronger correlation with human listening judgments.

\subsubsection{Data Collection}

Existing evaluation guidelines for audio separation are typically simplistic, focusing only on coarse criteria such as the relevance between the separated audio and the prompt, or the overall audio quality of the output \cite{AudioSep}. Such high-level and loosely defined objectives make the evaluation ambiguous. Scores collected under these settings often conflate multiple perceptual aspects and may be biased toward certain criteria depending on the raters’ individual interpretations, resulting in outcomes that are difficult to interpret consistently. 

In addition, existing studies rarely examine the difficulty of audio separation tasks themselves. Most prior work has focused solely on evaluating model outputs, without systematically analyzing how intrinsic factors such as the number of overlapping sources, loudness imbalance, or acoustic similarity between sounds affect human perception of task difficulty. Understanding separation difficulty is crucial, as it provides insights into the limitations and robustness of current models, facilitates the curriculum design for training and evaluation, and enables difficulty-aware benchmarking and adaptive model selection in real-world applications.

To better characterize the performance of audio separation models and the intrinsic difficulty of audio separation tasks, we introduce a new human annotation guideline, \textbf{SAM Audio Judge (SAJ)}, which defines nine perceptual dimensions.

The SAJ performance dimensions evaluate how well a model separates the target sounds:
\begin{itemize}
    \item \textbf{Recall}: Does the extracted audio contain all of the target sounds specified in the prompt?
    \item \textbf{Precision}: How effectively does the model remove non-target sounds from the extracted audio?
    \item \textbf{Faithfulness}: For target sounds present in the extracted audio, how similar do they sound to their counterparts in the original mixture?
    \item \textbf{Overall quality}: What is the overall perceptual quality of the model’s output?
\end{itemize}

In addition, the SAJ difficulty dimensions assess the complexity of the separation task itself:
\begin{itemize}
    \item \textbf{Counting}: How many non-target sounds are present in the source audio?
    \item \textbf{Overlapping}: To what extent do the target sounds overlap with non-target sounds?
    \item \textbf{Loudness}: How loud are the target sounds relative to the non-target sounds?
    \item \textbf{Confusion}: How easily can the non-target sounds be mistaken for the target sounds?
    \item \textbf{Overall difficulty}: Considering all the above factors, how difficult is it to extract the target sounds from the mixture?
\end{itemize}

\subsubsection{Data Annotation}

Based on the above definitions, we design an annotation task to collect SAJ data, where human raters evaluate the nine axes using a five-point Likert scale (1–5). We focus on \textbf{text-prompted} SAJ, the most commonly used scenario. We provide a comprehensive annotation guideline, which offers detailed explanations of each axis and specifies fine-grained aspects to consider during evaluation. An example of the annotation interface is shown in Appendix~\ref{appx:judgeprotocol}. To help raters calibrate their judgments, the guideline is supplemented with numerous audio examples and score references that illustrate what constitutes high or low scores along each dimension. 
We also design a rater qualification
program to ensure the selection of high-quality annotators. Details appear in Appendix~\ref{appx:judgeraters}.

\input{tables/saj_data}

\textbf{Paired data preparation.} We collect a comprehensive set of datasets spanning music, speech, and sound effects. To mitigate the mismatch between real-world and simulated data, we use both real mixtures and synthetic mixtures as input audio. Table~\ref{tab:saj_data} shows the detailed data statistics.
For each data source, we adopt either the original sound annotations (e.g., dog barking, man speaking, etc.) or the sound type predictions generated by PLM-Audio as text prompts, which is described in Section~\ref{sec:text_prompt_generation}. 
We adopt the same text prompt settings as the SAM Audio models, including speech extraction, speaker extraction, music extraction, instrument separation, and general sound event extraction. Following~\citep{jiarui2024dpmtse,wang2025soloaudio}, the text prompts are designed as single-sound descriptions, which may refer to either a specific sound source (e.g., dog barking, guitar) or a general sound category encompassing multiple sources (e.g., dog, music playing).
% }
For each modality, we gather outputs from various publicly available audio separation models as well as from our intermediate SAM Audio checkpoints, which are listed in Table~\ref{tab:saj_model}.
To establish a unified SAJ score that is calibrated across different audio modalities, we shuffle audio samples from all modalities during annotation. We also apply loudness normalization to eliminate potential confounding effects introduced by variations in audio volume. Finally, we collect three independent ratings per audio sample to reduce variance and improve reliability.

\input{tables/saj_model}

\subsubsection{SAJ Model Training}
\begin{figure}[h!]
  \centering
  \includegraphics[width=0.9\linewidth]{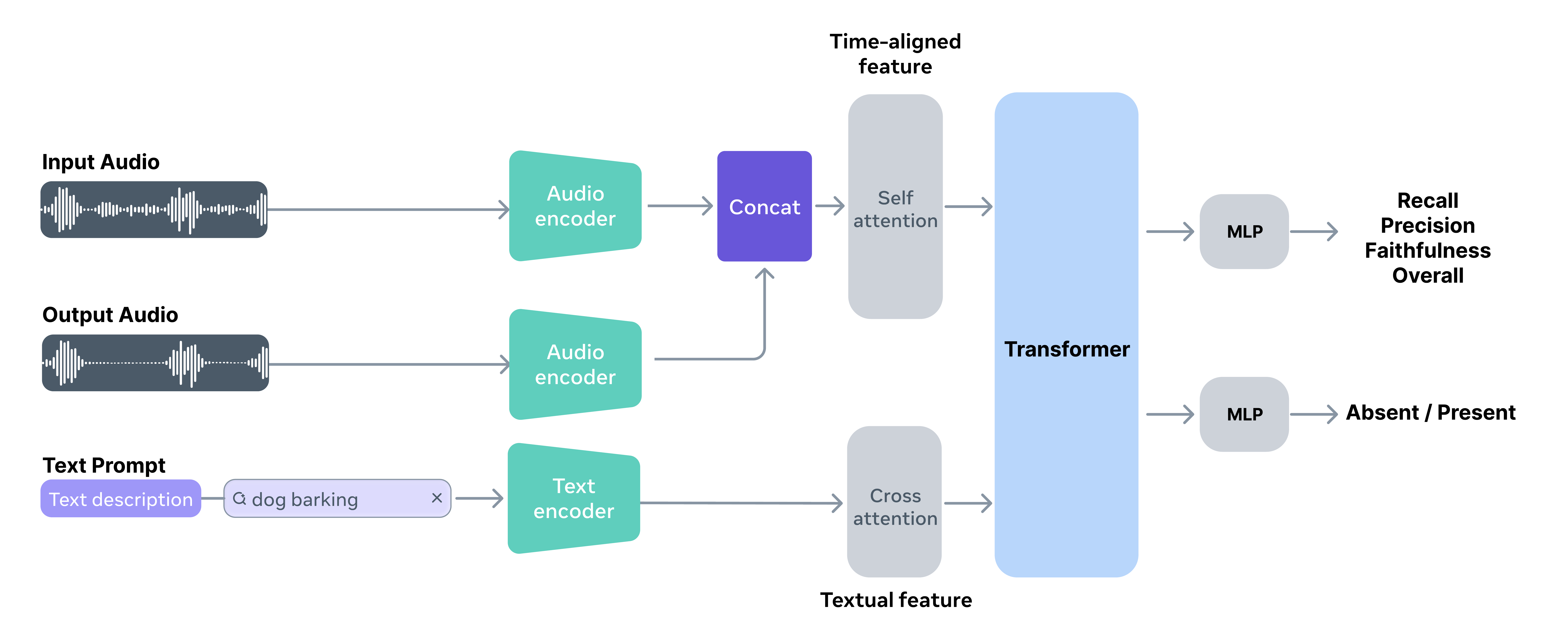}
  \caption{Diagram of SAM Audio Judge Model}
  \label{fig:saj_model}
\end{figure}

Our main SAJ model is designed to predict separation model performance, thus we only use the annotations along the performance dimension. 
As shown in Figure~\ref{fig:saj_model}, the SAJ model takes three inputs: the input audio, the output audio, and a text prompt. We use a pretrained audio encoder to extract audio features and a text encoder to extract text features. Both encoders are adopted from PE-AV~\citep{vyas2025peav}, which is trained through large-scale video-audio–text contrastive learning.

The text features are temporally aligned to match the length of the audio features and then fed into a Transformer to extract joint multimodal representations. Several linear layers are subsequently applied to predict the SAJ scores, including recall, precision, faithfulness, overall, and others.

In addition, we found that introducing a proxy task that predicts whether the output audio follows the text prompt~\citep{wang2022detect,wang2022improving}, significantly improves model performance. To this end, we pretrain the entire model on this text–audio alignment detection task using a large-scale simulated dataset that provides access to separated tracks within mixture audio. We alternate the output audio between the target sound and a random non-target sound from the same mixture to represent the presence or absence of the target sound. An additional linear layer is used to predict the presence or absence of the target sound described by the text prompt. After this pre-training stage, we finetune the SAJ model to predict the final SAJ scores. 

\subsection{Subjective Evaluation Protocol}

Given the limitations of objective metrics on in-the-wild content and the absence of reliable, general-purpose reference-free measures for separation quality, our benchmark adopts \text{human evaluation as the primary method}. We employ a \textbf{side-by-side Absolute Category Rating (ACR)} protocol with an \text{always-on preference tie-breaker}, a hybrid that yields both absolute quality signals and robust relative comparisons, while mitigating common evaluator biases.

Briefly, annotators are shown the source audio (and video, when applicable), the user prompt, and the extracted outputs from two models, and are asked to judge how well each output reflects the requested target sounds. The protocol begins by verifying whether the target sounds actually occur in the source audio, then assesses how much of the target content is present in the extracted audio, whether any portions are missing, and how similar the extracted target sounds are to their originals. Annotators also evaluate the presence and degree of non-target sounds, including whether they originate from the source audio or are artifacts introduced by the model. After answering these structured questions, raters assign a 1–5 Overall score reflecting fidelity to the prompt and acoustic faithfulness, followed—when model scores tie—by a forced-choice preference between the two outputs. The procedure applies consistently across text-, visual-, and span-prompted conditions.

Empirically, we find that this side-by-side ACR framework with an always-on preference tie-breaker offers clear advantages over alternative protocols. The tie-breaker improves inter-annotator agreement and yields sharper, more discriminative ACR deltas that align with expressed preferences. Side-by-side presentation also reduces uncertainty in score differences—narrowing confidence intervals by up to 20\% compared to single-stimulus rating, which translates to roughly 30\% cost savings for equivalent A/B sensitivity. Although handling time per item increases relative to pairwise-only preference evaluations, the protocol produces both absolute and relative judgments in a single pass. Finally, consistent with anchoring effects reported in prior work, absolute ACR scores remain context-dependent.

For each system pair, the subjective evaluation yields both an absolute score (OVR) and a net win rate (NWR) for system A vs.\ B across four dimensions: overall quality, coverage, correctness, and faithfulness. For our main result, we primarily report the overall OVR and NWR.

Full details on the subjective evaluation protocol appear in Appendix~\ref{sec:subjective_evaluation_design} and ~\ref{appx:humanevalprotocol}.

\subsection{Objective Metrics}

\textbf{Text prompting.}
For text-prompted separation, we evaluate model performance using the SAJ and CLAP~\citep{wu2023clap} scores\footnote{We use the checkpoint: \texttt{https://huggingface.co/lukewys/laion\_clap/blob/main/630k-best.pt}.}.  
SAJ measures separation quality across multiple dimensions and we report its overall score for simplicity. CLAP evaluates the semantic alignment between the separated audio and the input text prompt, serving as a proxy for perceptual correctness.

\textbf{Span prompting.}
For span-prompted separation, we reuse the ground-truth text description associated with the target region, enabling the use of the same SAJ and CLAP socres.  
In addition, we introduce the \textbf{SpanIoU} metric, which quantifies temporal alignment by measuring the intersection-over-union (IoU) between the predicted and reference spans. The predicted span is obtained by applying a simple voice activity detection (VAD) on the separated waveform, using the silence detection module in the \texttt{pydub} library~\citep{pydub}.

\textbf{Visual prompting.}
For visual-prompted separation, text-based metrics such as CLAP are less effective because textual descriptions are often insufficiently discriminative for visually localized regions.  
Instead, we measure audio-visual consistency using the ImageBind~\citep{girdhar2023imagebind} score, which captures the alignment between the separated audio and the corresponding visual mask region.

%% file: tables/saj_data.tex
\begin{table}[h!]
\centering
\adjustbox{max width=.4\textwidth}{%
\begin{tabular}{lccc}
\toprule
Split & Modality & Duration & Samples\\
\midrule
\multirow{3}{*}{Training Set} & Speech &59.31 hrs&13,149\\
&Music &133.64 hrs&26,101\\
&Sound &117.52 hrs&37,444\\
\midrule
\multirow{3}{*}{Test Set} & Speech & 6.38 hrs &2,311\\
&Music &9.32 hrs&3,367\\
&Sound &31.72 hrs&11,476
\\
\bottomrule
\end{tabular}
}
\caption{Audio samples and duration in SAM Audio Judge DataSet}
\label{tab:saj_data}

\end{table}

%% file: tables/saj_model.tex
\begin{table}[h!]
\centering
\adjustbox{max width=.7\textwidth}{%
\begin{tabular}{lc}
\toprule
Modality & Audio Separation Model\\
\midrule
Speech & SAM Audio, MossFormer2, Tiger, FastGeCo\\
Music & SAM Audio, AudioSep, FlowSep, ClapSep, SoloAudio, Demucs, Spleeter\\
Sound & SAM Audio, AudioSep, FlowSep, ClapSep, SoloAudio\\
\bottomrule
\end{tabular}
}
\caption{Model outputs used in SAM Audio Judge DataSet}
\label{tab:saj_model}
\end{table}

%% file: results.tex
\section{Experimental Setup}
\label{sec:experimental_setup}

\subsection{Training and inference configurations}

We train three \samaudio{} variants with parameter budgets of \text{500M}, \text{1B}, and \text{3B} (Table~\ref{tab:model_train_config}). 
Parameter counts exclude the PE visual encoder, the T5 text encoder, and the DAC-VAE audio codec.
Auxiliary alignment losses are injected at the 2nd, 4th, and 6th Transformer blocks for 500M, 1B and 3B models, respectively.

All models are optimized following the two-stage recipe of \citet{Polyak2024MovieGA}: 
a large-scale \emph{pre-training} stage, followed by a \emph{fine-tuning} stage on curated high-quality data. The model is pre-trained on synthetic audio mixtures from general video data in Table~\ref{tab:audio_data_source}. The remaining high-quality data, along with all pseudo-labeled data, is used exclusively during fine-tuning.
Because many training audios exceed typical context lengths, all clips are capped at 30~seconds and randomly chunked when longer. 
Training uses fully-sharded data parallelism to fit the model size.

\begin{table}[h!]
\centering
\begin{tabular}{lcccc}
\toprule
Model & Total params & Layers & Attn dim & FFN dim \\
\midrule
\samsmall  & 500M & 12 & 1{,}536 & 6{,}144 \\
\sammid    & 1B   & 16 & 2{,}048 & 8{,}192 \\
\samlarge  & 3B   & 22 & 2{,}816 & 11{,}264 \\
\bottomrule
\end{tabular}
\caption{\samaudio{} model configurations.}
\label{tab:model_train_config}
\end{table}

\paragraph{Pre-training.}
We use an effective batch size of 1{,}024 sequences, each truncated or padded to 30~seconds ($=750$ audio tokens). 
Models are trained for 500K updates with a constant learning rate of $1\times10^{-4}$, preceded by a 5K-step linear warmup.
AdamW is used throughout with weight decay~0.1 and \texttt{bf16} precision.
During pre-training, the auxiliary alignment loss is enabled with weight~1. Additionally, we apply conditioning dropout: the audio mixture, video, and text prompt are each independently dropped with probability~0.3.

\paragraph{Fine-tuning.}
Since fine-tuning data exhibit higher length variability, we adopt \emph{variable-length batching} with a per-batch token budget of \{96K, 120K, 144K\} tokens for the \{500M, 1B, 3B\} models, respectively.
Fine-tuning runs for 300K steps with a 5K warmup to a peak learning rate of $1\times10^{-4}$, kept constant thereafter.
An exponential-moving-average (EMA) checkpoint (decay~0.999) is maintained and used for inference.
The auxiliary loss is disabled during fine-tuning (weight~0), as we do not observe gains when training on clean outputs.  
During fine-tuning, we also disable conditioning dropout.

\paragraph{Inference.}
We use a 16-step midpoint ODE solver without classifier-free guidance, as CFG yielded no improvement in our setting.
We additionally apply candidate re-ranking with beam size 8:
for text prompting, a linear combination of SAM Audio Judge and CLAP scores (weights 1 and 5); 
for span prompting, span IoU; 
and for visual prompting, the ImageBind similarity between audio and masked region.
By default, we enable span prediction for text-only separation, as it brings overall gains (see Section~\ref{sec:exp_text_predspan}).  
Specifically, we employ the PE-A frame model~\citep{vyas2025peav} to estimate the temporal spans corresponding to the target sound given the text prompt. The predicted spans are then combined with the original text prompt and used as conditioning for \samaudio{}. We set the frame probability threshold to $0.3$, following the threshold setting in~\citet{vyas2025peav}.  

As \samlarge shows the best performance among the three variants, we adopt \samlarge for all key comparisons and analyses, denoting it simply as \samaudio{} in the remainder of the paper unless otherwise specified.

% 

% 

% % 

% 

% 

% 

\subsection{Tasks}
We benchmark \samaudio{} across a diverse set of separation tasks that reflect common real-world use cases, as mentioned in Section~\ref{section:evaluation}.
Each task is to evaluate the model’s ability to leverage text, video, and span prompts individually or jointly. 

\textbf{Text-prompting tasks.}
We evaluate \samaudio{} on a broad range of text-prompted source separation tasks:
\begin{enumerate}
    \item \textbf{General sound event extraction:} Extract any target sound event described by a text query (e.g., ``dog barking'' or ``glass shattering''). This task tests the model’s ability to handle extraction of general audio events.
    \item \textbf{Speech extraction:} Separate all speech from a noisy audio mixture. The number of speakers is unconstrained, and the goal is to recover the complete speech track regardless of speaker count or background noise.
    \item \textbf{Speaker extraction:} Isolate speech from a specific speaker given a text prompt describing speaker attributes such as gender or age (e.g., ``female speaking''). This task evaluates fine-grained speaker conditioning. We primarily focus on gender-based separation.
    \item \textbf{Music extraction:} Separate music from mixed audio, including cases where music is accompanied by speech or sound effects.
    \item \textbf{Instrument separation in the wild:} Extract a single instrument stem (e.g., ``piano,'' ``drums'') from full music audio. This benchmark includes both clean studio recordings and noisy in-the-wild music mixtures.
    \item \textbf{Professional instrument separation:} Separate stems from professionally recorded music (e.g., studio tracks). Unlike the general instrument separation benchmark, this focuses on high-quality, multi-track recordings and uses MUSDB18~\citep{musdb18} as the standard benchmark.
\end{enumerate}

\textbf{Visual-prompting tasks.}
Visual prompts consist of a pair of video masks and the raw video clip. These tasks evaluate the model’s ability to perform instance-level source separation conditioned on visual information:
\begin{enumerate}
    \item \textbf{General sound event separation:} Extract the sound associated with the highlighted region of interest (e.g., isolating the sound of car honking).
    \item \textbf{Instrument separation:} In a music video, extract the audio of the instrument corresponding to the highlighted mask, regardless of whether the mixture is noisy.
    \item \textbf{Speaker separation:} In a multi-speaker video, separate the speech of the highlighted speaker.
\end{enumerate}

\textbf{Span-prompting tasks.}
Span-prompting tasks mirror the text-prompting tasks. Instead of text descriptions, text model is conditioned on timespans that specify when the target source is active. 

Within each modality, we filter out samples with ambiguous prompts that could map to multiple plausible sources (e.g., videos with multiple visually indistinguishable instruments or overlapping events that are not temporally resolvable). 
% .
Each sample in our evaluation set are of $\sim10s$, which is to facilitate subjective evaluation. As the original music in MUSDB span several minutes, we randomly extract 10s chunks of the original full-mix audio for professional instrument separation.

% .

\subsection{Baselines}

We compare \samaudio{} against a broad set of baselines across all prompting modalities. For a fair comparison, we evaluate each model on tasks it is designed to handle, using its native input format where available.

\subsubsection{Text-prompting baselines}
Most existing research on text-guided audio separation focuses on general audio events, where broad categories such as \textit{speech} and \textit{music} are treated as single classes. We select four representative and recent open-domain baselines for comparison: AudioSep~\citep{AudioSep}, FlowSep~\citep{flowsep}, SoloAudio~\citep{wang2025soloaudio}, and CLAPSep~\citep{CLAPSep}. These models are designed to handle general audios and are trained on large-scale audio-text datasets. Although they are not specialized for individual domains, their pretraining data often include samples from speech and music, allowing a comparison to \samaudio{} across general and specialized tasks.

Beyond these open-domain systems, we also include a range of specialized baselines targeting specific audio domains. Unlike general-purpose models, these systems do not support free-form text prompting; instead, they decompose an audio mixture into a fixed ontology of stems (e.g., vocals, drums, bass). For a fair comparison, we extract the separated stem that corresponds to the target event type and evaluate it against \samaudio{}. The specialized baselines are outlined below.

\textbf{Instrument separation}
We evaluate against Demucs~\citep{rouard2022hybrid}, Spleeter~\citep{hennequin2020spleeter}, AudioShake~\citep{audioshake}, MoisesAI~\citep{moisesai}, LalalAI~\citep{lalalai}, and FADR~\citep{fadr}. Demucs and Spleeter are publicly available models, while the others are proprietary systems accessed via public APIs. All baseline models are limited to a small set of supported stems (see Table~\ref{tab:baseline_instrument_info}). For the professional instrument separation benchmark, we adopt MUSDB~\citep{musdb18}, which only include vocals, drums and bass separation. In contrast, for instrument separation \textit{in the wild}, we exclude Demucs and Spleeter since their supported stem vocabulary covers only a small fraction of required instruments. For the remaining baselines, we compare only on examples where the target instrument is within the supported list.

\begin{table*}[h]
\centering
\small
\adjustbox{max width=\textwidth}{
\begin{tabular}{l l}
\toprule
\textbf{Model} & \textbf{Supported Instruments} \\
\midrule
Demucs~\citep{rouard2022hybrid} & vocals, drums, bass \\
Spleeter~\citep{hennequin2020spleeter} & vocals, drums, bass, piano \\
AudioShake~\citep{audioshake} & vocals, drums, guitar, bass, wind, piano \\
MoisesAI~\citep{moisesai} & vocals, guitar, bass, drums, piano, wind, strings \\
LalalAI~\citep{lalalai} & vocals, drums, bass, guitar, piano, synthesizer, strings, wind \\
FADR~\citep{fadr} & vocals, drums, piano, guitars, strings, wind \\
\midrule
\samaudio{} & open-vocabulary \\
\bottomrule
\end{tabular}
}
\caption{Supported instrument types for each baseline.}
\label{tab:baseline_instrument_info}
\end{table*}

\textbf{Speech separation}
Speech separation aims at removing background noise of a speech recording. We compare \samaudio{} against LalaAI~\citep{lalalai}, ElevenLabs~\citep{elevenlabs}, Auphonic~\citep{auphonic}, and AudioShake~\citep{audioshake}. To ensure fair comparison, we disable post-processing or enhancement modules in the baselines, as our evaluation focuses on \textit{separation fidelity}—the model’s ability to isolate speech content—rather than perceptual enhancement or reverb suppression.

\textbf{Music separation}
For music separation, we benchmark against MoisesAI~\citep{moisesai} and AudioShake~\citep{audioshake}, both of which can isolate or remove background music. Similar to speech separation, we disable any additional enhancement modules. We report results both for \textit{music extraction} (isolating music) and \textit{music removal} (removing music while preserving other sounds).

\textbf{Speaker separation}
Speaker separation aims to isolate speech from specific individuals. Most specialized systems rely on fixed-stem decomposition and do not natively support prompting for arbitrary speakers. To enable comparison under the prompted setting, we select the separated stem with the highest CLAP similarity score~\citep{wu2023clap}.
We evaluate public models such as Mossformer2~\citep{zhao2024mossformer2}, Tiger~\citep{xu2024tiger}, and FastGeCo~\citep{wang2024noise}, as well as the proprietary AudioShake model~\citep{audioshake}.

\subsubsection{Visual-prompting baselines}

Compared to text-guided separation, visual-prompted audio separation is much less explored. No commercial models are available to our knowledge, thus we focus on public research models. We evaluate \samaudio{} against two general-purpose visual separation models, DAVIS-Flow~\citep{huang2025davis} and CLIPSep~\citep{dong2023clipsep}, across general sound, speaker, and instrument separation tasks. In addition, we include the instrument-specific checkpoint\footnote{We use the DAVIS-Flow checkpoints trained on AVE and MUSIC respectively for general and instrument-specific separation.} of DAVIS-Flow (denoted as \textit{DAVIS-Flow(Music)}) for the visual-prompted instrument separation benchmark.

Among visual separation tasks, speaker separation has been studied more extensively~\citep{Wu2019TimeDA,Pan2025PlugandPlayCF,li2024audio,IIANet}. We therefore compare to two strong baselines, IIANet~\citep{IIANet} and AV-Mossformer2~\citep{zhao2025clearvoice}, which achieve state-of-the-art results on a recent visual-prompted speech separation benchmark~\citep{visual_target_clearvoice}. These models rely on a preprocessing pipeline involving face detection and lip-motion extraction; we follow their official preprocessing setup and feed masked videos accordingly. In practice, about 20\% of our evaluation video fail in AV-Mossformer2 preprocessing stage, and these samples are excluded from comparison with \samaudio{}.

\subsubsection{Span-prompting baselines}
To our knowledge, there are no public separation models supporting span prompting. Therefore we use our text-prompting model as a baseline, comparing span-only and joint text+span conditioning against text-only separation to quantify the benefits of span prompting.

\section{\samaudio{} Results}
\label{sec:results}

In this section, we present the main results of \samaudio{}, covering text, visual, and span prompting, sound removal, model latency, and long-form separation. Detailed ablation studies are deferred to Appendix~\ref{sec:sam_audio_ablation_study}.

\subsection{Text-prompted separation}

\input{tables/text_main_result}

Table \ref{tab:res_main_text} presents quantitative comparisons between \samaudio and a range of public text-prompted separation models. The baselines fall into two broad categories: general models that aim to handle a wide variety of separation tasks (such as AudioSep, FlowSep, SoloAudio, and CLAPSep), and specialized models optimized for specific domains like speech or music (e.g., MossFormer2, Demucs, Spleeter). 
We further show the net win rate against the baselines in Figure~\ref{fig:nwr_text_baselines}.
In Table~\ref{tab:res_main_text}, we additionally show the overall subjective score (i.e., OVR) aside from the objective metrics. 
As our evaluation protocol follows a pairwise comparison setup, the final OVR score is obtained by averaging the overall preference scores across all pairwise comparisons involving \samaudio{}.

Overall, \samaudio consistently outperforms all prior models by a substantial margin across nearly all categories. In general sound event separation, \samaudio achieves roughly a $\sim36\%$ net win rate over one of the best public general sound separation model (SoloAudio~\citep{wang2025soloaudio}). 
In specialized domains such as instrument or speaker separation, we observe that general-purpose models like FlowSep~\citep{flowsep} or AudioSep~\citep{AudioSep} fail to reach competitive quality. Their mean judge scores remain below 3, significantly trailing specialized systems such as Demucs~\citep{rouard2022hybrid}. Yet even in such domains, \samaudio surpasses specialized models (e.g., NWR of \samaudio{} vs. Demucs = 17.6\%).  
While proprietary systems generally outperform the OSS counterparts, \samaudio{} remains superior in most settings. On instrument separation for professional audios (MUSDB~\citep{musdb18}), \samaudio{} achieves an overall subjective score of 4.45 compared to 4.28 from AudioShake~\citep{audioshake} (a relative gain of $\sim4$\%), and in speaker separation, it achieves 4.15 versus 3.51, corresponding to a net win rate improvement of $\sim 39$\%. Through unified training, \samaudio{} generalize across domains and achieves SoTA performance.

\subsection{Visual-prompted separation}

\input{tables/visual_main_result}
We show the comparison between \samaudio{} and existing visual-prompted separation models in Table~\ref{tab:res_main_visual}. Compared to text-prompted separation, there are substantially fewer general-purpose visual separation systems available publicly.
% ;   

\input{tables/visual_sample_figure}

Across all settings, \samaudio{} achieves stronger improvements over prior work. On average, \samaudio{} outperforms DAVIS by a large margin, achieving net win rates ranging from {5\% to 48\%} depending on the separation task.
Similar to the trends observed in text-prompted separation, we find that general visual models struggle on specialized domains such as \textit{instrument} and \textit{speaker} separation. In contrast, \samaudio{} maintains and surpasses the best specialized baselines by approximately {25\%} on speaker separation and {5\%} on instrument separation.  

Additionally, we notice that visual prompting yields notably lower overall subjective scores compared to text prompting. The advantage of text prompt stems primarily from the availability of a much larger pool of high-quality text–based training data, whereas video-based supervision is generally noisier due to errors in visual masks and the frequent presence of off-screen sounds in training. Beyond scale, text also tends to provide more specific cues about the target source. For instance, a prompt such as \textit{"man shuffling"} directly points to a unique sound event, whereas a visual mask of a person is inherently ambiguous since it can be associated with multiple sounds. 

Nevertheless, visual prompting plays a complementary role in scenarios where text alone is insufficient. A common example arises in conversational scenes where multiple people of the same gender are speaking. A text prompt such as \textit{"male speech"} cannot disambiguate between the two male speakers, but visual masks can localize the target speaker and enables separation effectively. In such cases, visual cues provide instance-level grounding that is difficult to achieve with text alone (see Figure~\ref{fig:video_prompt_samples}).

\input{tables/span_main_result}

\subsection{Span-prompted separation}
\label{exp:text_span}

Table~\ref{tab:res_main_span} and Figure~\ref{fig:nwr_text_span_comparison} compare models conditioned on text, span, and text+span inputs. To better ablate the effect of ground-truth span, no predicted span is used in text prompting. Using span prompts alone does not consistently improve performance, as the target and distractor sounds may co-occur throughout the same temporal regions. This effect is particularly evident for long-duration or ambient sounds such as speech and music, where span-only models exhibit large performance degradations (NWR of $-16\%$ to $-49.6\%$ relative to text-only baselines). In contrast, for short-duration and well-localized sounds, where temporal cues are more discriminative, span-only conditioning yields noticeable gains (NWR $+26\%$ over text-only in sound).

Despite these fluctuations, combining text and span inputs consistently improves performance across all domains, achieving NWR between $+12.9\%$ and $+39.0\%$. These results show that temporal localization from span prompts complements the semantic information in text, using both would enable more precise separation.

\subsection{Span prediction boosts text-prompted separation}
\label{sec:exp_text_predspan}

Leveraging predicted spans as additional input is our default choice for text prompting. Here, we compare using vs. not using span prediction across all separation tasks. The comparison is shown in Table~\ref{tab:res_abl_span_pred}. Note that the OVR scores for \textit{text + predicted span} differ from those in Table~\ref{tab:res_main_text}, as the absolute ratings depend on the specific baseline used in each pairwise evaluation under our human evaluation protocol.

Incorporating predicted spans boosts performance in majority of the domains, including general SFX, speech, speaker, and music, where temporal cues play a critical role in disambiguating target sounds. 
We noticed a minor degradation in professional instrument separation, likely because many MUSDB instrument stems span the entire segment, leaving limited room for temporal cues to provide additional benefit.
There is only a small gap between using predicted spans and ground-truth spans (see Table~\ref{tab:res_main_span}), which shows the robustness of \samaudio{} to temporal inaccuracies in span estimation.
Importantly, span prediction improves separation quality without costly human annotations, allowing \samaudio{} to separate precisely at scale.

\input{tables/abl_effect_span}

\subsection{Sound Removal}

\samaudio{} outputs both a \emph{target} and a \emph{residual} audio. While earlier sections focus on evaluating target quality, we now evaluate the residual, corresponding to the \emph{removal} task—removing the prompt-specified sound from the mixture. We use text-prompted music removal as a representative removal task.

Among existing baselines, only MoisesAI~\citep{moisesai} and AudioShake~\citep{audioshake} support explicit sound removal. Therefore, we compare \samaudio{} against these two systems. As shown in Table~\ref{tab:ovr_music_removal} and Figure~\ref{fig:nwr_music_removal}, \samaudio{} outperforms both systems. The trends mirror the music extraction results in Figure~\ref{fig:nwr_text_baselines}, suggesting the high correlation of extraction and removal modes. 
The high OVR score by \samaudio{} further shows the model's ability to cleanly suppress target sources.

\input{tables/music_removal_comparison}

\subsection{Latency}

\begin{wrapfigure}{r}{0.5\textwidth}
    \centering
    % \vspace{-10pt}
    % \vspace{3pt}
    \includegraphics[width=\linewidth]{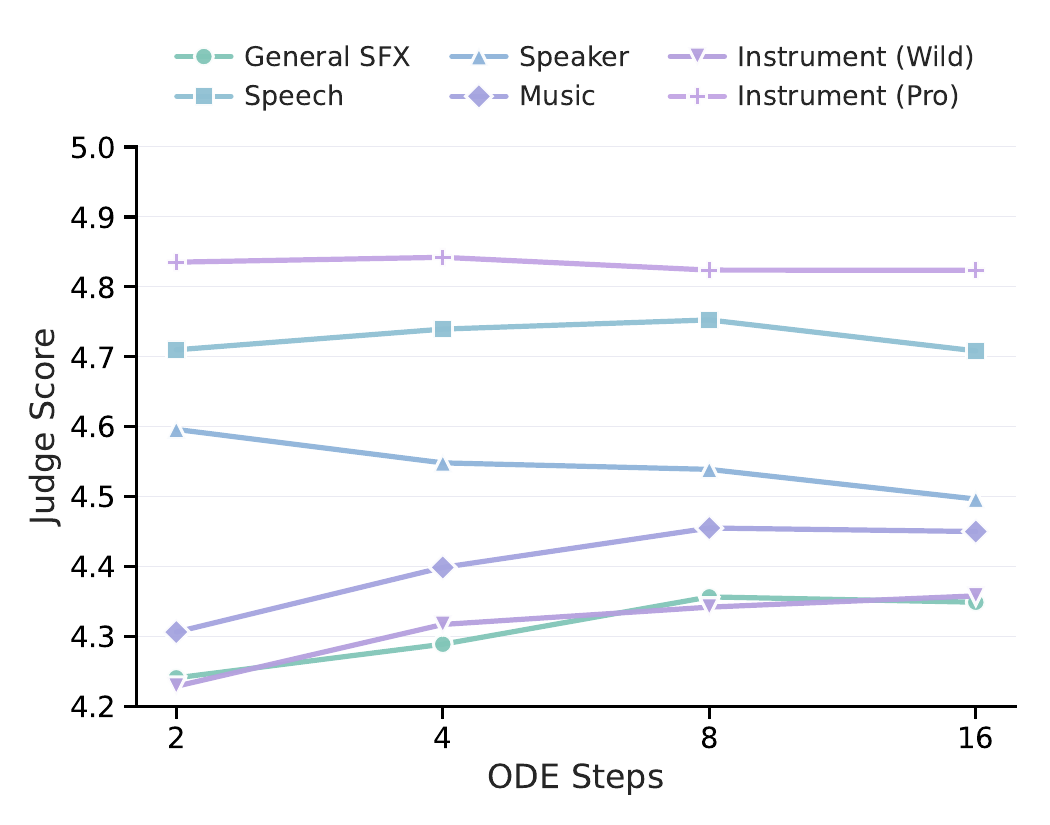}
    \vspace{-10pt}
    \caption{Effect of varying ODE steps under the midpoint solver. Fewer steps reduce computation at a modest cost in quality.}
    \label{fig:res_abl_ode_step}

\end{wrapfigure}

Our default inference configuration uses 16 ODE steps with the midpoint solver. For text prompting, each separation on \samaudio{}-Large takes approximately \text{7.3\,s} for a 10-second input on one A100 GPU, including \text{6.5\,s} for model forward computation, \text{0.1\,s} for span prompting, and \text{0.5\,s} for judge reranking.

We further study the trade-off between inference cost and output quality by varying the number of ODE steps. As shown in Figure~\ref{fig:res_abl_ode_step}, increasing ODE steps generally improves performance across tasks as expected. Nonetheless, the model achieves surprisingly competitive results even with as few as two ODE steps (e.g., speech separation).
Qualitatively, we find a larger performance gap between 16 and 2 ODE steps for speaker or instrument separation. In contrast, for general sound effects that are often short and sparse, lower NFEs still yield acceptable perceptual quality. We hypothesize that the input audio mixture provides strong conditioning signal, enabling separation with fewer refinement steps compared to fully generative tasks such as TTS. Overall, fewer NFEs offer a favorable trade-off between speed and quality for many practical separation scenarios for \samaudio{}.

\subsection{Long-form Audio Separation}

Most samples in our evaluation set are around 10 seconds. To further assess the performance of \samaudio{} on longer inputs, we evaluate its performance on long-form separation using the multi-diffusion approach described in Section~\ref{sec:method}. Specifically, we adopt a 20-second window with a 5-second context overlap. For comparison, we consider two baselines:  
(a) \textbf{chunk-wise separation}, where the audio is divided into 20-second segments that are processed independently and stitched together, and  
(b) \textbf{one-shot separation}, where the entire audio is processed in a single forward pass.

\begin{wraptable}{r}{0.4\textwidth}
% \vspace{-10pt}
\centering
\adjustbox{max width=0.35\textwidth}{
\begin{tabular}{lccc}
\toprule
Method & SAJ & CLAP \\
\midrule
One-shot & 3.48 & 0.26  \\
Chunk-wise & 3.57 & 0.24  \\
Multi-diffusion & \textbf{3.67} & \textbf{0.27} \\
\bottomrule
\end{tabular}
}
\caption{Comparison of methods for long-form separation.}
\label{tab:res_longform_sep_comparison}

% \vspace{-5pt}
\end{wraptable}

%  and  
% 

We curate an internal test set of 50 1-minute audio samples, to evaluate long-horizon separation quality. As shown in Table~\ref{tab:res_longform_sep_comparison}, the one-shot model exhibits a noticeable degradation on long recordings, which aligns with expectations given that most training samples are shorter than 30 seconds. The chunk-wise baseline mitigates this issue but introduces audible discontinuities at segment boundaries. In contrast, the proposed multi-diffusion strategy maintains high perceptual quality across the full sequence and achieves the best judge scores.
% ,

% 

% 

\subsection{\samaudio{} Judge Results}
\label{sec:exp_sam_audio_judge}

\subsubsection{Subjective score correlation}

We compare the proposed SAM Audio Judge model with several representative baseline systems covering diverse evaluation paradigms:
\begin{itemize}
    \item CLAP~\citep{wu2023clap}: A large-scale contrastive audio–text model trained on millions of audio–caption pairs. We use its cosine similarity between the text prompt and output audio embeddings as a proxy for perceptual alignment. This represents the current standard for audio–language correspondence evaluation.
    % .
    \item SDR Estimator~\citep{dang2023using}: A regression model trained to estimate SDR without access to ground-truth references. It reflects the performance of traditional distortion-based metrics that focus on signal fidelity rather than perceptual judgment. We trained the SDR estimator using the same architecture as the SAJ model, except that its training target was the SDR value. We used a balanced training dataset covering target audio levels from $-25$ dB to $25$ dB, totaling $496$ hours across speech, music, and sound effects. The SDR estimator achieved PPC scores of 0.923, 0.681, and 0.665 on speech, music, and sound effects, respectively.

    \item Gemini-2.5-pro~\citep{comanici2025gemini}: A large multimodal LLM capable of reasoning over both text and audio inputs. We prompt it to rate the separated audio quality according to the same evaluation axes used in SAJ, representing a language-model-based perceptual evaluation baseline. We also provide several examples with different scores to enable few-shot learning. Our prompts could be found in Appendix~\ref{appx:gemini}.
\end{itemize}
These baselines cover the main paradigms of 
signal processing-based separation measures and general-purpose multimodal LLM-based metrics, allowing for a comprehensive comparison with our task-specific SAJ model.
We report Pearson (PCC) and Spearman (SRCC) correlation between automatic metrics and human ratings.

\input{tables/saj_com}
As shown in Table~\ref{tab:res_saj_com}, the proposed SAM Audio Judge model consistently outperforms all baselines across the three modalities, \textit{i.e.} speech, music, and sound effects, under both PCC and SRCC. SAJ achieves markedly higher correlations with human ratings, reaching PCCs of 0.883, 0.815, and 0.815 for speech, music, and sound, respectively, and SRCCs of 0.817, 0.714, and 0.781. 
% . 
In contrast, baseline models such as CLAP and Gemini-2.5-pro show moderate correlations, while distortion-based metric (SDR Estimator) fails to capture perceptual quality, often yielding low or even negative correlations. The consistent performance of SAJ across modalities highlights its ability to generalize beyond speech to more complex acoustic domains such as music and environmental sounds. These results confirm that the proposed SAJ model effectively captures human perceptual judgment by leveraging joint audio–text representations and text-conditioned pretraining, offering a more reliable and fine-grained evaluation framework than existing baselines. Appendix~\ref{appx:judgediff} presents the automatic fine-grained performance analysis enabled by SAJ.

\input{tables/saj_reranking}
\subsubsection{Judge as a reranker}
To evaluate the impact of different rerankers, we apply them to the SAM Audio model, which generates eight candidate separation outputs for each input mixture. Each reranker is responsible for selecting the best candidate according to its scoring mechanism, and the selected outputs are then evaluated on the test set (see Table~\ref{tab:saj_data}). We compare three configurations: Judge, CLAP, and their combination (CLAP w/ Judge). The combined reranker computes a linear combination of the two scores with a weighting ratio of 5:1 (CLAP : Judge).

Table~\ref{tab:saj_reranking} presents a comparison of NWR between different reranking configurations across five separation tasks: speech, speaker, music, instrument, and general sound separation. Overall, we observe that integrating the SAM Audio Judge model as a reranker consistently improves or stabilizes performance relative to using CLAP alone.

When comparing Judge vs. CLAP, the Judge reranker achieves higher NWR in most categories (e.g., 0.17 vs. 0.19 in speech and 0.15 in sound separation), indicating its stronger ability to align selection decisions with perceptual quality.
After combining both signals (CLAP w/ Judge vs. CLAP), the hybrid reranker further improves NWR on music (0.14) and instrument (0.03) separation, suggesting that CLAP score can be complementary to SAJ.
Finally, when CLAP w/ Judge is evaluated against Judge directly, the NWR drops generally compared to its evaluation against CLAP (e.g., 0.00 in speech, 0.02 in speaker, and 0.10–0.15 in music/instrument), showing that SAJ score already captures most of the reranking signal.

In summary, the results suggest that the Judge-based reranker is highly effective across audio domains. Combining judge and CLAP produces largest gains, particularly in speech and general sound separation, where reranking with CLAP alone tends to underperform.

%% file: tables/text_main_result.tex
\begin{table*}[h]
 \setlength{\tabcolsep}{3pt}
    \centering
    \adjustbox{max width=1\textwidth, center}{%
    \begin{tabular}{lcc 
    cc>{\columncolor{gray!15}}c 
    cc>{\columncolor{gray!15}}c 
    cc>{\columncolor{gray!15}}c 
    cc>{\columncolor{gray!15}}c 
    cc>{\columncolor{gray!15}}c 
    cc>{\columncolor{gray!15}}c 
    }
    \toprule
     \multicolumn{3}{c}{} &
     \multicolumn{3}{c}{\textbf{General SFX}} &
     \multicolumn{3}{c}{\textbf{Speech}} &
     \multicolumn{3}{c}{\textbf{Speaker}} &
     \multicolumn{3}{c}{\textbf{Music}} &
     \multicolumn{3}{c}{\textbf{Instr(wild)}} &
     \multicolumn{3}{c}{\textbf{Instr(pro)}} \\
    \cmidrule(lr){4-6} \cmidrule(lr){7-9} \cmidrule(lr){10-12}
    \cmidrule(lr){13-15} \cmidrule(lr){16-18} \cmidrule(lr){19-21}
    \textbf{Model} & \textbf{OSS} & \textbf{Promptable} &
    SAJ & CLAP & OVR &
    SAJ & CLAP & OVR &
    SAJ & CLAP & OVR &
    SAJ & CLAP & OVR &
    SAJ & CLAP & OVR &
    SAJ & CLAP & OVR \\
    \midrule
    MossFormer2~\citep{zhao2024mossformer2} & \cmark & \xmark & - & - & - & - & - & - & 2.43 & 0.14 & 2.54 & - & - & - & - & - & - & - & - & - \\
    Tiger~\citep{xu2024tiger} & \cmark & \xmark & - & - & - & - & - & - & 2.47 & 0.15 & 2.50 & - & - & - & - & - & - & - & - & - \\
    Fast-GeCo~\citep{wang2024noise} & \cmark &\xmark & - & - & - & - & - & - & 2.66 & 0.16 & 2.71 & - & - & - & - & - & - & - & - & - \\
    Demucs~\citep{rouard2022hybrid} & \cmark & \xmark& - & - & - & - & - & - & - & - & -  & - & - & - & - & - &  & 4.48 & 0.15 & 4.26 \\
    Spleeter~\citep{hennequin2020spleeter} & \cmark & \xmark & - & - & - & - & - & - & - & - & -  & - & - & - & - & - & - & 4.26 & 0.11 & 3.90 \\
    FlowSep~\citep{flowsep} & \cmark & \cmark & 2.36 & 0.21 & 2.65 & 2.18 & 0.20 & 2.14 & 1.85 & 0.09 & 2.13 & 2.73 & 0.18 & 2.90 & 2.37 & 0.10 & 2.69 & 2.13 & -0.01 &  2.02 \\
    AudioSep~\citep{AudioSep} & \cmark & \cmark & 2.63 & 0.25 & 2.88 & 2.93 & 0.28 & 2.85 & 2.50 & 0.17 & 2.79 & 3.47 & 0.27 & 3.51 & 2.16 & 0.13 & 2.59 & 2.34 & 0.04 & 2.45  \\
    CLAPSep~\citep{CLAPSep} & \cmark & \cmark & 2.68 & 0.23 & 2.92 & 2.30 & 0.22 & 2.47 & 2.80 & 0.17 & 2.79 & 2.48 & 0.04 & 2.97 & 2.47 & 0.14 & 2.81 & 2.48 & 0.04 &  2.56 \\ 
    SoloAudio~\citep{wang2025soloaudio} & \cmark & \cmark & 3.29 & 0.25 & 2.97 & 3.45 & 0.30 & 3.32 & 2.26 & 0.19 & 2.45 & 2.68 & 0.21 & 2.47 & 2.92 & 0.13 & 2.71 & 2.65 & 0.01 & 2.30  \\
    AudioShake~\citep{audioshake} & \xmark &\xmark & - & - & - & 3.90 & 0.28 & 3.95 & 3.28 & 0.14 & 3.51 & 3.22 & 0.29 & 3.37 & 3.37 & 0.29 & 3.43 & 3.87 & 0.29 & 4.28  \\
    MoisesAI~\citep{moisesai} & \xmark & \xmark & - & - &  & - & - &  & - & - & - & 3.79 & 0.27 & 3.90 & 3.03 & 0.29 & 3.12 & 3.78 & 0.28 & 4.22   \\
    FADR~\citep{fadr} & \xmark & \xmark & - & - & - & - & - & - & - & - & - & - & - & - & 2.44 & 0.19 & 2.45 & 3.63 & 0.25 & 3.92 \\
    LalalAI~\citep{lalalai} & \xmark & \xmark & - & - & - & 3.77 & 0.33 & 3.92 & - & - & - & - & - & - & 3.07 & 0.25 & 3.03 & 3.83 & 0.27 & 4.18 \\
    Auphonic~\citep{auphonic} & \xmark & \xmark & - & - & - & 4.32 & 0.27 & 4.08 & - & - & - & - & - & - & - & - & - & - & - & - \\
    ElevenLabs~\citep{elevenlabs} & \xmark &\xmark & - & - & - & 3.79 & 0.25 & 3.72 & - & - & - & - & - & - & - & - & - & - & - & - \\
    \midrule
    \samaudio & \cmark & \cmark & \textbf{4.35} & \textbf{0.31} & \textbf{3.59} & \textbf{4.67} & \textbf{0.35} & \textbf{4.29} & \textbf{4.51} & \textbf{0.18} & \textbf{4.15} & \textbf{4.45} & \textbf{0.26} & \textbf{4.05} & \textbf{4.32} & \textbf{0.31} & \textbf{4.00} & \textbf{4.82} & \textbf{0.28} & \textbf{4.45}  \\ 
    \bottomrule
    \end{tabular}
    } 
    \caption{Comparison against text-prompted baselines. --: not applicable. OVR: overall subjective score.}
    \label{tab:res_main_text}

\end{table*}

\begin{figure*}[htp]
    \centering
    \includegraphics[width=\textwidth]{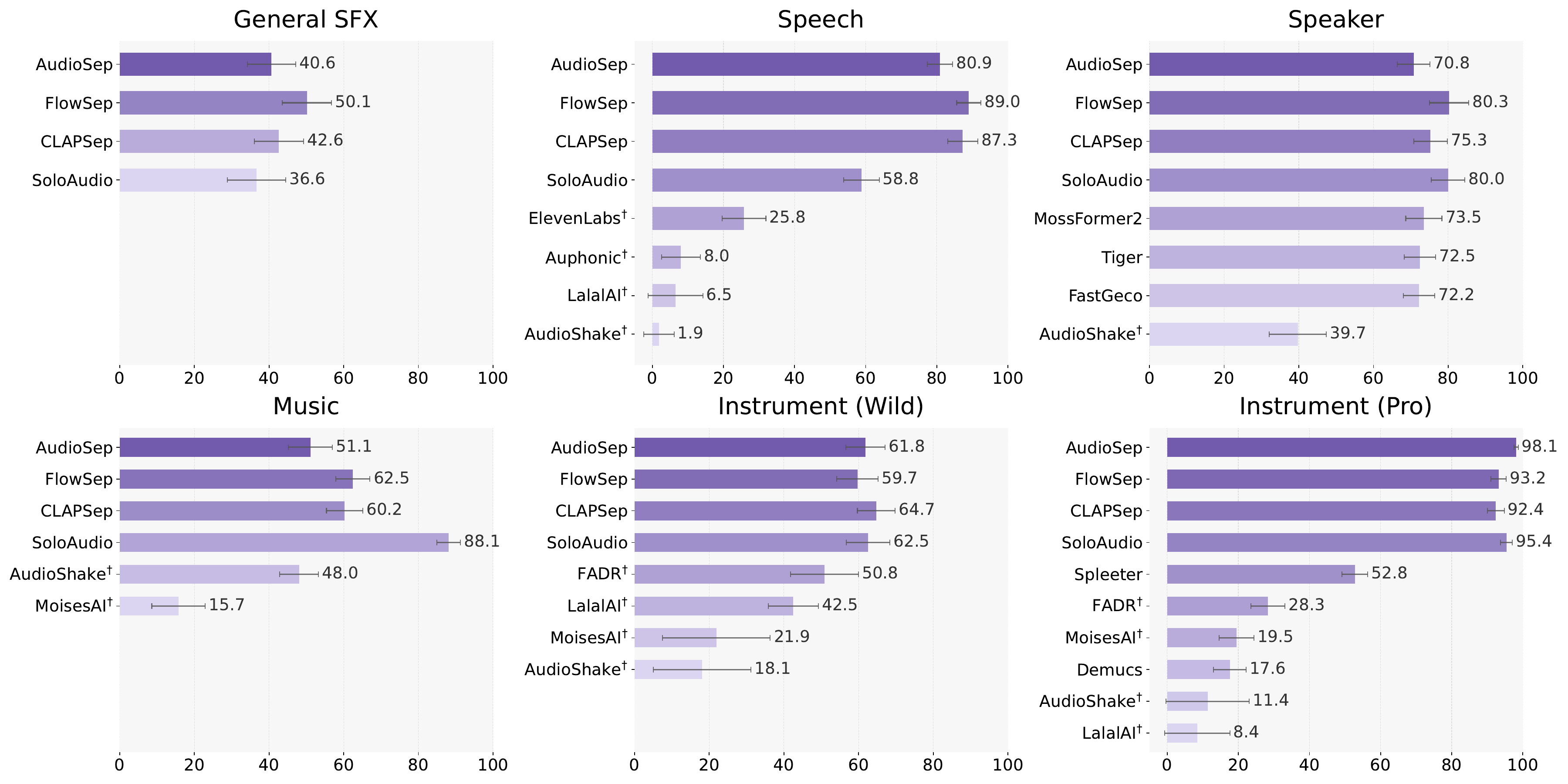}
    \caption{Net Win Rate (\%) of \samaudio against SoTA separation models in text-prompted tasks. $\dagger$: proprietary models}
    \label{fig:nwr_text_baselines}

\end{figure*}

%% file: tables/visual_main_result.tex
\begin{table*}[h]
\centering

\adjustbox{max width=.7\textwidth}{%
\begin{tabular}{lc c>{\columncolor{gray!15}}c c>{\columncolor{gray!15}}c c>{\columncolor{gray!15}}c}
\toprule
 & & \multicolumn{2}{c}{\textbf{General SFX}}
 & \multicolumn{2}{c}{\textbf{Speaker}}
 & \multicolumn{2}{c}{\textbf{Instr (wild)}} \\
\cmidrule(lr){3-4}\cmidrule(lr){5-6}\cmidrule(l){7-8}
\textbf{Model} & \textbf{Generic} & IB & OVR & IB & OVR  & IB & OVR  \\
\midrule
AV-MossFormer2~\citep{zhao2025clearvoice} & \xmark & -  & - & 0.20 & 2.62 & - & -  \\
IIANet~\citep{IIANet}  & \xmark       & - & - & 0.16 & 2.41  & - & -  \\
ClipSep~\citep{dong2023clipsep} & \cmark       & 0.16 & 1.53 & 0.14 & 1.47  & 0.15 & 1.12 \\
DAVIS-Flow~\citep{huang2025davis} & \cmark         & 0.14 & 1.96 & 0.13 & 1.97  & 0.13 & 2.08 \\
DAVIS-Flow (Music)~\citep{huang2025davis} & \xmark   & - & - & - & - & 0.13 & 2.40   \\
%   \\
\samaudio{} & \cmark & \textbf{0.25} & \textbf{2.61}  & \textbf{0.24} & \textbf{3.07}  & \textbf{0.24} & \textbf{2.56}  \\
\bottomrule
\end{tabular}
}
\caption{Comparison against visual-prompted baselines. --: Not applicable. OVR: overall subjective score. }
\label{tab:res_main_visual}

\end{table*}

\begin{figure*}[htp]
    \centering
    \includegraphics[width=\textwidth]{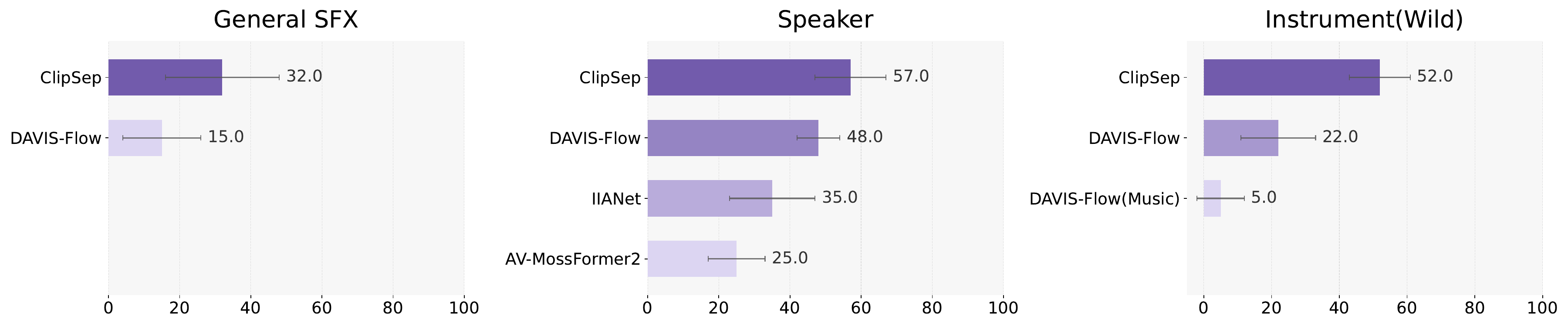}
    \caption{Net Win Rate (\%) of \samaudio against SoTA separation models in visual-prompted tasks}
    \label{fig:nwr_visual_baselines}
\end{figure*}

%% file: tables/visual_sample_figure.tex
% 

\begin{figure*}[htp]
    \centering

    % --- Sample 1 ---
    \includegraphics[width=\textwidth]{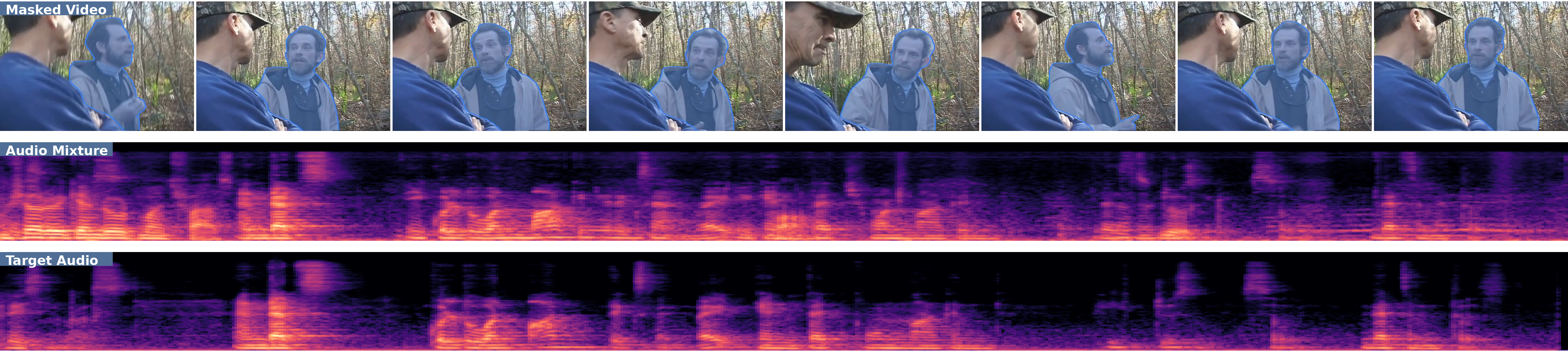}
    \vspace{4pt} % small gap

    % 

    % --- Sample 3 ---
    \includegraphics[width=\textwidth]{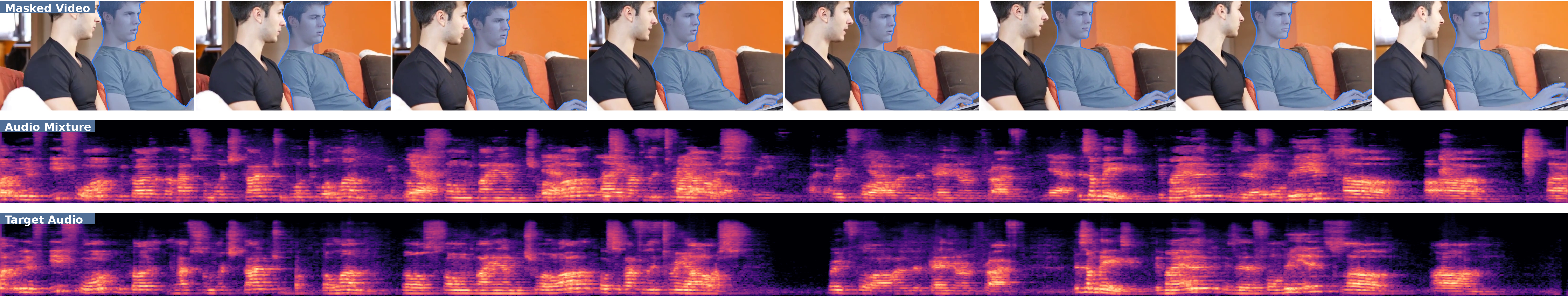}
    \caption{Visual-prompted samples where text descriptions are ambiguous. 
    Each example shows (\textbf{top}) masked video, (\textbf{middle}) input mixture spectrogram, 
    and (\textbf{bottom}) separated target spectrogram. The target speaker is highlighted in each video.}
    \label{fig:video_prompt_samples}

\end{figure*}

%% file: tables/span_main_result.tex
\begin{table*}[h]
    \centering
    \adjustbox{max width=0.9\textwidth, center}{%
    \begin{tabular}{c cc>{\columncolor{gray!15}}c cc>{\columncolor{gray!15}}c cc>{\columncolor{gray!15}}c cc>{\columncolor{gray!15}}c cc>{\columncolor{gray!15}}c}
    \toprule
    \multirow{2}{*}{\shortstack{\textbf{Prompt} \\ \textbf{Modality}}}  &
      \multicolumn{3}{c}{\textbf{General}} &
      \multicolumn{3}{c}{\textbf{Speech}} &
      \multicolumn{3}{c}{\textbf{Speaker}} &
      \multicolumn{3}{c}{\textbf{Music}} &
      \multicolumn{3}{c}{\textbf{Instr(wild)}} \\
    \cmidrule(lr){2-4}\cmidrule(lr){5-7}\cmidrule(lr){8-10}\cmidrule(lr){11-13}\cmidrule(lr){14-16}
     &
      SAJ & CLAP & OVR &
SAJ & CLAP & OVR &
SAJ & CLAP & OVR &
SAJ & CLAP & OVR &
SAJ & CLAP & OVR \\
    % 

    % 

    % 

    %  \\
    \midrule
     text         & 4.11 & 0.31 & 3.32 & 4.59 & 0.33 & 4.18 & 4.08 & 0.17 & 3.63 & 4.30 & 0.28  & 4.09 & \textbf{4.45} & 0.30 & 3.78 \\
    span         &3.27 & 0.30 & 3.54 & 3.37 & 0.28 & 3.65 & 3.26 & 0.26 & 4.08 & 2.18 & -0.04 & 2.57  & 1.81 & 0.01 & 2.20 \\
    text + span  & \textbf{4.25} & \textbf{0.31} & \textbf{4.04} & \textbf{4.66} & \textbf{0.35} & \textbf{4.33} & \textbf{4.51} & \textbf{0.18} & \textbf{4.22} & \textbf{4.38} & \textbf{0.27} & \textbf{4.19} & 4.42 & \textbf{0.32} & \textbf{3.88} \\

    \bottomrule
    \end{tabular}
    }
    \caption{Comparison between span prompting and text prompting for \samaudio{}. OVR: overall subjective score. For all the metrics below, higher is better.}
    \label{tab:res_main_span}

\end{table*}

\begin{figure*}[htp]
    \centering
    \includegraphics[width=\linewidth]{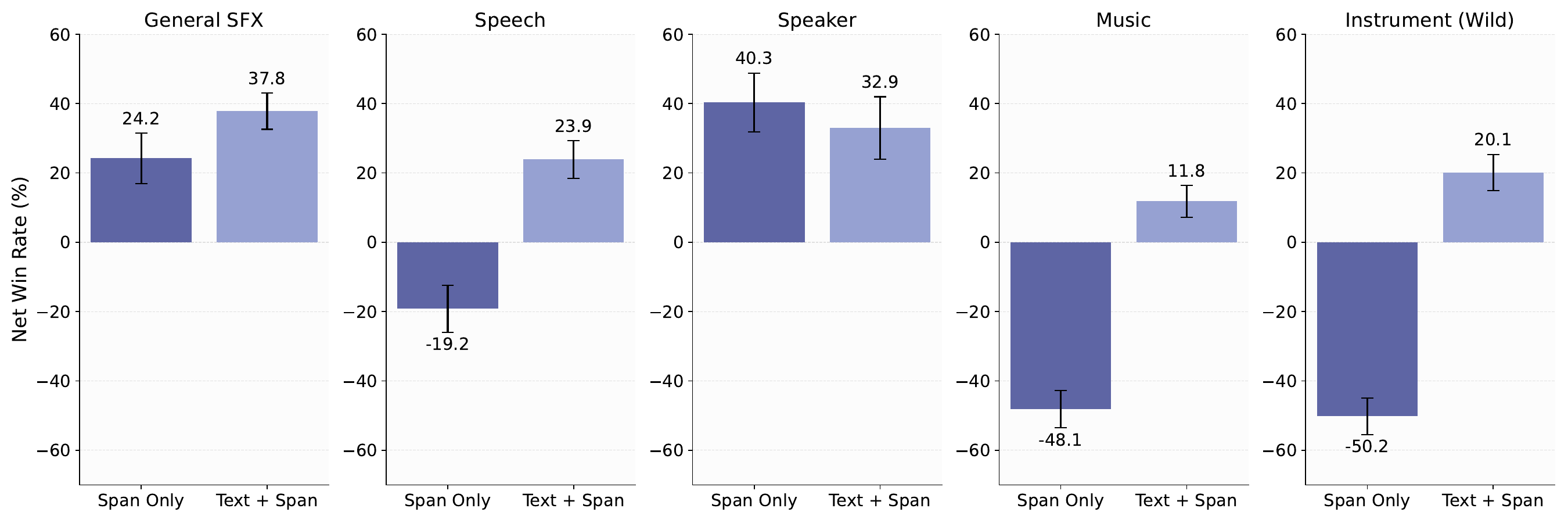}
        \caption{Net Win Rate (\%) of \samaudio{} with text \& span / span as input against a text-only model}
    \label{fig:nwr_text_span_comparison}
\end{figure*}

%% file: tables/abl_effect_span.tex
\begin{figure}[htp]
    \centering
    % --- Left: Table ---
    \begin{minipage}[t]{0.45\textwidth}
\vspace{0pt} % 
        \centering
        \adjustbox{max width=\textwidth}{
        \begin{tabular}{ccccc}
        \toprule
             Task & w/ Pred Span & SAJ & CLAP & OVR  \\
        \midrule
             \multirow{2}{*}{General SFX} & \xmark & 4.11 & \textbf{0.31} & 3.36 \\
             & \cellcolor{gray!15}\cmark  & \cellcolor{gray!15}\textbf{4.35} & \cellcolor{gray!15}\textbf{0.31} & \cellcolor{gray!15}\textbf{3.89} \\
        \midrule
             \multirow{2}{*}{Speech} & \xmark & 4.59 & 0.33 & 4.17 \\
             & \cellcolor{gray!15}\cmark & \cellcolor{gray!15}\textbf{4.67} & \cellcolor{gray!15}\textbf{0.35} & \cellcolor{gray!15}\textbf{4.22} \\
        \midrule
             \multirow{2}{*}{Speaker} & \xmark & 4.08 & 0.17 & 3.62 \\
             & \cellcolor{gray!15}\cmark & \cellcolor{gray!15}\textbf{4.51} & \cellcolor{gray!15}\textbf{0.18} & \cellcolor{gray!15}\textbf{4.01} \\
        \midrule
             \multirow{2}{*}{Music} & \xmark & 4.30 & \textbf{0.28} & \textbf{4.16} \\
             & \cellcolor{gray!15}\cmark & \cellcolor{gray!15}\textbf{4.45} & \cellcolor{gray!15}0.26 & \cellcolor{gray!15}4.12 \\
        \midrule
             \multirow{2}{*}{Instr(wild)} & \xmark & \textbf{4.45} & 0.30 & 3.70 \\
             & \cellcolor{gray!15}\cmark & \cellcolor{gray!15}4.32 & \cellcolor{gray!15}\textbf{0.31} & \cellcolor{gray!15}\textbf{3.88} \\
        \midrule
             \multirow{2}{*}{Instr(pro)} & \xmark & \textbf{4.83} & \textbf{0.28} & \textbf{4.16} \\
             & \cellcolor{gray!15}\cmark & \cellcolor{gray!15}4.82 & \cellcolor{gray!15}\textbf{0.28} & \cellcolor{gray!15}4.12 \\
        \bottomrule
        \end{tabular}
        }
        \captionof{table}{Using vs. not using predicted span for text-prompting. 
        OVR: overall subjective score. For all the metrics below, higher is better.}
        \label{tab:res_abl_span_pred}

    \end{minipage}%
    \hfill
    \begin{minipage}[t]{0.5\textwidth}
\vspace{0pt} % 
        \centering
        \includegraphics[width=\linewidth]{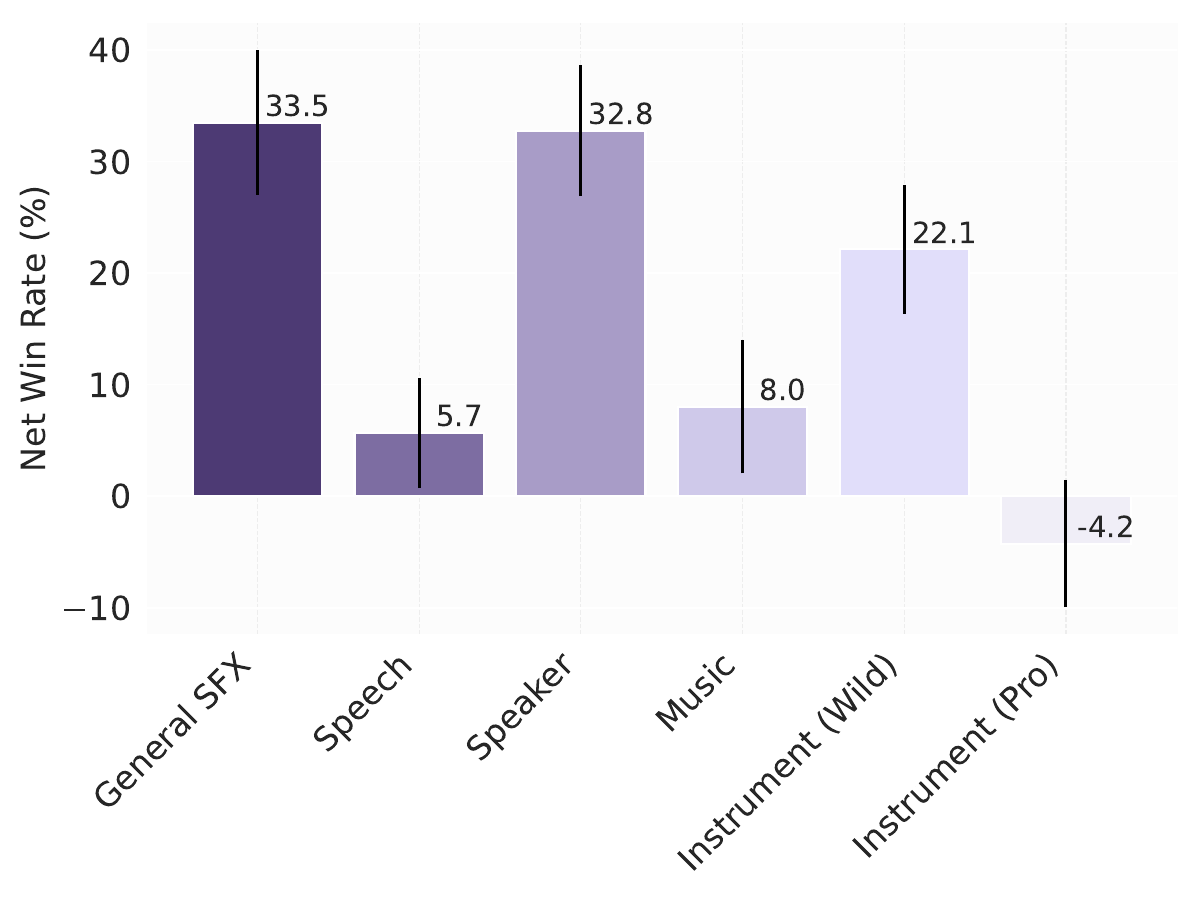}
        \captionof{figure}{Net Win Rate (\%) of \samaudio{} using predicted span against not using predicted span for text prompting.}
        \label{fig:nwr_text_span_comparison}
    \end{minipage}
\end{figure}

%% file: tables/music_removal_comparison.tex
\begin{figure*}[htp]
\centering

\begin{minipage}[c]{0.45\textwidth}
    \centering
    \adjustbox{max width=\linewidth}{
    \begin{tabular}{l>{\columncolor{gray!15}}c} 
    \toprule
    Model & OVR \\
    \midrule
    AudioShake~\citep{audioshake} & 3.75 \\
    MoisesAI~\citep{moisesai} & 4.00 \\
    \samaudio{} & \textbf{4.05} \\
    \bottomrule
    \end{tabular}
    }
    \captionof{table}{Comparison against baselines for music removal}
    \label{tab:ovr_music_removal}
\end{minipage}
\hfill % 
\begin{minipage}[c]{0.5\textwidth}
    \centering
    \includegraphics[width=\linewidth]{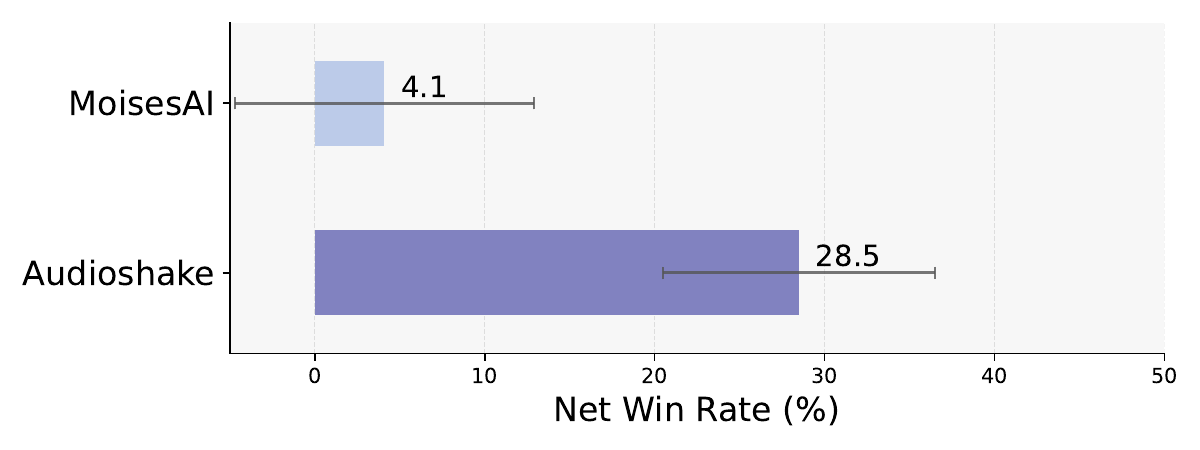}
    \captionof{figure}{Net Win Rate (\%) of \samaudio{} over baselines in music removal}
    \label{fig:nwr_music_removal}
\end{minipage}

\end{figure*}

%% file: tables/saj_com.tex
\begin{table*}[h]
\centering
\adjustbox{max width=\textwidth}{%
\begin{tabular}{l| cccc |cccc |cccc}
\toprule
 & \multicolumn{4}{c|}{\textbf{Speech}}
 & \multicolumn{4}{c|}{\textbf{Music}}
 & \multicolumn{4}{c}{\textbf{Sound}} \\
% }
Model
& Overall & Recall & Precision & Faithfulness
& Overall & Recall & Precision & Faithfulness
& Overall & Recall & Precision & Faithfulness \\
\midrule
\rowcolor{gray!15}\multicolumn{13}{c}{Pearson Correlation Coefficient (PCC)} \\
\midrule
CLAP &  0.490&  0.431& 0.283 &  0.477&  0.487&  0.416&  0.385&  0.432&  0.367&  0.431&  0.283&  0.418\\
SDR Estimator        & 0.336 & 0.004 & 0.403 & 0.055 &  0.369&  0.157&  0.388&  0.182&  0.181&  0.040&  0.222&  0.055\\
Gemini-2.5-pro         &  0.487& 0.498 &  0.169&  0.430 &0.351&0.287&0.115&0.303 &0.462&0.493&0.192&0.369\\
\textbf{SAM Audio Judge}        & \textbf{0.883} & \textbf{0.943} & \textbf{0.841} & \textbf{0.891} & \textbf{0.815} & \textbf{0.858} & \textbf{0.766} & \textbf{0.791} & \textbf{0.815} & \textbf{0.837} & \textbf{0.775} & \textbf{0.818} \\
\midrule
\rowcolor{gray!15}\multicolumn{13}{c}{Spearman Rank Correlation Coefficient (SRCC)} \\
\midrule
CLAP &  0.380&  0.291& 0.325 &  0.273&  0.285&  0.293&  0.199&  0.296&  0.493&  0.376& 0.388 & 0.406 \\
SDR Estimator        & 0.338 & 0.000 & 0.395 & 0.079 & 0.390 & 0.203 & 0.375 & 0.210 & 0.173 & 0.053 & 0.220 & 0.073 \\
Gemini-2.5-pro        &0.495&0.361&-0.015&0.117 &0.338&0.232&0.010&0.008 &0.390&0.324&-0.006&0.180
\\
\textbf{SAM Audio Judge}        & \textbf{0.817} & \textbf{0.573} & \textbf{0.774} & \textbf{0.573} & \textbf{0.714} & \textbf{0.569} & \textbf{0.658} & \textbf{0.476} & \textbf{0.781} & \textbf{0.660} & \textbf{0.734} & \textbf{0.607} \\
\bottomrule
\end{tabular}
}
\caption{Comparison Between SAM Audio Judge Model and Baselines}
\label{tab:res_saj_com}

\end{table*}

%% file: tables/saj_reranking.tex
\begin{table}[h!]
\centering
\adjustbox{max width=0.7\textwidth}{%
\begin{tabular}{l|l|ccccc}
\toprule
\textbf{Reranker A} & \textbf{Reranker B}
& \textbf{Speech} & \textbf{Speaker} & \textbf{Music} & \textbf{Instr(wild)} & \textbf{General} \\
\midrule
SAJ & CLAP & 0.17 & 0.19 & 0.09&0.01&0.15\\
CLAP w/ SAJ &  CLAP & 0.18 & 0.18 & 0.14& 0.03&0.06\\
CLAP w/ SAJ &  SAJ & 0.00 & 0.02& 0.10&0.15&0.07\\
\bottomrule
\end{tabular}
}
\caption{Net Win Rate Comparison Between Rerankers (Reranker A vs. Reranker B)}
\label{tab:saj_reranking}

\end{table}

%% file: conclusion.tex
\section{Conclusion}
\label{sec:conclusion}

We presented \samaudio{}, a general-purpose audio separation model that supports multimodal prompting and achieves state-of-the-art performance across various audio separation tasks. \samaudio{} advances universal audio separation by scaling both data and model capacity with flow matching.
To mitigate the scarcity of ground-truth stems, we developed scalable data construction pipelines, including domain-aware synthetic mixing and a large pseudo-labeling process that bootstraps stems from natural recordings using intermediate \samaudio{} checkpoints. These strategies provide broad coverage without requiring manual stem annotation. We further introduced visual and span prompting as complementary modalities to text prompting. Notably, span prompting substantially improves text-based separation and enables practical iterative refinement.

To support future research, we additionally release \samaudiobench{} and \samaudiojudge{}. \samaudiobench{} offers a carefully balanced benchmark with human-annotated text, visual, and span prompts. \samaudiojudge{} provides a reference-free metric with significantly higher correlation to human perception than existing alternatives, and can also be used as a post-processing module to improve separation quality.

Despite its strong performance, our results also reveal several limitations. In particular, visual prompting is noticeably less effective than text prompting, and general sound effects remains more challenging than specialized domains such as speech. Addressing these gaps will likely require stronger audio–visual grounding and better modeling of complex, multi-source acoustic scenes, which we leave for future work.

%% file: acknowledgement.tex
\section*{Acknowledgment}
The authors would like to thank Peng-Jen Chen, Dangna Li, Robin San Roman, Leying Zhang, Carleigh Wood, Andrew Westbury, George Orlin, Anushka Sagar, Vivian Lee, Cynthia Gao, Ida Cheng, Roman Rädle, Victor Loh, Alex He, Dex Honsa, Eric Gan, Kei Koyama, Kevin Ngo, Meng Wang, Michelle Chan, Phillip Thomas, Andrew Huang, Robbie Adkins, Jason Holland, Josh Terry, Ben Samples, Karla Martucci, Bruno Nakano, Yoko Kristiansen, Ashley Gabriel, Athyu Eleti, Andy Bass, Helen Klein, Emma Leibman Baker, Britt Montalvo,  Nikhila Ravi, and Manohar Paluri for their inspiring discussions and timely support throughout this work. 

%% file: appendix.tex
\input{humanevalappendix}

\newpage

\input{samaudiobenchappendix}

\input{judgeappendix}

\input{ablation_study}

%% file: humanevalappendix.tex
\section{Subjective Evaluation}
\subsection{Design}
\label{sec:subjective_evaluation_design}
Our protocol is designed to meet four requirements: (i) \emph{ecological validity} in complex, real audio/video; (ii) \emph{sensitivity} to small deltas for ablations; (iii) \emph{comparability} across prompt modalities (text, visual masklets, temporal spans) and domains (speech, music, instruments, general sounds); and (iv) \emph{operational feasibility} given the scale and diversity of test items. These criteria led us away from classic single-stimulus MOS/CMOS and MUSHRA-style multi-stimulus tests for this task, and toward a side-by-side ACR hybrid.

\subsubsection{Why not MUSHRA, MOS, or CMOS?}

\textbf{MUSHRA (ITU-R BS.1534)} excels when a high-quality \emph{reference} and \emph{anchors} are available for each item, enabling multi-stimulus ratings with hidden references and low anchors~\cite{ITU_BS1534}. In our setting, however, (a) we often lack any trustworthy reference extract from real mixtures; (b) anchor selection is non-trivial for heterogeneous targets (speech, instruments, events) and can induce \emph{anchor and reference-matching biases}—higher-quality references can depress scores for otherwise good outputs; scores can drift toward “similarity to reference” rather than perceived quality~\cite{Varadhan2024MUSHRAAnchors}; and (c) multi-stimulus sessions amplify \emph{range equalization} and \emph{session drift} effects~\cite{Cooper2023RangeEqualization}. Combined with significant \emph{cost and time} overhead, these make MUSHRA mismatched to our in-the-wild, multi-domain separation benchmark.

\textbf{MOS/ACR (ITU-T P.800)} offers simple single-stimulus absolute ratings on a 5-point scale~\cite{ITU_P800}. While operationally easy, single-stimulus MOS has well-known shortcomings for our use case: (i) limited sensitivity for small model deltas in ablation studies; (ii) susceptibility to \emph{anchoring} and \emph{range} biases—absolute scores drift with context over a session~\cite{Tversky1974Anchoring,Cooper2023RangeEqualization}; and (iii) no direct handle on \emph{relative} model ranking without substantially increasing sample sizes.

\textbf{CMOS/CCR (paired comparisons)} addresses relative sensitivity by asking evaluators to express preference (and sometimes strength) between two stimuli~\cite{ITU_P800}. Pure preference designs, however, provide \emph{no} insight into absolute performance thresholds (e.g., “good enough” for deployment), and can be brittle when different error modes produce similar overall impressions.

\subsubsection{Our Hybrid: Side-by-Side ACR with Preference Tie-Breaker}

We therefore adopt a \textbf{side-by-side ACR} protocol with an \textbf{always-on preference} question:

\begin{itemize}
  \item \textbf{Presentation}: Two model outputs (A, B) for the \emph{same} item are presented \emph{side-by-side}, with the target specification (text prompt, SAM masklets, and/or temporal spans) and the original mixture for reference context.
  \item \textbf{Absolute ratings}: Evaluators assign \emph{independent} 5-point ACR scores to A and B for three dimensions aligned to separation goals: \emph{Recall} (how much target was extracted), \emph{Precision} (how much non-target leaked in), and \emph{Faithfulness} (similarity to how the target sounded in-context). We also collect an \emph{Overall} score for both model outputs.
  \item \textbf{Preference tie-breaker}: Evaluators \emph{always} answer a direct preference question (“Which extracted audio do you prefer, A or B?”), even when ACR scores differ. This “forced” tie-breaker increases inter-annotator agreement (IAA) and sharpens delta estimates. Conditionally asking this question only when evaluators provide identical ACR scores for A and B may induce evaluators to provide different scores more frequently to avoid the extra conditional question (though this hypothesis was not directly tested).
  \item \textbf{Modality coverage}: The same protocol applies across prompt modalities—text, visual (SAM masklets), and temporal spans—allowing controlled ablations (e.g., text-only vs.\ text+visual vs.\ visual-only) on the \emph{same} in-the-wild items.
\end{itemize}

This hybrid provides \emph{the best of both worlds}: \emph{absolute} quality signals for deployment-readiness and \emph{relative} signals for ablations and model selection.

\subsubsection{Empirical Observations}

Across multiple studies, we observe:

\begin{itemize}
  \item \textbf{Inter-annotator agreement improves} with the preference tie-breaker: preference agreement and the consistency of absolute deltas increase relative to ACR-without-tie-breaker and pairwise-only preference protocols, measured in terms of Gwet's AC2 \citep{gwet2014ac2}.
  \item \textbf{Sharper deltas}: The presence of the tie-breaker leads evaluators to produce \emph{larger} ACR score deltas on average (approximately +0.15 points per item), aligning absolute scores with expressed preference and increasing sensitivity for near-tied comparisons.
  \item \textbf{Lower uncertainty on deltas}: Side-by-side judgment reduces confidence interval (CI) width for differences between models by up to \(\sim\)20\% versus single-stimulus rating, implying \(\sim\)30\% cost savings to achieve the same sensitivity in A/B testing, where CI width is measured by bootstrap.
  \item \textbf{Handling time trade-off}: Side-by-side ACR roughly doubles per-item handling time relative to pairwise-only (more questions per item, 2 minute handling vs 1 minute handling for CMOS), but yields both absolute and relative signals in a single assignment. 
  \item \textbf{Anchoring effects persist}: Absolute scores depend on comparison context—presenting an output alongside a higher-quality counterpart depresses its ACR scores, consistent with broader anchoring and reference-matching literature~\cite{Tversky1974Anchoring,Varadhan2024MUSHRAAnchors,Cooper2023RangeEqualization}. We therefore emphasize comparative statistics (win rate, ACR deltas) over raw absolute means.
\end{itemize}

\subsubsection{Operational Details}

\paragraph{Instructioning and training.} Appendix~\ref{appx:humanevalprotocol} contains the full evaluator instructions and examples. Evaluators are trained to:
(i) attend to the provided prompts (text/masklets/spans),
(ii) separate “precision” (non-target leakage) from “recall” (missed target),
(iii) judge “faithfulness” relative to the target’s timbral/temporal character in the mixture (not similarity to any external reference).

\paragraph{Randomization and blinding.} Model identities are blinded; A/B order is randomized per item; items are batched to avoid long runs from one domain/modality. % 

\paragraph{Quality controls.} We leverage several tools for quality controls; (i) vendor-side quality assurance protocols, where a core group of trusted and trained evaluators check evaluations from a sample of annotators, (ii) Bayesian modeling of rater quality: we leverage CLARA \citep{Nguyen2020CLARA} to obtain posteriors over rater confusion matrices and remove raters with anomalously high aggregate measures of deviation from ground truth.  %}

Given anchoring risks, we emphasize \emph{comparative} statistics over raw absolute means, and caution against over-interpreting cross-experiment shifts in absolute levels.

\subsubsection{Positioning vs.\ Standards and Literature}

Our protocol inherits \emph{absolute} rating interpretability from ACR/MOS~\cite{ITU_P800} while retaining \emph{relative} sensitivity akin to CMOS/CCR~\cite{ITU_P800}, but avoids the \emph{reference/anchor} dependencies and multi-stimulus biases of MUSHRA~\cite{ITU_BS1534,Varadhan2024MUSHRAAnchors}. It is aligned with recent large-scale audio separation evaluations that highlight the \emph{disconnect} between reference-based metrics (SDR/SI-SDR/SAR) and human perception across stems~\cite{stoter2024bakeoff}, and with broader findings on evaluator behavior (\emph{anchoring}, \emph{range equalization}, \emph{session drift})~\cite{Tversky1974Anchoring,Cooper2023RangeEqualization}. % 

\paragraph{Limitations.} Side-by-side ACR remains subject to context effects; absolute scores should be interpreted cautiously. Developing better reference-free objective metrics that correlate with human judgments—especially for vocals and complex mixtures—remains an open area~\cite{stoter2024bakeoff}.

\subsection{Subjective evaluation protocol\label{appx:humanevalprotocol}}
Below we present the content of the instructions presented to annotators when evaluating audio separation models.
\input{humanevalprotocol}

%% file: humanevalprotocol.tex
\subsection*{Audio Separation}
You will be evaluating models that perform audio separation.
\paragraph{What is audio separation?}
When using an audio separation model, users provide the model with either an audio track or a video with sound, and descriptions of which parts of the sound they want the model to extract (for example ``guitar'' or ``dog barking''). The model then extracts the part of the audio that the user requests without any other sounds.
\subsection*{Key Terms}
\begin{description}
  \item[Prompt] The user description of the target audio (e.g., ``dog barking'' or the visually highlighted dog in the video).
  \item[Source audio] The original video or audio provided to the model.
  \item[Target sounds] The portions of the source audio that the user is requesting.
  \item[Non-target sounds] All other sounds in the source audio not requested by the user.
  \item[Extracted audio] The audio that the model produces from the source audio and the prompt.
\end{description}
\subsection*{What is ``distortion''?}
When we say a sound is ``distorted'', we are referring to how a particular sound may have changed from the way it sounds in the source audio.
\paragraph{Distortions}
Distortions are modifications of sounds that originate from a discernable source in the source audio. These modifications are not the result of mixing with other sounds present in the source audio.
If you can't tell if a sound is a distortion or simply the result of mixing with other sounds in the source audio, you may consider it to be a distortion.
\paragraph{Examples of distortions}
\begin{itemize}
  \item Different levels of ``bass'' or ``treble'': the target sound may sound like it is more or less ``bass''-y, or there are more high pitched sounds.
  \item Pitch differences: if the sound is higher or lower pitched than in the source audio.
  \item Static / artifacts / other noises with no discernable origin in the source audio: crackling, static, popping noises, etc. that are not present in the source audio, or are hard to tell.
  \item Garbling: if, for instance, speech sounds garbled when it wasn't in the source audio.
\end{itemize}
\subsection*{Overview}
\paragraph{Model inputs and outputs that you will be presented with}
In this evaluation you will be presented with:
\begin{itemize}
  \item The source audio (and possibly video).
  \item The prompt, which will be in one of three forms described later in this document.
  \item The extracted audio from two separate models.
\end{itemize}
First you will be asked several questions about how well the extracted audio of each model matches the target sounds that the user requested in their prompt.
After answering these questions, you will then give the extracted audio an overall score based on its:
\begin{enumerate}
  \item Faithfulness to the prompt: how well did the model follow the user's instructions about which sounds to extract?
  \item Faithfulness to the sound of the target audio: does the extracted target sound actually sound like the target sound in the source audio?
\end{enumerate}
\subsection*{The types of prompts you may see}
\begin{table}[h]
\centering
\begin{tabularx}{\textwidth}{l l X}
\toprule
\textbf{Prompt type} & \textbf{Example} & \textbf{Description} \\
\midrule
Text-based & --- & A text description of the target sound(s) that the user wants the model to extract from the source audio. \\
Video-based & --- & A highlighted portion of the source video that is creating the target sounds that the user wants the model to extract from the source audio. \\
Span-based & --- & Highlighted \textcolor{yellow}{yellow} segments of the audio where the target sound that the user wants the model to extract \textbf{is} present; highlighted \textcolor{red}{red} segments of the audio where the target sound is \textbf{not} present; regions of the audio without spans may or may not contain the target sound. \\
\bottomrule
\end{tabularx}
\end{table}
\subsection*{What you will be annotating}
\paragraph{Survey questions for each model's extracted audio}
We will ask you to provide specific information about different aspects of each model's performance on a particular set of inputs.
\medskip
\noindent First, we will ask a question to understand if the target sounds that are requested in the prompt are actually present in the source audio. \textit{This first question does not have anything to do with model outputs!}
\begin{description}
  \item[Q1.] Does the source audio contain all of the target sounds requested in the prompt?
  \begin{itemize}
    \item Yes
    \item No
    \item Possibly, I cannot tell
  \end{itemize}
\end{description}
If the answer to the above question is ``Yes,'' you will then be asked the following question:
\begin{description}
  \item[Q2.] What portion of the target sound(s) is/are present in the extracted audio, regardless if other non-target sounds or distortions are present? Consider portion in terms of (1) duration and (2) number of target sounds.
  \begin{itemize}
    \item All
    \item Most (a majority of the target sound is present in extracted audio but some is missing)
    \item Some (a minority of the target sound is present in extracted audio but most is missing)
    \item None
  \end{itemize}
\end{description}
If the answer to Q2 is not ``All'', you will be asked about what aspects of the target sound(s) are missing:
\begin{description}
  \item[Q2a.] Choose one or more options to clarify what aspects of the target sound(s) may be missing from the extracted audio:
  \begin{itemize}
    \item One or more of the target sounds requested in the prompt is completely absent.
    \item One or more target sounds are missing for part of their duration.
    \item When present in the extracted audio, one or more complex target sounds are only partially extracted (ignoring distortions) [e.g., if ``music'' is the target sound and extracted audio is missing an instrument].
    \item Other (please specify).
  \end{itemize}
\end{description}
If your answer to Q2 is not ``None'', we will also ask about similarity:
\begin{description}
  \item[Q3.] For target sounds that are present in the extracted audio, how similar do they sound to how they sounded in the source audio?
  \begin{itemize}
    \item Exactly the same
    \item Some distortions / artifacts
    \item Moderate distortions / artifacts
    \item Sounds completely different, barely recognizeable
  \end{itemize}
\end{description}
Afterwards, you will be asked:
\begin{description}
  \item[Q4.] Are there distinguishable non-target sounds present in the extracted audio?
  \begin{itemize}
    \item Yes
    \item No
  \end{itemize}
  \item[Q5.] How many non-target sounds present in the extracted audio are also present in the source audio?
  \begin{itemize}
    \item All
    \item Some
    \item None
  \end{itemize}
  \item[Q6.] How well do you feel the model removed non-target sounds from the extracted audio?
  \begin{itemize}
    \item Almost perfectly: the vast majority of non-target sounds present in the source audio are filtered out of the extracted audio, no additional non-target sounds are added.
    \item Adequately: Most non-target sounds present in the source audio are filtered out of the extracted audio.
    \item Poorly: Some filtering but most non-target sounds present in the source audio are also mostly present in the extracted audio.
    \item Very poorly: no non-target sounds are filtered out, and additional non-target sounds may be present.
  \end{itemize}
  \item[Q7.] Are there any non-target sounds from the source audio that appear completely unfiltered in the extracted audio?
  \begin{itemize}
    \item Yes, more than one
    \item Yes, one
    \item No [Any non-target audios in the extracted audio have been at least partially filtered from the source audio]
  \end{itemize}
\end{description}
\subsection*{Overall score for each model's extracted audio}
After you provide responses to the above questions, we will ask you to provide an overall score for how well each model performed, on a scale from 1 to 5. This should ideally incorporate and be consistent with all of the information you provided above but just as importantly we would like you to use your own sense of judgement.
\begin{table}[h]
\centering
\begin{tabularx}{\textwidth}{l X}
\toprule
\textbf{Score} & \textbf{Criteria} \\
\midrule
5: Perfect & All target sounds are present in their entirety, and sound identical to the way they sound in the source audio. No non-target sounds present in extracted audio. \\
4: Good & Only minor issues with extracted audio (e.g., minor portions of target audio may be missing or slightly distorted, a very small amount of non-target sounds may be present). \\
3: OK & Some serious issues with extracted audio: target audio is maybe half present and/or non-target sounds are about half present. \\
2: Poor & Many serious issues with extracted audio: some target audio in the extracted audio but heavily distorted, missing, and several non-target sounds present. \\
1: Terrible & Extracted audio is completely incorrect; none of the target audio is in the extracted audio, or, if present, none of the non-target audio is filtered. \\
\bottomrule
\end{tabularx}
\end{table}
\paragraph{Which model's extracted audio do you prefer?}
If your scores for each model output are the same, we will also ask you to pick which model performed better in your opinion. Use your best judgement about which output you prefer, even if neither model performed well.
\begin{itemize}
  \item Group 1
  \item Group 2
  \item Can't decide
\end{itemize}
\subsection*{Frequently asked questions}
\begin{description}
  \item[Distinct types of sounds:] you may be asked to provide input on the number of distinct target or non-target sounds. In some cases this may be simple (the audio has a dog barking and a cat meowing --- there are two distinct sounds here), however in many cases this task may be more ambiguous. For instance, in an audio where two people are talking but there is a rock song playing in the background --- is that three sounds: one for each person speaking and one for the music? Or do we need to count each instrument that makes up the rock song as a distinct sound?
  \begin{itemize}
    \item If the prompt asks for a specific instrument or song component you may consider each instrument of the music in the source audio to be an individual target or non-target sound.
    \item Background noise: for sounds of an environment (street sounds, honking horns, cars passing etc.; or sounds of a cafe) that are not easily discernible, these can be considered a single non-target ``background'' sound.
  \end{itemize}
\end{description}

%% file: samaudiobenchappendix.tex
\section{\samaudiobench \label{appx:samaudiobenchcharacterization}}

\subsection{Characterization and Statistics}
\begin{figure}
\centering
\includegraphics[width=0.9\textwidth]{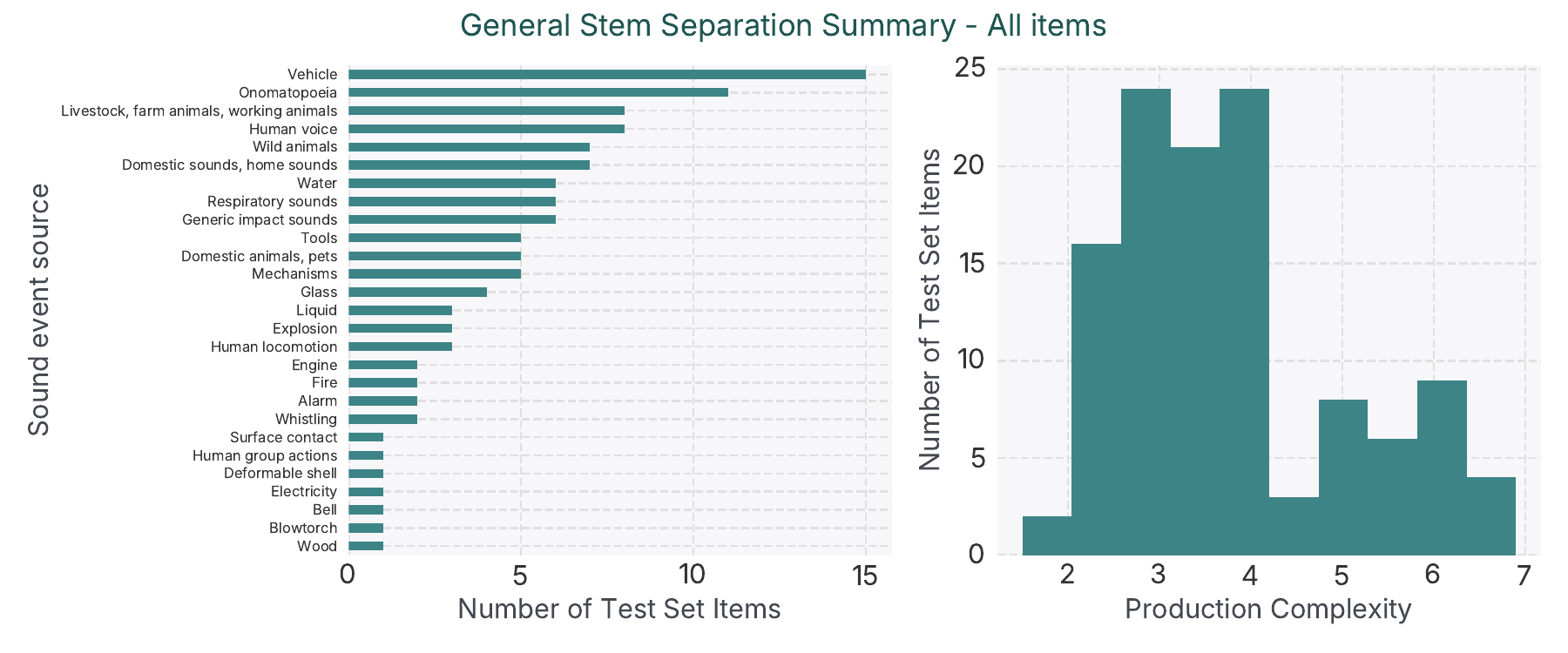}
\includegraphics[width=0.9\textwidth]{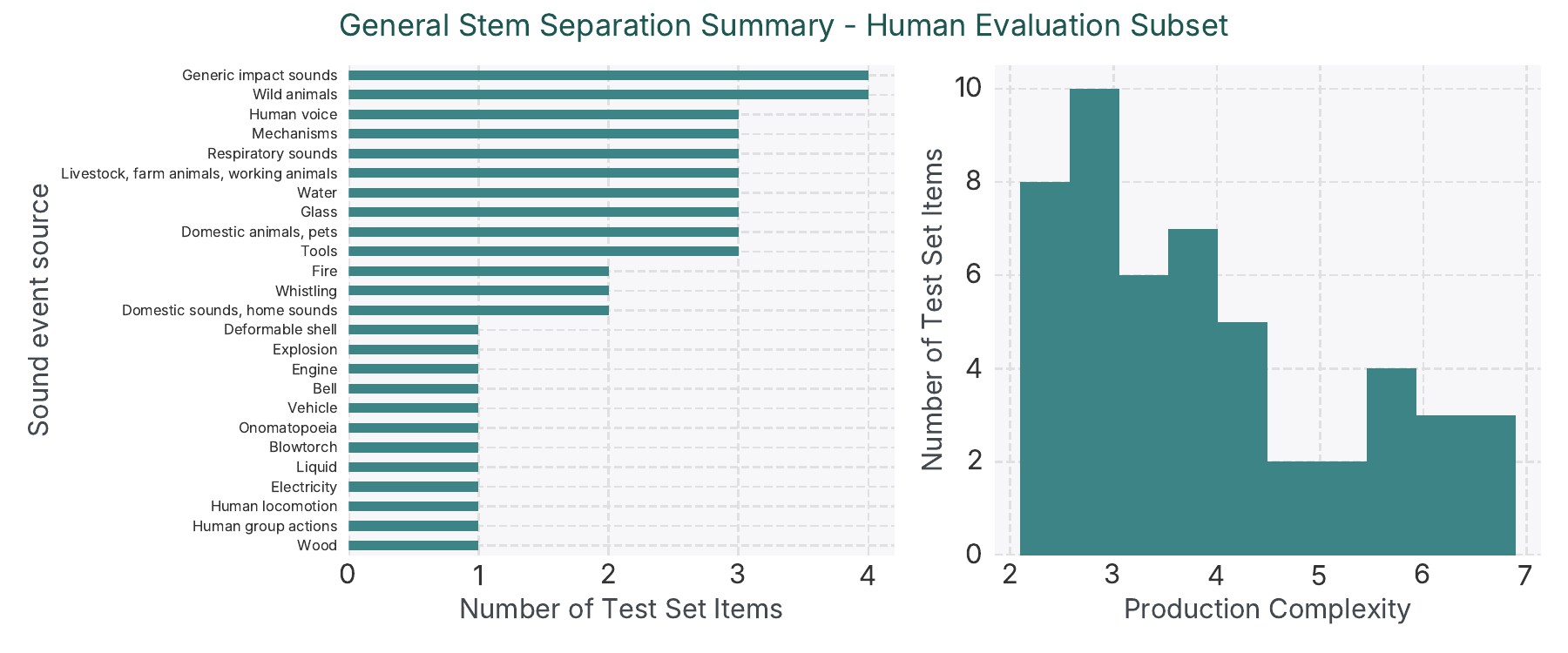}
\caption{\label{fig:samaudiobench_sounds} Statistics on \textbf{General} sound separation task. \textbf{Left}: distribution of target sounds using the second level of the Audioset hierarchy. \textbf{Right}: Distribution of the audio production complexity for each benchmark item. \textbf{Top}: Summary for all samples in \samaudiobench, \textbf{bottom}: summary for samples in the subset used for human evaluations.}
\end{figure}

\begin{figure}
\centering
\includegraphics[width=0.9\textwidth]{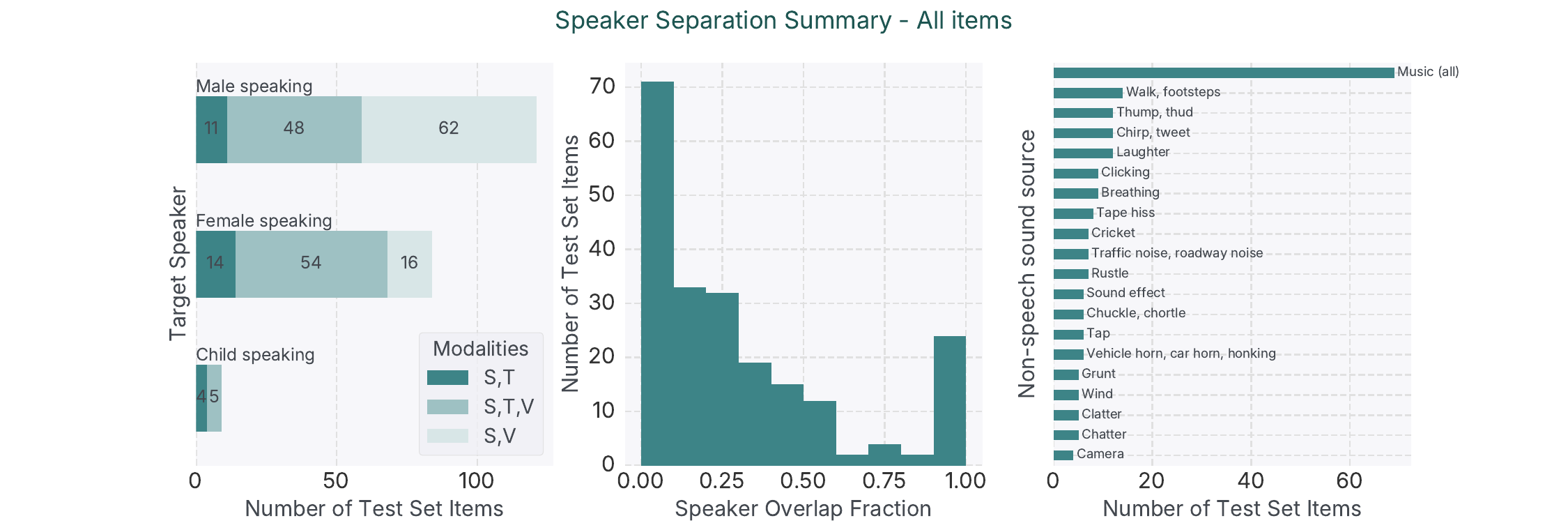}
\includegraphics[width=0.9\textwidth]{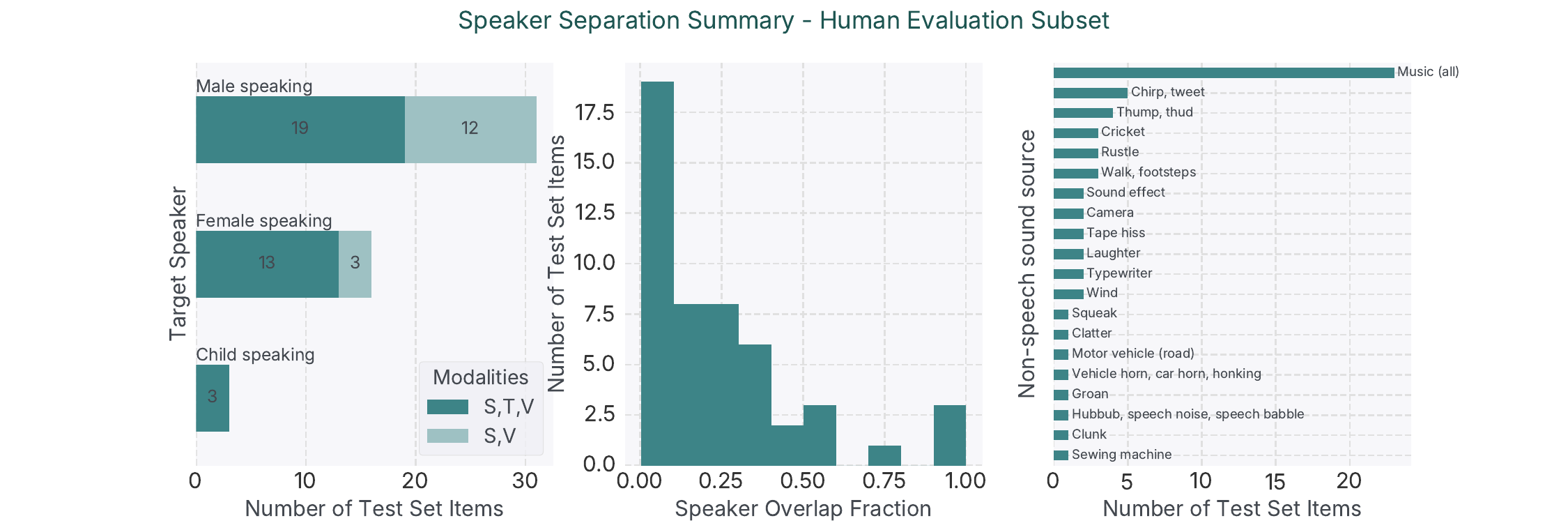}
\caption{\label{fig:samaudiobench_summary_speakersep} Statistics on the speaker separation task for \samaudiobench. \textbf{Left}: is the distribution of target speaker type and modalities available for prompting (some items may not have visual or text prompts available). \textbf{Middle}: distribution of ``speaker overlap'' -- the fraction of target speaker duration during which there is other speech occurring. \textbf{Top}: Summary for all samples in \samaudiobench, \textbf{bottom}: summary for samples in the subset used for human evaluations.}
\end{figure}

Figures~\ref{fig:samaudiobench_sounds} and~\ref{fig:samaudiobench_summary_speakersep} provide an overview of the data distribution for the instrument and speaker separation tasks in \samaudiobench{}.  
For instruments, the benchmark is dominated by a handful of common sources, with a long tail of less frequent instruments, and most audios contain only two or three active instruments.  
For speakers, the dataset spans multiple speaker types and prompting modalities, and exhibits substantial variation in overlap between the target and interfering speakers.  
Non-speech distractors are diverse, ranging from environmental sounds to background music, which reflects the complexity of real-world audio mixtures.

\begin{figure}
\centering
\includegraphics[width=0.9\textwidth]{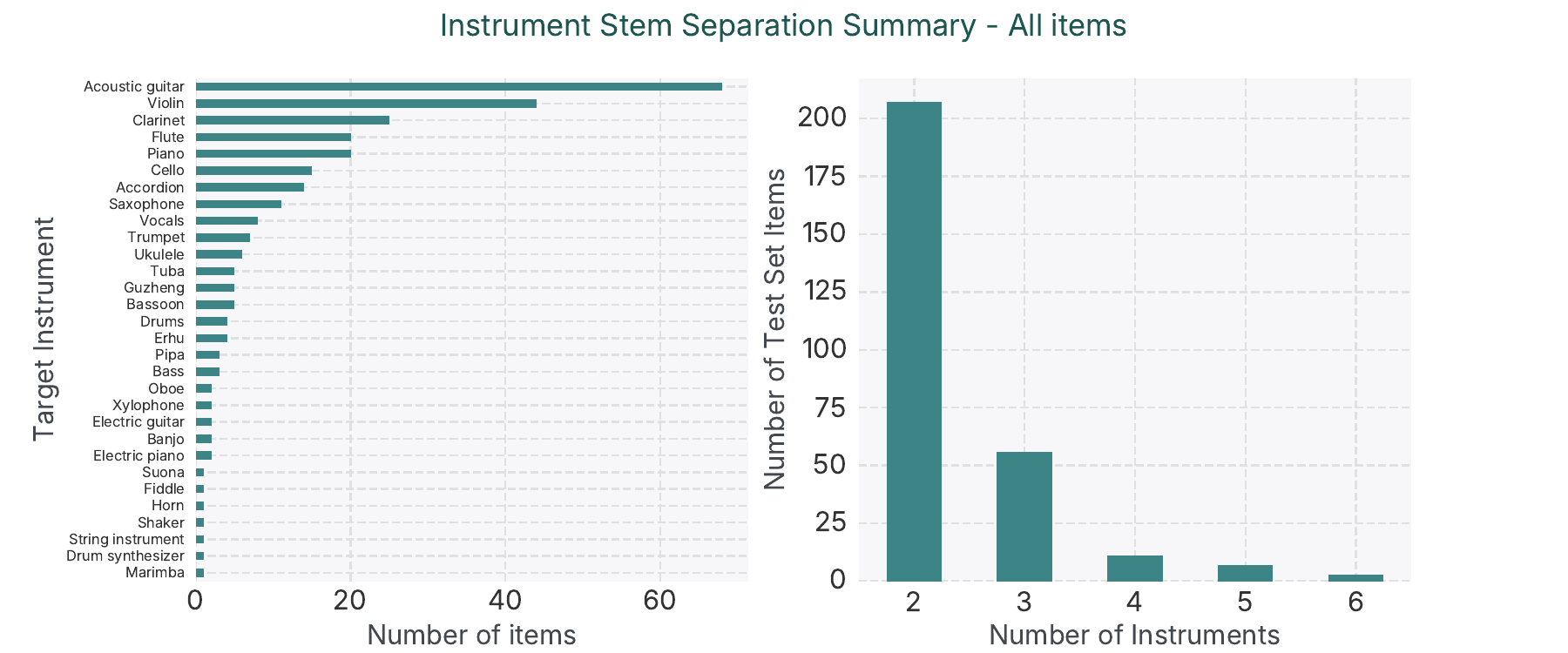}
\includegraphics[width=0.9\textwidth]{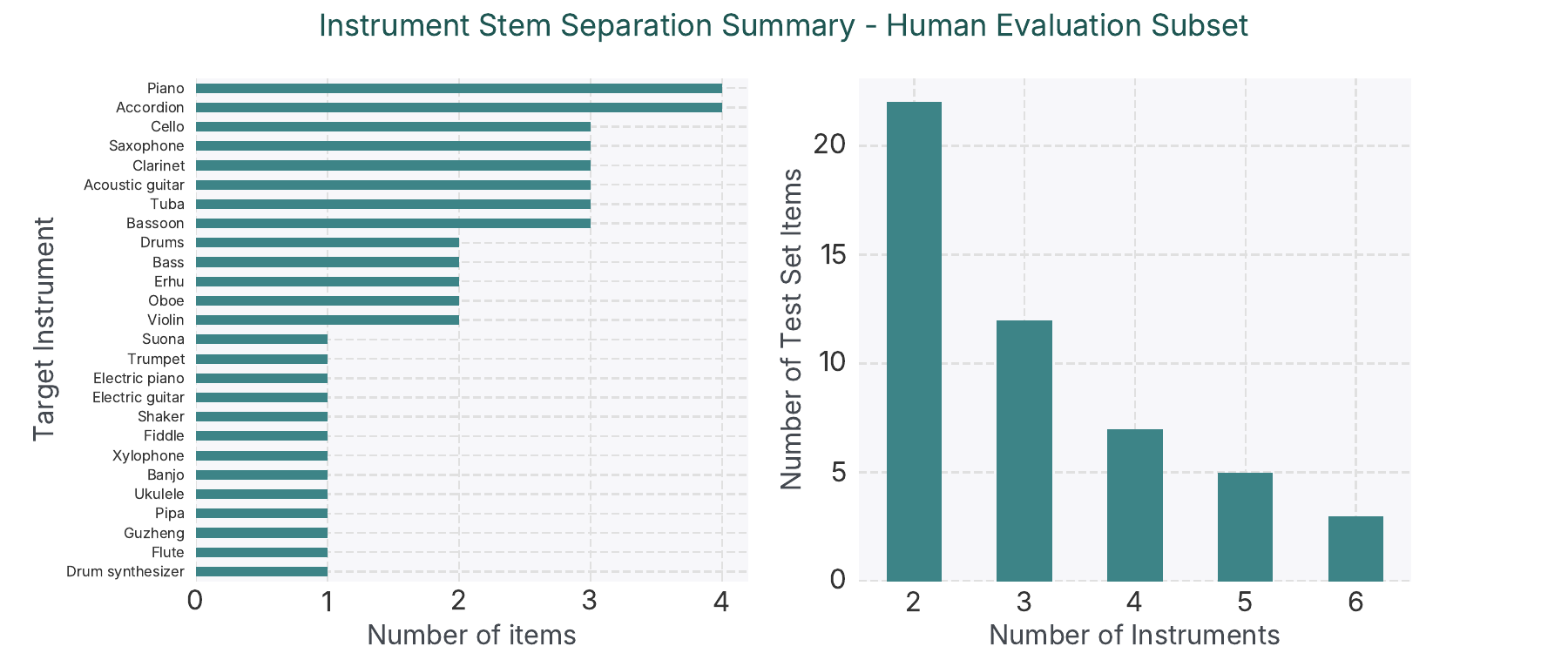}
\caption{\label{fig:samaudiobench_summary_inst} Statistics on the instrument stem separation task in \samaudiobench. \textbf{Left}: histogram of target instruments. \textbf{Right}: histogram of the number of instruments playing in the video for each item in the test set. \textbf{Top}: Summary for all samples in \samaudiobench, \textbf{bottom}: summary for samples in the subset used for human evaluations.}
\end{figure}

%% file: judgeappendix.tex
\newpage
\section{SAM Audio Judge}
\subsection{Data Annotation protocol\label{appx:judgeprotocol}}
Below we present the content of the instructions presented to annotators when annotating the SAM Audio Judge scores.
\input{judgeprotocol}

\subsection{High quality raters selection in SAM Audio Judge\label{appx:judgeraters}}
\input{judgerater}

\input{geminiprompt}

\subsection{SAJ for Automatic and Fine-grained Performance analysis\label{appx:judgediff}}
\input{judgediff}

\input{tables/judge_diff}

%% file: judgeprotocol.tex
\subsection*{Audio Separation}
You will be evaluating models that perform audio separation. \paragraph{What is audio separation?}
When using an audio separation model, users provide the model with either an audio track or a video with sound, and descriptions of which parts of the sound they want the model to extract (for example ``guitar'' or ``dog barking''). The model then extracts the part of the audio that the user requests without any other sounds.
\subsection*{Key Terms}
\begin{description}
  \item[Prompt] The user description of the target audio (e.g., ``dog barking'').
  \item[Source audio] The original video or audio provided to the model.
  \item[Target sounds] The portions of the source audio that the user is requesting.
  \item[Non-target sounds] All other sounds in the source audio not requested by the user.
  \item[Extracted audio] The audio that the model produces from the source audio and the prompt.
\end{description}
\subsection*{What is ``distortion''?}
When we say a sound is ``distorted'', we are referring to how a particular sound may have changed from the way it sounds in the source audio.
\paragraph{Distortions}
Distortions are modifications of sounds that originate from a discernable source in the source audio. These modifications are not the result of mixing with other sounds present in the source audio.
If you can't tell if a sound is a distortion or simply the result of mixing with other sounds in the source audio, you may consider it to be a distortion.
\paragraph{Examples of distortions}
\begin{itemize}
  \item Different levels of ``bass'' or ``treble'': the target sound may sound like it is more or less ``bass''-y, or there are more high pitched sounds.
  \item Pitch differences: if the sound is higher or lower pitched than in the source audio.
  \item Static / artifacts / other noises with no discernable origin in the source audio: crackling, static, popping noises, etc. that are not present in the source audio, or are hard to tell.
  \item Garbling: if, for instance, speech sounds garbled when it wasn't in the source audio.
\end{itemize}
\subsection*{Overview}
\paragraph{Model inputs and outputs that you will be presented with}
In this evaluation you will be presented with:
\begin{itemize}
  \item The source audio (and possibly video).
  \item The prompt, which is a text description of the target sound(s) that the user wants the model to extract from the source audio.
  \item The extracted audio from two separate models.
\end{itemize}
First you will be asked several questions about how well the extracted audio of each model matches the target sounds that the user requested in their prompt.
After answering these questions, you will then give the extracted audio an overall score based on its:
\begin{enumerate}
  \item Faithfulness to the prompt: how well did the model follow the user's instructions about which sounds to extract?
  \item Faithfulness to the sound of the target audio: does the extracted target sound actually sound like the target sound in the source audio?
\end{enumerate}
\subsection*{What you will be annotating}
First, we will ask three questions (Q1-Q3) to assess whether the target sounds specified in the prompt are indeed present in the source audio, and whether any non-target sounds are also included. 

If the answers to Q1 and Q3 are both “Yes”, you will then be asked the four questions (Q4-Q8) to assess the difficulty of the audio separation task. \textit{Note: Q1-Q8 have nothing to do with model outputs!}

After that, we will ask you questions (Q9-Q12) to assess the quality of the model’s extracted audio.
\textit{Note: Q9-Q12 need model outputs!}

\begin{description}
  \item[Q1.] Is it clear to you which sounds in the source audio the prompt is referring to?
  \begin{itemize}
    \item Yes
    \item No
  \end{itemize}
\end{description}

\begin{description}
  \item[Q2.] (Follow-up to Q1 if answer is No)
  \begin{itemize}
    \item There are no sounds in the source audio that match the prompt description.
    \item I'm not sure what the prompt is supposed to sound like.
    \item The prompt is ambiguous: there are several sounds present in the source audio that the prompt could be referring to. It is not clear which one is the target sound.
  \end{itemize}
\end{description}

\begin{description}
  \item[Q3.] Does the source audio contain any non-target sounds not requested in the prompt?
  \begin{itemize}
    \item Yes
    \item No
    \item Possibly, I cannot tell
  \end{itemize}
\end{description}

If the answer to Q3 is “Yes”,  you will then be asked the following questions:

\begin{description}
  \item[Q4.] How many non-target sounds are there in the source audio? \\
  Guidance: A non-target sound is any clearly audible sound that is not requested in the prompt. Multiple instances of the same type of sound (e.g., repeated coughs) count as one. Sounds that are not identifiable but clearly come from different sources (e.g., speech vs. machinery) should be counted separately. Completely unidentifiable or ambiguous background sounds should be grouped and counted as a single non-target sound.
  \begin{itemize}
    \item 1
    \item 2
    \item 3
    \item 4
    \item 5+
  \end{itemize}
\end{description}

\begin{description}
  \item[Q5.] How much do the target sound(s) overlap with the non-target sounds?
  \begin{itemize}
    \item When the target sounds are audible, no non-target sounds are audible at the same time.
    \item There is some slight overlap (< 25$\%$) of target sounds with non-target sounds.
    \item For about half of their duration the target sounds overlap with one or more non-target sounds.
    \item Target sounds are overlapped by one or more target sounds for most of their duration ($\sim$75$\%$)
    \item All target sounds overlap completely with one or more non-target sounds.
    \item The target sounds are ambiguous, so I cannot tell.
  \end{itemize}
\end{description}

\begin{description}
  \item[Q6.] How loud are the target sound(s) compared to the non-target sounds? \\
  Guidance: Rate the overall perceived loudness of the target sound(s) relative to the non-target sounds throughout the audio. If the relative loudness varies over time, focus on the dominant trend or the most prominent portion of the target sound(s).
  \begin{itemize}
    \item Much louder than non-target sounds.
    \item Slightly louder than non-target sounds.
    \item About the same loudness.
    \item Slightly quieter than non-target sounds.
    \item Much quieter than non-target sounds.
    \item The target sounds are ambiguous, so I cannot tell.
  \end{itemize}
\end{description}

\begin{description}
  \item[Q7.] How confusing non-target sounds are with the target sound(s)?  \\
  Guidance: Confusion may arise when non-target sounds share similar acoustic characteristics (e.g., pitch, timbre, timing) with the target, or occur at similar moments. For example, If the prompt is "a baby crying" and there is a cat meowing in the background that briefly resembles a baby cry, that would be slightly or moderately confusing depending on how similar it sounds.
  \begin{itemize}
    \item Not confusing (Non-target sounds are completely distinct from the target sounds.)
    \item Slightly confusing (Non-target sounds are mostly distinguishable but may briefly resemble the target sounds.)
    \item Moderately confusing (Some parts of the non-target sounds resemble the target sounds, causing occasional confusion.)
    \item Very confusing (Non-target sounds are often hard to distinguish from the target sounds.)
    \item Extremely confusing (Non-target sounds are nearly indistinguishable from the target sounds.)
  \end{itemize}
\end{description}

\begin{description}
  \item[Q8.] Considering all the factors listed in Q4-Q7, how difficult do you think it is to extract the target sound(s) from the source audio?  \\
  Guidance: Confusion may arise when non-target sounds share similar acoustic characteristics (e.g., pitch, timbre, timing) with the target, or occur at similar moments. For example, If the prompt is "a baby crying" and there is a cat meowing in the background that briefly resembles a baby cry, that would be slightly or moderately confusing depending on how similar it sounds.
  \begin{itemize}
    \item 1: Very easy (Target sounds are loud, distinct from and barely overlap with the non-target sounds. There are very few non-target sounds)
    \item 2: Easy
    \item 3: Medium
    \item 4: Hard
    \item 5: Very hard (Target sounds are quiet, similar to and significantly overlap with the non-target sounds. There are many non-target sounds)
  \end{itemize}
\end{description}
\textit{The following questions need model outputs!}
Next,  you will then be asked the following question:
\begin{description}
  \item[Q9.] What portion of the target sound(s) is/are present in the extracted audio, regardless if other non-target sounds or distortions are present? Consider portion in terms of (1) duration and (2) number of target sounds. 
  \begin{itemize}
    \item All
    \item Most (a majority of the target sound is present in extracted audio but some is missing)
    \item Some (half of the target sound is present in extracted audio but half is missing)
    \item Few (a minority of the target sound is present in extracted audio but most is missing)
    \item None
  \end{itemize}
\end{description}
If your answer to “what portion of the target sounds is/are present in the extracted audio” is not “None”, we will also ask about how similar the target sounds in the extracted audio sound to the way they sound in the source audio.
\begin{description}
  \item[Q10.] For target sounds that are present in the extracted audio, how similar do they sound to how they sounded in the source audio?
  \begin{itemize}
    \item Exactly the same
    \item Minor distortions / artifacts
    \item Moderate distortions / artifacts
    \item Serious distortions / artifacts
    \item Sounds completely different, barely recognizeable
  \end{itemize}
\end{description}
Afterwards, you will be asked
\begin{description}
  \item[Q11.] How well do you feel the model removed non-target sounds from the extracted audio?
  \begin{itemize}
    \item Perfectly: No non-target sounds are present in the extracted audio.
    \item Almost perfectly: Most non-target sounds in the source audio are filtered out, with only minimal non-target sounds remaining.
    \item Reasonably: Some non-target sounds are removed from the source audio, but a noticeable portion remains in the extracted audio.
    \item Poorly: Only limited filtering occurs; most non-target sounds in the source audio are still present in the extracted audio.
    \item Very poorly: No non-target sounds are removed, and additional non-target sounds may be introduced during extraction.
  \end{itemize}
\end{description}

\subsection*{Q12. Overall score}
After you provide responses to the above questions, we will ask you to provide an overall score for how well the model performed, on a scale from 1 to 5. This should ideally incorporate and be consistent with all of the information you provided above but just as importantly we would like you to use your own sense of judgement.
\begin{table}[h]
\centering
\begin{tabularx}{\textwidth}{l X}
\toprule
\textbf{Score} & \textbf{Criteria} \\
\midrule
5: Perfect & All target sounds are present in their entirety, and sound identical to the way they sound in the source audio. No non-target sounds present in extracted audio. \\
4: Good & Only minor issues with extracted audio (e.g., minor portions of target audio may be missing or slightly distorted, a very small amount of non-target sounds may be present). \\
3: OK & Some serious issues with extracted audio: target audio is maybe half present and/or non-target sounds are about half present. \\
2: Poor & Many serious issues with extracted audio: some target audio in the extracted audio but heavily distorted, missing, and several non-target sounds present. \\
1: Terrible & Extracted audio is completely incorrect; none of the target audio is in the extracted audio, or, if present, none of the non-target audio is filtered. \\
\bottomrule
\end{tabularx}
\end{table}

\subsection*{Frequently asked questions}
\begin{description}
  \item[Distinct types of sounds:] you may be asked to provide input on the number of distinct target or non-target sounds. In some cases this may be simple (the audio has a dog barking and a cat meowing --- there are two distinct sounds here), however in many cases this task may be more ambiguous. For instance, in an audio where two people are talking but there is a rock song playing in the background --- is that three sounds: one for each person speaking and one for the music? Or do we need to count each instrument that makes up the rock song as a distinct sound?
  \begin{itemize}
    \item If the prompt asks for a specific instrument or song component you may consider each instrument of the music in the source audio to be an individual target or non-target sound.
    \item Background noise: for sounds of an environment (street sounds, honking horns, cars passing etc.; or sounds of a cafe) that are not easily discernible, these can be considered a single non-target ``background'' sound.
  \end{itemize}
  \item[Speech prompts:] you may see some prompts like “one of the men speaking”, “one of the women speaking”, “one of the people talking”. In these cases, please focus on a single, consistent speaker throughout the audio.
    \begin{itemize}
    \item For example, if the prompt is “one of the men speaking” and the extracted audio contains two men speaking, the overall score should be rated low due to inconsistency with the prompt.
    \item As there may be multiple candidates, you could choose “The target sounds are ambiguous, so I cannot tell.” for Q5 and Q6.
  \end{itemize}
  \item[Volume:] is also an important factor that influences the scores. 
    \begin{itemize}
    \item For Q10, changing the volume of the target sound(s) should be considered as a distortion.
    \item For Q11, reducing the volume of non-target sounds should be considered as a valid form of removal.
  \end{itemize}
  \item[Distortion:] If the quality of the extracted audio sounds better than that in the source audio, it also should be considered as a distortion when answering Q10.
\end{description}

%% file: judgerater.tex
We design a rater qualification program to ensure the selection of high-quality annotators. Specifically, we curate a 40-sample golden set, with scores labeled by experts and treated as ground truth. We then invite outsourced vendors to annotate this golden set and qualify their annotators using a Bayesian rater modeling approach. This model estimates a posterior distribution over each annotator’s confusion matrix, capturing their reliability and bias across rating values. From this posterior, we derive the expected Mean Absolute Deviation (MAD) of each annotator’s scores, which naturally accounts for partially completed tasks, item-score uncertainty, and imbalances in score distributions.
Annotators with the lowest expected MAD on the overall score are selected as qualified raters. While we do not directly compare against expert annotations, we use them as a spot check to ensure that the selected annotators exhibit reasonable consensus and consistent rating behavior. In the end, we successfully recruit 128 qualified raters, representing diverse subjective perspectives from the general public.

%% file: geminiprompt.tex
\subsection{Gemini-2.5-pro Prompts for Judge Evaluation\label{appx:gemini}}

\begin{tcolorbox}[
  enhanced,
  colback=white,
  colframe=black!35,
  boxrule=0.6pt,
  arc=3pt,
  left=6pt,right=6pt,top=6pt,bottom=6pt
]
\ttfamily
You will be evaluating models that perform audio separation.

When using an audio separation model, users provide the model with either an audio track, and descriptions of which parts of the sound they want the model to extract (for example ``guitar'' or ``dog barking''). The model then extracts the part of the audio that the user requests without any other sounds.

The user description of the target audio (e.g. ``dog barking'' or the visually highlighted dog in the video) is referred to as the prompt. The original video or audio provided to the model is the source video/audio. The portions of the source audio that the user is requesting are known as the target sounds. All other sounds in the source audio not requested by the user are known as non-target sounds. The audio that the model produces from the source audio and the prompt is referred to as the extracted audio.

\textbf{Task:}
\begin{itemize}
  \item The user provides:
  \begin{enumerate}
    \item A prompt, which is a text description of the target sound(s) that the user wants the model to extract from the source audio
    \item Two audio files:
    \begin{itemize}
      \item The source audio (contains the target and other sounds)
      \item The extracted audio from the separation model
    \end{itemize}
  \end{enumerate}
\end{itemize}

Please evaluate the extracted audio compared with the mixture and the target description, along the following four dimensions:
\begin{itemize}
  \item \textbf{Recall}: What portion of the target sound(s) is/are present in the extracted audio, regardless if other non-target sounds or distortions are present? Consider portion in terms of (1) duration and (2) number of target sounds.
  \item \textbf{Precision}: For target sounds that are present in the extracted audio, how similar do they sound to how they sounded in the source audio?
  \item \textbf{Faithfulness}: How well do you feel the model removed non-target sounds from the extracted audio?
  \item \textbf{Overall}: Please provide an overall score for the model’s performance.
\end{itemize}

For each dimension, provide a score from \textbf{1 (very poor) to 5 (excellent)}:
\begin{itemize}
  \item \textbf{1: Terrible}: Extracted audio is completely incorrect; none of the target audio is in the extracted audio, or, if present, none of the non-target audio is filtered.
  \item \textbf{2: Poor}: Many serious issues with extracted audio: some target audio in the extracted audio but heavily distorted, missing, and many non-target sounds present.
  \item \textbf{3: OK}: Some serious issues with extracted audio: target audio is maybe half present and/or non-target sounds are about half present.
  \item \textbf{4: Good}: Only minor issues with extracted audio (e.g. minor portions of target audio may be missing or slightly distorted, a very small amount of non-target sounds may be present).
  \item \textbf{5: Perfect}: All target sounds are present in their entirety, and sound identical to the way they sound in the source audio. No non-target sounds present in extracted audio.
\end{itemize}

We also provide 5 reference examples that illustrate what score 1,2,3,4,5 means.
\end{tcolorbox}

%% file: judgediff.tex
We could use SAJ to automatically sample evaluation cases at different difficulty levels for any target audio concept list, enabling a fully automatic and fine-grained performance analysis pipeline. To support this, we train a text-prompted SAJ model to predict the intrinsic difficulty of a separation task. Because human annotations show that levels 4 and 5 are rare, we merge them into a four-level scale (1–4) and balance the dataset across these levels.

Unlike the standard SAJ model, which predicts separation performance using both mixture and output audio, the difficulty model only takes the mixture audio and text prompt as input, allowing it to estimate task difficulty before running any separation system.

We apply the predicted difficulty levels to the SAJ test set and compute human ratings for the corresponding SAM Audio outputs. As shown in Table~\ref{tab:diff}, human-annotated performance monotonically decreases as difficulty increases: Level 1 cases are easiest, while Level 4 shows clear degradation. This enables automatic curation of evaluation subsets at different difficulty levels, supporting fine-grained, concept-specific robustness analysis.

%% file: tables/judge_diff.tex
\begin{table}[h!]
\centering

\adjustbox{max width=.5\textwidth}{%
\begin{tabular}{ccccc}
\toprule
\textbf{Difficulty Level} & \textbf{Overall} & \textbf{Recall} & \textbf{Precision}& \textbf{Faithfulness}\\
\midrule
1&3.716&4.084& 3.984& 3.963\\
2&3.327&3.995&3.537&3.798\\
3&3.272&4.022&3.326&3.807\\
4&2.894&3.791&3.024&3.528\\
\bottomrule
\end{tabular}
}
\caption{Human-rated separation quality across different task difficulty levels.}
\label{tab:diff}
\end{table}

%% file: ablation_study.tex
\newpage
\section{\samaudio{} Ablation Study}
\label{sec:sam_audio_ablation_study}

\subsection{Effect of Model Scale}
Tables~\ref{tab:abl_scale_text} and \ref{tab:abl_scale_visual}, together with Figure~\ref{fig:nwr_scale}, compare models of three sizes—500M, 1B, and 3B parameters.  
Across the scale comparison, we disable span prediction.
Overall, the 3B model achieves the strongest performance across most tasks, though in certain cases such as general SFX separation it performs on par with or slightly below the smaller models.

Scaling model capacity provides the greatest benefit in specialized domains.  
For instrument separation, for example, \samlarge{} achieves substantial gains over the smaller variants: on instrument-in-the-wild separation, it outperforms \sammid{} by 23\% NWR and \samsmall{} by 20\%.  
These results suggest that larger models better capture the fine-grained acoustic cues required for high-fidelity separation in structured domains such as musical instruments, while smaller models remain competitive for broader sound categories.

\input{tables/abl_scale_ft}

\subsection{Effect of Auxiliary Loss}

\input{tables/abl_loss}

Table~\ref{tab:res_abl_loss} reports the effect of incorporating the representation alignment loss during \samaudio{} pre-training. We compare two 3B pre-trained checkpoints: (1) a baseline model without the auxiliary loss, and (2) a model trained with the auxiliary loss using $\lambda = 1.0$.  
We evaluate both checkpoints on the general SFX test set under text- and visual-prompted separation, as the pre-trained model alone underperforms on specialized domains such as speech and music.

As shown in Table~\ref{tab:res_abl_loss}, adding the AED-based alignment objective yields consistent improvements in both settings, with a larger gain in the text-prompted case (over 20\% relative improvement in text alignment) compared to visual prompting ($\sim$5\% relative improvement).  
This ablation is performed only during pre-training, as we do not apply the auxiliary loss in fine-tuning. We hypothesize that the benefit arises from the noisier audio target in pre-training corpus, where the alignment loss helps the model learn intermediate semantic representations beneficial for separation.

\subsection{Effect of fine-tuning}

\input{tables/abl_pt_ft}

Table~\ref{tab:pt-ft-results} compares the 3B model after pre-training only with the same model further fine-tuned on curated separation data. In addition to standard metrics, we also report the PC score from Audiobox-Aesthetics~\citep{tjandra2025meta}, which serves as a proxy for audio cleanness.

Overall, fine-tuning leads to consistent gains, though the magnitude varies across tasks.  
For text-prompted separation, the largest improvements appear in instrument extraction, speech extraction, and speaker separation. These tasks benefit from the availability of high-quality datasets containing professionally recorded stems, providing clean and reliable supervision. By contrast, improvements in general sound event separation and music separation are smaller. Large-scale pre-training already covers a wide distribution of sound events and mixtures, including many with background music, which leaves less room for fine-tuning to provide additional gains.

For visual-prompted separation, the visual alignment score (IB) remain relatively stable for general SFX, mirroring the trend seen in text-prompted separation.  Fine-tuning data provide broader coverage for music and speech videos than for general sound events. Large-scale audio–visual pre-training already establishes strong correspondences between visual regions and audio, explaining the smaller incremental benefit for SFX.

Finally, we observe a substantial improvement in audio cleanness across tasks, according to the PC metric. Fine-tuned models consistently produce cleaner separated audios with fewer artifacts, owing to the clean audio targets used during fine-tuning.

\subsection{Effect of using pseudo-labeled audio stem data}
Table~\ref{tab:res_data_abl_pl} shows a comparison of using or not using pseudo-labeled audio data in the fine-tuning stage. 
Details of the pseudo-labeled data are summarized in Table~\ref{tab:audio_pl_data_source}. The baseline model is trained using only fully real triplets and synthetic mixtures. As shown in Table~\ref{tab:res_data_abl_pl}, incorporating pseudo-labeled data yields consistent gains for both text and visual prompting. Notably, the largest improvements are observed in AES-PC for general sound, indicating that pseudo-labeled audio helps the model learn to produce cleaner separation stems.

\input{tables/abl_text_pl}

%% file: tables/abl_scale_ft.tex
\begin{table*}[h]
 \setlength{\tabcolsep}{3pt}
    \centering
    \adjustbox{max width=1\textwidth, center}{%
    \begin{tabular}{l 
    cc>{\columncolor{gray!15}}c 
    cc>{\columncolor{gray!15}}c 
    cc>{\columncolor{gray!15}}c 
    cc>{\columncolor{gray!15}}c 
    cc>{\columncolor{gray!15}}c 
    cc>{\columncolor{gray!15}}c 
    }
    \toprule
     \multicolumn{1}{c}{} &
     \multicolumn{3}{c}{\textbf{General SFX}} &
     \multicolumn{3}{c}{\textbf{Speech}} &
     \multicolumn{3}{c}{\textbf{Speaker}} &
     \multicolumn{3}{c}{\textbf{Music}} &
     \multicolumn{3}{c}{\textbf{Instr(wild)}} &
     \multicolumn{3}{c}{\textbf{Instr(pro)}} \\
    \cmidrule(lr){2-4} \cmidrule(lr){5-7} \cmidrule(lr){8-10}
    \cmidrule(lr){11-13} \cmidrule(lr){14-16} \cmidrule(lr){17-19}
    \textbf{Model} &
    SAJ & CLAP & OVR &
    SAJ & CLAP & OVR &
    SAJ & CLAP & OVR &
    SAJ & CLAP & OVR &
    SAJ & CLAP & OVR &
    SAJ & CLAP & OVR \\
    \midrule

\samsmall & \textbf{4.25} & 0.30 & \textbf{3.62} 
          & 4.55 & \textbf{0.35} & 3.99 
          & 3.89 & \textbf{0.17} & 3.12 
          & 4.32 & \textbf{0.28} & 4.11 
          & 4.27 & \text{0.27} & 3.56 
          & 4.78 & \textbf{0.30} & 4.24 \\

\sammid & \text{4.23} & 0.28 & 3.28 
        & \textbf{4.61} & \text{0.33} & \textbf{4.25} 
        & \text{3.94} & 0.15 & \text{3.57} 
        & 4.26 & 0.27 & 3.87 
        & \text{4.33} & \text{0.29} & \textbf{3.66} 
        & 4.78 & \textbf{0.30} & 4.27 \\

\samlarge & 4.11 & \textbf{0.31} & 3.50 
          & 4.59 & 0.33 & 4.03 
          & \textbf{4.08} & \textbf{0.17} & \textbf{3.60} 
          & \textbf{4.30} & \textbf{0.28} & \textbf{4.22} 
          & \textbf{4.45} & \textbf{0.30} & \textbf{3.66} 
          & \textbf{4.83} & 0.28 & \textbf{4.49} \\

    \bottomrule
    \end{tabular}
    } 
    \caption{Comparison of \samaudio{} of different scales in text prompting. --: not applicable. OVR: overall subjective score.}
    \label{tab:abl_scale_text}

\end{table*}

\begin{table*}[h]
\centering

\adjustbox{max width=.5\textwidth}{%
\begin{tabular}{l c>{\columncolor{gray!15}}c c>{\columncolor{gray!15}}c c>{\columncolor{gray!15}}c}
\toprule
 & \multicolumn{2}{c}{\textbf{General SFX}}
 & \multicolumn{2}{c}{\textbf{Speaker}}
 & \multicolumn{2}{c}{\textbf{Instr (wild)}} \\
\cmidrule(lr){2-3}\cmidrule(lr){4-5}\cmidrule(l){6-7}
\textbf{Model} & IB & OVR & IB & OVR  & IB & OVR  \\
\midrule

\samsmall & 0.24 & 2.62 & \text{0.23} & 2.79 & 0.21 & 2.25 \\
\sammid & \textbf{0.25} & \textbf{2.63} & 0.24 & \textbf{3.25} & \text{0.22} & \textbf{2.76} \\
\samlarge & \textbf{0.25} & 2.61 & \textbf{0.24} & 2.95 & \textbf{0.24} & 2.58 \\

\bottomrule
\end{tabular}
}
\caption{Comparison of \samaudio{} of different scales in visual prompting. --: not applicable. OVR: overall subjective score.}
\label{tab:abl_scale_visual}

\end{table*}

\begin{figure*}[htp]
    \centering
    \includegraphics[width=\linewidth]{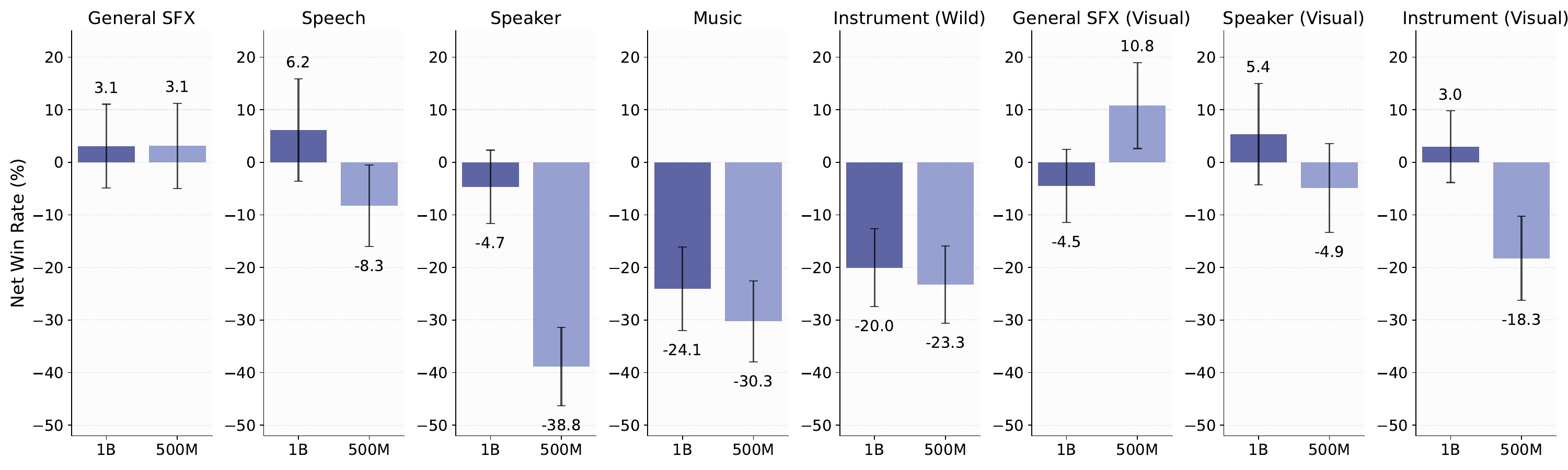}
    \caption{Net Win Rate (\%) of \sammid and \samsmall against \samlarge}
    \label{fig:nwr_scale}
\end{figure*}

%% file: tables/abl_loss.tex
\begin{table}[h!]
\centering
\adjustbox{max width=.5\textwidth}{%
\begin{tabular}{l cc c}
\toprule
 & \multicolumn{2}{c}{\textbf{General SFX (Text)}} & \textbf{General SFX (Visual)} \\
\cmidrule(lr){2-3} \cmidrule(lr){4-4}
Auxiliary target & SAJ & CLAP & IB \\
\midrule
None & 2.48 & 0.24 & 0.17 \\
Target AED & \textbf{3.18} & \textbf{0.29} & \textbf{0.18} \\
\bottomrule
\end{tabular}
}
\caption{Effect of AED-based representation alignment loss in pre-training}
\label{tab:res_abl_loss}
\end{table}

%% file: tables/abl_pt_ft.tex
\begin{table}[t]
\centering
\small
\setlength{\tabcolsep}{5pt}
\adjustbox{max width=1.0\textwidth}{%

\begin{tabular}{llccccccccc}
\toprule
 & & \multicolumn{6}{c}{\textbf{Text}} & \multicolumn{3}{c}{\textbf{Visual}} \\
\cmidrule(lr){3-8}\cmidrule(l){9-11}
\textbf{Separation Task} & \textbf{Stage} &
General SFX & Speech & Speaker & Music & Instr(wild) & Instr(pro) &
General SFX & Speaker & Instr(wild) \\
\midrule
SAJ ($\uparrow$) & PT
  & 3.93 & 2.90 & 3.28 & 4.14& 3.17 & 3.36
  & - & - & - \\
& FT
  & \textbf{4.14} & \textbf{4.55}& \textbf{4.07}& \textbf{4.38} &\textbf{4.48} &\textbf{4.82}
  & - & - & - \\
  \midrule
CLAP ($\uparrow$)  & PT
  & \textbf{0.31} & 0.28 & \textbf{0.23} & \textbf{0.31}& 0.21 &0.15
  & - & - & - \\
   & FT
  &0.30 & \textbf{0.36}& 0.21& \textbf{0.31} & \textbf{0.30}& \textbf{0.29}
  & - & - & - \\
  \midrule
IB ($\uparrow$)  & PT
  & - & - & - & - & - & -
  & \textbf{0.24}& 0.22 & 0.20\\
   & FT
  & - & - & - & - & - & -
  & \textbf{0.24}& \textbf{0.24}&\textbf{0.22}\\
  \midrule
AES-PC ($\downarrow$) & PT
  &2.57 & 2.91 & 2.98 & 4.53& 3.54&4.72
  & 3.09& 3.49 & 3.69\\
& FT
  & \textbf{2.08} & \textbf{1.89} & \textbf{1.94} & \textbf{4.44} &\textbf{3.16} & \textbf{3.24}
  & \textbf{2.22}& \textbf{2.20}& \textbf{3.26} \\
\bottomrule
\end{tabular}
}
\caption{Comparison of pre-trained vs fine-tuned results across audio separation tasks. --: not applicable.}
\label{tab:pt-ft-results}

\end{table}

%% file: tables/abl_text_pl.tex
\begin{table}[h!]
\centering
\small
\setlength{\tabcolsep}{5pt}
\adjustbox{max width=1.0\textwidth}{%

\begin{tabular}{llccccccccc}
\toprule
 & & \multicolumn{6}{c}{\textbf{Text}} & \multicolumn{3}{c}{\textbf{Visual}} \\
\cmidrule(lr){3-8}\cmidrule(l){9-11}
\textbf{Separation Task} & \textbf{Setting} &
General SFX & Speech & Speaker & Music & Instr(wild) & Instr(pro) &
General SFX & Speaker & Instr(wild) \\
\midrule
SAJ ($\uparrow$) & w/o PL
  & 4.02 & 4.50 & 4.06 & 4.34& 4.34 & 4.80
  & - & - & - \\
& PL
  & \textbf{4.14} & \textbf{4.55}& \textbf{4.07} & \textbf{4.38} &\textbf{4.48} &\textbf{4.82}
  & - & - & - \\
  \midrule
CLAP ($\uparrow$)  & w/o PL
  & \textbf{0.30} & 0.34 & \textbf{0.21} & \textbf{0.31}& 0.29 &0.28
  & - & - & - \\
   & PL
  &\textbf{0.30} & \textbf{0.36}& \textbf{0.21}& \textbf{0.31} & \textbf{0.30}& \textbf{0.29}
  & - & - & - \\
  \midrule
 IB ($\uparrow$)  & w/o PL
  & - & - & - & - & - & -
  & 0.23& 0.21 & \textbf{0.22}\\
   & PL
  & - & - & - & - & - & -
  & \textbf{0.24}& \textbf{0.24}&\textbf{0.22}\\
  \midrule
AES-PC ($\downarrow$) & w/o PL
  &2.36 & 1.95 & 2.02 & \textbf{4.38}& 3.19&\textbf{3.24}
  & 2.28 & 2.29 & \textbf{3.25}\\
& PL
  & \textbf{2.08} & \textbf{1.89} & \textbf{1.94} & 4.44 &\textbf{3.16} & \textbf{3.24}
  & \textbf{2.22} & \textbf{2.20}& 3.26 \\
\bottomrule
\end{tabular}
}
\caption{Effect of using pseudo-labeled audio stem data}
\label{tab:res_data_abl_pl}

\end{table}